\documentclass[
  aps,
  prx,
  twocolumn,
  superscriptaddress,
  longbibliography,
  amsmath,amssymb,
]{revtex4-2}

\usepackage{graphicx}
\graphicspath{{Figures/}}
\usepackage{bm}
\usepackage{dcolumn}
\usepackage{booktabs}
\usepackage{siunitx}
\usepackage[table]{xcolor}
\usepackage{hyperref}
\hypersetup{colorlinks=true, citecolor=blue, linkcolor=blue, urlcolor=blue}
\usepackage{xurl}% break long URLs anywhere to avoid column overflow
\usepackage{placeins}% \FloatBarrier to keep appendix figures before \bibliography
\usepackage{capt-of}% \captionof for the inline (non-floating) appendix figures
\makeatletter
\@booleanfalse\footinbib@sw
\let\frontmatter@footnote@produce\frontmatter@footnote@produce@footnote
\makeatother

\begin{document}

\title{Learning Standard Model structure from LHC data with Riemannian flow matching}

\author{Midori Kato}
\email{midori.kato.73@gmail.com}
\author{Kevin A.\ Urqu\'ia-Calder\'on}
\email{kevin.urquia@nbi.ku.dk}
\author{Inar Timiryasov}
\email{inar.timiryasov@nbi.ku.dk}
\author{Oleg Ruchayskiy}
\email{oleg.ruchayskiy@nbi.ku.dk}
\affiliation{Niels Bohr Institute, University of Copenhagen, Jagtvej 155A, DK-2200, Copenhagen, Denmark}

\date{\today}
\begin{abstract}
In this work we demonstrate that a single transformer-based
generative model can capture Standard Model structure spanning five
decades of invariant mass, from the sub-GeV regime to the TeV
continuum, a range that no single Monte Carlo sample covers.
To achieve this we design \textsc{ShellFlow}, a Riemannian conditional flow
matching model that, given the recorded event composition,
generates each particle on its on-shell
manifold. Its only physics priors are the on-shell condition and
the invariant-mass formula. The model is trained
on $\sim 10^{9}$ real $pp$ collision events from the ATLAS Open
Data 13~TeV release and told nothing else. From a single training
run, the model learns to reproduce all of the following:
intra-particle kinematics, the dilepton resonances ($J/\psi$,
$\Upsilon$, $Z$) at their PDG positions, the leptonic Weinberg
angle, the $W$ and top-quark masses, and inter-particle
correlations that enter no training objective. A substantial fraction of the Standard Model is thus learnable directly from recorded collision data.
\end{abstract}

\maketitle

%======================================================================
\section{Introduction}
\label{sec:introduction}

\subsection*{Motivation and contribution}

The Large Hadron Collider (LHC) produces enormous datasets, driven both by the need to access rare physics processes and by the five-sigma significance threshold that the field imposes on discovery claims. 
Interpretation of these datasets currently relies on
Monte Carlo (MC) event generators such as
\textsc{Pythia}~\cite{Sjostrand:2014zea},
\textsc{Herwig}~\cite{Bellm:2015jjp},
\textsc{Sherpa}~\cite{Sherpa:2019gpd},
\textsc{MadGraph5\_aMC@NLO}~\cite{Alwall:2014hca, Frixione:2002ik},
\textsc{Powheg}~\cite{Alioli:2010xd},
\textsc{Whizard}~\cite{Kilian:2007gr}, and
\textsc{EvtGen}~\cite{Lange:2001uf}, tuned per analysis to a
specific physics channel or BSM signal hypothesis.
Detector-level MC simulation is computationally expensive, and the
corresponding analyses target one channel at a time, making it
difficult to assess which of the many proposed BSM scenarios is in
fact compatible with the data.

Machine learning has been applied to collider analyses for decades,
beginning with multilayer perceptrons as event-level
classifiers~\cite{Denby:1988hw} and growing into the graph neural
networks and transformers that now drive state-of-the-art
tracking~\cite{Shlomi:2020gdn, ExaTrkX:2021abe}, reconstruction, and
jet tagging~\cite{Qu:2022mxj}. A continuously updated
bibliography of this field is maintained in
Ref.~\cite{Feickert:2021livingreview}.

Three directions of current research are directly relevant for the
present work. The first is generative modelling of collider data.
Deep generative networks were introduced for fast calorimeter
simulation, first with GANs~\cite{Paganini:2018calogan}, then with
normalizing flows~\cite{Krause:2023caloflow} and diffusion
models~\cite{Mikuni:2022caloscore, Amram:2023calodiffusion}. They
have been compared in a community benchmark~\cite{Krause:2024avx}
and now run in production in the ATLAS fast simulation
chain~\cite{ATLAS:2022atlfast3}. The same tools are applied to
event generation itself~\cite{Butter:2019cae, Butter:2023mlreview},
using diffusion models~\cite{Leigh:2024pcjedi, Mikuni:2023fpcd,
Leigh:2023pcdroid}, autoregressive
transformers~\cite{Butter:2025jetgpt}, and most recently flow
matching~\cite{Buhmann:2023epicfm, Favaro:2024rle, Vaselli:2024vrx}. Throughout this
line of work the generator stands in for a simulator, and it is
both trained on and judged against Monte Carlo samples. Closest in
concept to the geometry used here, Ref.~\cite{Bogorad:2026oxa}
constrains the generative trajectories themselves to relativistic
phase space, imposing exact energy--momentum conservation on a
global phase-space manifold, whereas \textsc{ShellFlow} factorises
the geometry into a product of per-particle mass shells.

The second direction is foundation models for particle physics,
which aim to replace the per-analysis pipeline with a single
pre-trained model~\cite{Hallin:2025ywf, Barman:2025wfb}. Early
self-supervision for jets learned representations by comparing
augmented views of the same jet~\cite{Dillon:2021gag}.
Current examples include masked pretraining on particle
sets~\cite{Golling:2024mpm}, generative models that transfer across
tasks~\cite{Birk:2024omnijet, Mikuni:2025omnilearn}, a
foundation-model framework trained on over $10^{9}$
jets~\cite{Bhimji:2025isp}, models of
complete events~\cite{Hsu:2026sww}, sequence models of detector
hits~\cite{Huang:2024voo}, and Lorentz-equivariant
transformers~\cite{Spinner:2024hjm}. Nearly all of these models are
trained on simulated samples, and they are demonstrated through
transfer to downstream tasks, whereas the model of the present work
is a generator judged by reconstruction-level closure against the
data it was trained on. The two closest precedents each
satisfy one half of what we do here: the Aspen Open Jets
project~\cite{Amram:2025aoj} pretrains a foundation model on
$1.8 \times 10^{8}$ individual jets from CMS Open Data (recorded
data, but individual jets rather than complete events), while a
point-cloud diffusion model for heavy-ion
collisions~\cite{OmanaKuttan:2024mwr} generates complete events,
but is trained on transport-model simulation.

The third direction learns from recorded collider data directly,
with as little reference to simulation as possible. Classification
without labels~\cite{Metodiev:2017vrx} established that classifiers
can be trained on mixed samples of real events carrying no
per-event labels, and turned data-vs-data discrimination into a
search strategy~\cite{Collins:2018epr}. Resonant anomaly detection
extends this by fitting generative density models
(ANODE~\cite{Nachman:2020lpy}, CATHODE~\cite{Hallin:2021wme},
CURTAINs~\cite{Raine:2022hht}) to recorded sideband data and
sampling background templates from them in a single search channel.
Simulation-assisted and diffusion-based variants refine the
template construction~\cite{Andreassen:2020nkr, Golling:2022nkl,
Sengupta:2023vtm}. Beyond the resonant setting, normalizing flows
have been used to sharpen autoencoder-based
detection~\cite{Jawahar:2021vyu}, and anomaly-preserving
contrastive embeddings extend model-independent searches to
complete events~\cite{Metzger:2025ecl}.
The programme is organised around community
challenges~\cite{Kasieczka:2021xcg}, is reviewed in
Refs.~\cite{Karagiorgi:2021ngt, Belis:2023mqs}, and has been
deployed on recorded
data by both ATLAS and CMS~\cite{ATLAS:2020iwa, ATLAS:2023ixc,
CMS:2024nsz}. These density models are, however, deliberately
narrow, restricted to a handful of engineered observables, a
single search channel, and the background alone.

The present work occupies the intersection that these three
directions leave open. We are not aware of a previous model that
generates the complete final-state kinematics of reconstructed
events, across all processes at once, after training on recorded
data alone. The event composition, particle types and charges, and
MET are supplied as conditioning inputs
(Sec.~\ref{sec:method_dataset}).

We train \textsc{ShellFlow}, a Riemannian Conditional Flow
Matching~\cite{Chen:RiemannianFM2024} generator with a transformer
backbone~\cite{vaswani2017attention}, directly on the union of the
2-to-4 lepton and 1LMET30 skims of the
ATLAS Open Data 13~TeV $pp$
release~\cite{ATLAS:opendata:1LMET30, ATLAS:opendata:2to4lep},
comprising $\sim 8.2 \times 10^{8}$ recorded collision events. The training distribution is real
collider data, and no Monte-Carlo events are used. The only physics
priors are the on-shell constraint $E^{2} = m^{2} + |\mathbf{p}|^{2}$
and the invariant-mass formula, embedded by routing the flow on a
product of on-shell per-particle manifolds. Both are
Lorentz-invariant statements that hold for every particle and every
process, and they therefore impose no bias toward any specific
channel or specific resonance. The model receives no channel label
and no objective attached to a specific task during training. The
only selections upstream of training are the Open Data skim
definitions and the multiplicity cut of
Sec.~\ref{sec:method_dataset}, neither of which is tied to the
analyses below. 
The same single training run is evaluated across five orders of
magnitude in invariant mass, from the sub-GeV light-meson bound
states through the charmonium and bottomonium families to the
electroweak resonances and the top quark. Downstream observables are extracted
by applying standard ATLAS analysis selections to the trained
generator at evaluation time.

The principal result of this work is that a single self-supervised
generator can internalise nontrivial Standard Model relationships
from real collider data, without a channel label or a per-process
objective. The evidence takes two forms, and the Results section is
organised around them. The first is fidelity to the training
distribution. The single-particle kinematic marginals of the six
reconstructed object types are reproduced with high accuracy,
including the geometric crack regions of the ATLAS detector, and
the opposite-sign same-flavour (OSSF) dilepton invariant-mass
spectrum is captured across five decades in $m_{\ell\ell}$ with the
$J/\psi(1S)$, $\psi(2S)$, $\Upsilon$ family, and $Z$ peaks all
sitting at their Particle Data
Group~\cite{ParticleDataGroup:2024cfk} positions (cf.\ ATLAS
measurements of the same resonances~\cite{ATLAS:2011vbz,
ATLAS:2012aa, ATLAS:2017rue}). The second is agreement on
quantities that no training objective targets. Several
inter-particle observables that appear in neither the primary nor
the auxiliary loss, in particular the dilepton cone distance
$\Delta R_{\ell\ell}$, the dilepton-to-MET azimuthal opening
$\Delta\phi(\ell\ell, \mathrm{MET})$, the transverse mass $m_T$, the
scalar transverse-momentum sum $H_T$, and the leading and sub-leading
lepton $p_T$, agree with the truth distributions to within the
statistical resolution of the validation sample. The supervised
inclusive mass densities constrain at most low-order combinations
of these quantities (the pair mass, for instance, couples
$\Delta R_{\ell\ell}$ to the lepton momenta), so their simultaneous
agreement reflects the model's internal representation of the joint
event structure. The forward-backward asymmetry of the
Drell--Yan continuum yields a Weinberg angle~\cite{Weinberg:1967tq}
$\sin^{2}\theta_w$ that
reproduces the truth-sample leading-order fit within statistical
uncertainty in both flavour channels, with the offset from the
PDG~\cite{ParticleDataGroup:2024cfk} effective leptonic value
attributable to the LO template rather than to the generator, and
event-level transverse-momentum conservation closes to the
precision of the visible-momentum prediction. Together these
closures provide a starting point for searches for physics beyond
the Standard Model and for event-generation tasks that have so far
required dedicated analyses tied to a single channel.

The remainder of the paper presents the results
(Sec.~\ref{sec:results}) before the methodology
(Sec.~\ref{sec:method}), reflecting that the central interest of
this work lies in what the model recovers from real collider data.
Section~\ref{sec:discussion} discusses the design choices that proved
necessary, the limitations of the current model, and the outlook,
and Sec.~\ref{sec:conclusion} concludes.

\subsection*{How to read this paper}

The paper is written for two audiences with different natural
entry points.

\emph{For the reader whose primary interest is the physics.}
We recommend proceeding directly to Sec.~\ref{sec:results}, which
opens with Table~\ref{tab:result_summary}. That table maps every
physics observable examined in this paper to the figure, table, or
equation that establishes its closure between the generative model
output and the real data, and records the supervision channel
through which each structure could have been influenced during
training, separating what is imposed by the model design from what
is recovered from the data.
Figure~\ref{fig:dilepton} is the central result of the paper:
the OSSF dilepton invariant-mass spectrum
from generated events, in which the $J/\psi$, $\Upsilon$, and
$Z$ resonances appear at their PDG positions across five decades
in $m_{\ell\ell}$ although none of these masses was supplied to
the model.

\emph{For the reader whose primary interest is the methodology.}
The technical construction of \textsc{ShellFlow} is collected in
Sec.~\ref{sec:method}: the Riemannian conditional flow matching
objective on which the training rests, the per-particle on-shell
manifolds and their charts that constrain the generator geometry,
the dual-head transformer architecture with its primary and
auxiliary loss terms, and the ATLAS Open Data training pipeline
used to produce the weights whose outputs are analysed in
Sec.~\ref{sec:results}.

%======================================================================
\section{Results}
\label{sec:results}

Table~\ref{tab:result_summary} maps each physics signal probed in
this section to the figure, table, or equation in which it appears.
The body of this section then describes each entry in detail, while
the architectural and training choices that make these closures
possible are deferred to Sec.~\ref{sec:method}.

\begingroup
\definecolor{groupbg}{RGB}{226,232,240}%   slate-100, group header band
\definecolor{constraintfg}{RGB}{30,64,175}% slate-blue, imposed-constraint text
\newcommand{\imposed}[1]{\textcolor{constraintfg}{#1}}%
\newcommand{\grouphdr}[1]{\rowcolor{groupbg}\multicolumn{4}{l}{\textit{#1}}}%
\makeatletter\def\table@hook{\small}\makeatother
\begin{table*}[!tbp]
  \caption{\label{tab:result_summary}%
    \textbf{Result summary.} Each row maps a physics signal probed
    in this section to the figure, table, or equation where the
    corresponding closure (or imposed constraint) is shown.
    \imposed{Rows in blue marked with $^{\ast}$ are constraints
    imposed by the model design}. All other rows are recovered from
    the data. The Supervision column records the training channel
    that could have influenced each structure (see text). Angular
    structure and analysis selections are never supervised.
    Reference values are taken from PDG~\cite{ParticleDataGroup:2024cfk}.
    The full set of truth-vs.-generated validation figures is
    available at \url{https://hep-ssl-webapp.pages.dev/}.}
  \begin{ruledtabular}
    \begin{tabular}{l l l l}
      \textbf{Physics} & \textbf{Content} & \textbf{Result} & \textbf{Supervision} \\
      \hline
      \grouphdr{1. Kinematic constraints \& conservation laws} \\
      \imposed{\quad On-shell relation$^{\ast}$}
         & \imposed{$m^{2} = E^{2} - |\mathbf{p}|^{2}$ \quad (by the manifold design)}
         & \imposed{Eq.~\eqref{eq:on_shell}}
         & \imposed{imposed} \\
      \imposed{\quad Invariant-mass formula$^{\ast}$}
         & \imposed{$M_K^{2} = (\textstyle\sum_i E_i)^{2} - |\textstyle\sum_i \mathbf{p}_i|^{2}$ \quad (by loss design)}
         & \imposed{Eq.~\eqref{eq:mK_decomp}}
         & \imposed{imposed} \\
      \quad Intra-particle kinematics
         & coordinate marginals $(p_T,\, \eta,\, \phi,\, E,\, m)$ per type
         & Fig.~\ref{fig:single_particle_muon}
         & \hyperref[eq:cfm_primary]{primary loss} \\
      \quad $p_T$ conservation
         & $\textstyle\sum_i \mathbf{p}_{T,i} + \mathbf{p}_T^{\mathrm{miss}} \to 0$
         & Fig.~\ref{fig:conservation}
         & none (MET cond.) \\
      \quad Inter-particle kinematics
         & $\Delta R$, $\Delta\phi$, $m_T$, $\Sigma H_T$, lead./sub-lead.\ lepton $p_T$
         & Fig.~\ref{fig:other_metrics_grid}
         & none \\
      \grouphdr{2. OSSF dilepton resonances} \\
      \quad Resonance peaks
         & $J/\psi$, $\Upsilon$, $Z \to \ell\ell$
         & Fig.~\ref{fig:dilepton}
         & \hyperref[eq:aux_loss]{aux.\ mass density} \\
      \quad Resonances and angular relations
         & $m_{\ell\ell}$ vs.\ $\Delta\phi$
         & Fig.~\ref{fig:dilepton_resonance_and_angular_relation}
         & \hyperref[eq:aux_loss]{aux.\ (mass axis only)} \\
      \grouphdr{3. Electroweak measurement} \\
      \quad Weinberg angle
         & $\sin^{2}\theta_w$ from the Drell--Yan $A_{\mathrm{FB}}$
         & Fig.~\ref{fig:weinberg}, Table~\ref{tab:weinberg}
         & none \\
      \grouphdr{4. Multi-jet \& parton kinematics} \\
      \quad Leading-dijet correlation
         & $(m_{jj},\, \bar\eta)$ joint density; envelope $x_{1,2} \leq 1$
         & Fig.~\ref{fig:dijet_panel}
         & \hyperref[eq:aux_loss]{aux.\ ($m_{jj}$ only)} \\
      \quad Three-particle invariant masses
         & $m_{\ell\ell\ell}$, $m_{\gamma\ell\ell}$, $m_{\ell\ell j}$, $m_{jjj}$, $m_{\ell jj}$, system $p_T$
         & Fig.~\ref{fig:three_particle_grid} (App.~\ref{app:threebody})
         & \hyperref[eq:aux_loss]{aux.\ mass density} \\
      \grouphdr{5. Heavy-particle mass} \\
      \quad Top quark
         & $t \to W^{+}b \to jjb$, $\bar t \to W^{-}\bar b \to \ell\bar\nu\bar b$
         & Fig.~\ref{fig:heavy_particle_mass}
         & \hyperref[eq:aux_loss]{aux.\ mass density} \\
      \quad $W$, $Z$
         & $W \to jj$, $Z \to \ell\ell$
         & Fig.~\ref{fig:heavy_particle_mass}, Table~\ref{tab:top_mass}
         & \hyperref[eq:aux_loss]{aux.\ mass density} \\
      \quad Higgs
         & $H \to 4\ell$
         & Fig.~\ref{fig:higgs_m4l}
         & \hyperref[eq:aux_loss]{aux.\ mass density} \\
    \end{tabular}
  \end{ruledtabular}
\end{table*}

\endgroup

Each truth-vs.\ generated comparison in the figures below is
quantified by three scalar agreement metrics, annotated on each
panel and reported throughout this section. The
1-Wasserstein distance
$W_{1}(p,q) = \int |F_{p}-F_{q}|$~\cite{Kantorovich:1942masses,
Vaserstein:1969markov} measures the integrated
cumulative-distribution gap between the truth and the generated
distributions and is sensitive to global shape distortions. The
symmetrised Kullback--Leibler (KL)
divergence~\cite{Kullback:1951on, Jeffreys:1948theory}
$\mathrm{SKL}(p,q) = \tfrac{1}{2}\,[\mathrm{KL}(p\|q) +
\mathrm{KL}(q\|p)]$\footnote{The directed divergence is
$\mathrm{KL}(p\|q) = \sum_{i} p_{i} \log(p_{i}/q_{i})$, evaluated
here on the common binning $\{i\}$ of the compared histograms, with
both densities normalised to unit sum.} measures
the binwise log-density mismatch and is sensitive to local under- or
over-production. The Bray--Curtis
dissimilarity $\mathrm{BC}(p,q) = \sum_{i} |p_{i} - q_{i}| /
\sum_{i} (p_{i} + q_{i}) \in [0, 1]$~\cite{Bray:1957ordination}
measures the bin-by-bin density overlap. All three vanish when the generated distribution matches
the truth exactly, so smaller values indicate better closure.
Systematic treatments of generative-model evaluation in collider
physics, including high-dimensional two-sample tests and
classifier-based diagnostics, are developed in
Refs.~\cite{Kansal:2022spb, Das:2023ktd, Grossi:2024axb}.
Each metric is computed on one generated sample. Regenerating the
sample five times with independent prior seeds, at fixed
conditioning, changes the dilepton metrics of this section by
0.5--3\%, at or below the statistical fluctuation of the truth
sample itself. The fitted quantities of Tables~\ref{tab:weinberg}
and~\ref{tab:top_mass} are more sensitive and carry this
regeneration spread as a second uncertainty.

We evaluate the trained model on five groups of observables,
following the grouping of Table~\ref{tab:result_summary}. The
first group establishes faithfulness of the intra-particle
kinematic marginals, of the event-level transverse-momentum
balance, and of a set of inter-particle observables that enter no
training loss. The second group examines the resonance structure
of the OSSF dilepton invariant-mass
spectrum. The third is an electroweak measurement, the Weinberg
angle extracted from the Drell--Yan forward--backward asymmetry.
The fourth covers multi-jet and parton-level structure, where the
leading dijet kinematics and the three-particle invariant masses
test whether pairwise and triple correlations combine into the
correct joint densities. The fifth reaches the heavy-particle
masses, with the hadronic and leptonic top quark peaks, the
accompanying hadronic $W$ and leptonic $Z$ peaks, and the Higgs
golden channel $H \to ZZ^{*} \to 4\ell$. The Supervision column
of Table~\ref{tab:result_summary} records what each of these
closures can and cannot inherit from training: the analysis
selections and the angular structure of every observable enter no
training objective, while the mass spectra are supervised, because
alongside its primary flow-matching objective the model carries an
auxiliary loss term on the inclusive invariant-mass density of
every pair, triplet, and quadruplet of generated particles (the
$K$-body masses, Sec.~\ref{sec:method_architecture}), so
the corresponding peak placements are fidelity rather than
emergence claims. Truth
distributions refer throughout to the 5\% validation split of the
joint sample, held out from training.

Each test below is structured as a closure between truth
(validation) and generated samples, processed through the same
analysis pipeline, and asks whether the generator preserves the
shape and the location of a given physical structure when the
standard ATLAS-style selection is applied to it. The claims of
this section are therefore claims of agreement, not measurements
of the underlying physical parameters, and they should not be read
as competing with the dedicated precision measurements of those
parameters. The model is
given no information about any specific resonance, channel, or
decay topology, so the structures recovered below must be encoded
in internal features of the trained network, activated by the data
themselves. What this section aims to establish is that this
encoding exists. The precision of any single closure matters less.

\subsection{Intra-particle kinematics}
\label{sec:results_singleparticle}

We compare truth and generated marginals in collider coordinates,
$(p_T, \eta, \phi, E)$ for the massless types and
$(p_T, \eta, \phi, E, m)$ for the massive types. The model reproduces
the truth distributions across all six particle types to high
accuracy, with only a handful of identifiable defects. The uniform
$\phi$ marginals follow directly from the model's internal
representation of each momentum direction as a point on the unit
sphere $S^{2}$ (Sec.~\ref{sec:method_manifolds}): a uniform
spherical prior projects to a uniform
marginal in $\phi$, so the model does not have to learn periodicity.
A shallow deficit of generated events at $\eta \approx 0$ is visible
in every type, and its origin is not established. By contrast, the
prominent twin gaps in the hadronic-$\tau$ $\eta$ marginal and the
sharp $\eta \approx 0$ gap in the muon $\eta$ marginal are
detector-level features. The former correspond to the transition
region between the barrel and endcap calorimeters,
$1.37 < |\eta| < 1.52$~\cite{ATLAS:2014qcj}; the latter to the
$|\eta| < 0.1$ region, where the muon spectrometer is only
partially instrumented to leave room for the cabling and services
of the inner detector and the calorimeters~\cite{ATLAS:2016lqx}.
The model reproduces both gaps from data rather than smoothing
them out.

\begin{figure*}[!tbp]
  \includegraphics[width=\textwidth]{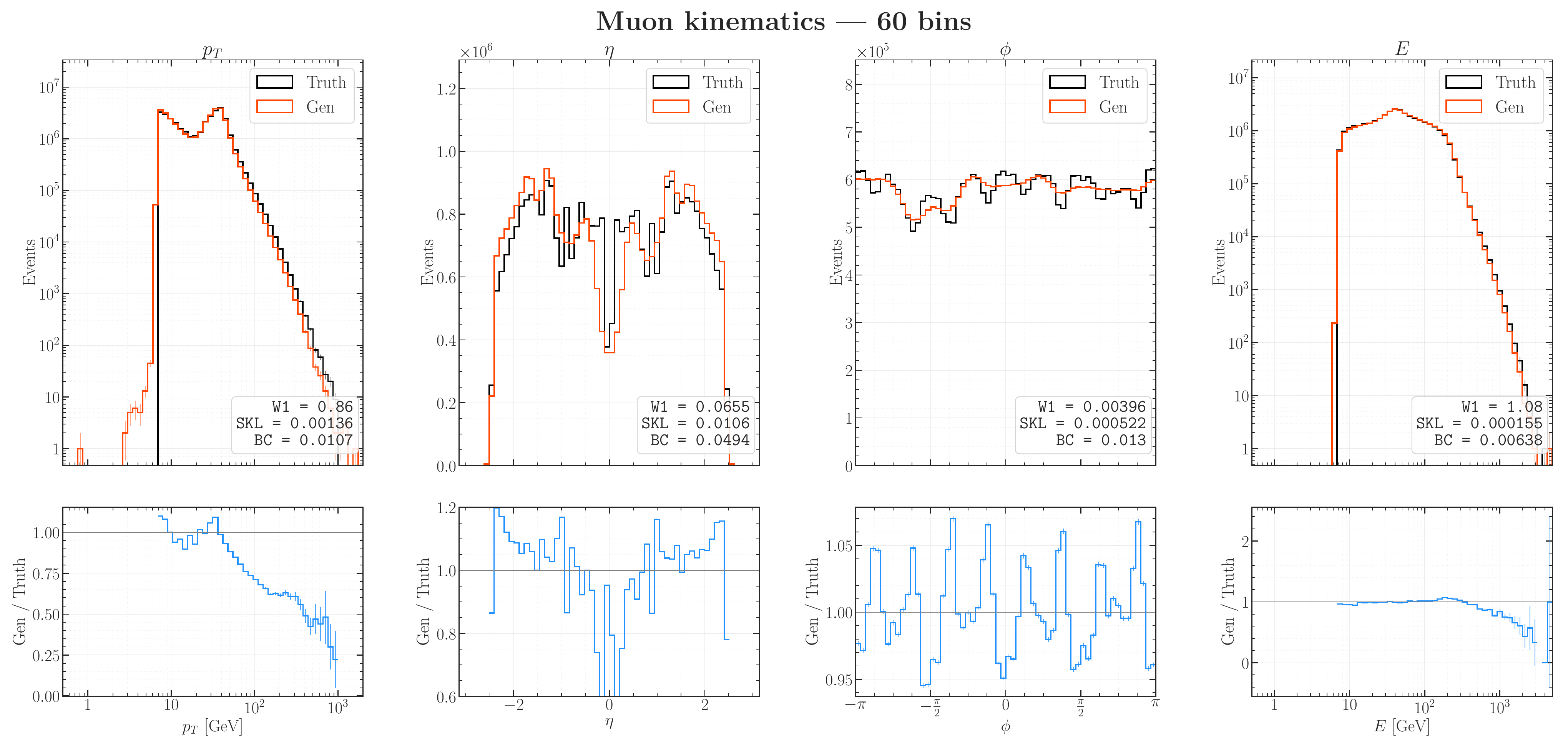}
  \caption{\label{fig:single_particle_muon}%
    \textbf{Muon kinematics.}
    Muon intra-particle marginals in collider coordinates
    $(p_T, \eta, \phi, E)$: truth compared with the
    generated distribution. \emph{Selection:} all reconstructed
    muons in the validation sample. The agreement metrics
    $W_{1}$, SKL, and BC are defined at the start of
    Sec.~\ref{sec:results}. The corresponding marginals for the
    other five reconstructed object types are collected in
    App.~\ref{app:collider}.}
\end{figure*}

Electrons and muons, the primary objects in the 2-to-4 lepton dataset,
fit the truth distributions tightly across all four variables, with
$W_{1}$, SKL, and BC scores reaching the smallest
numerical values of any particle type.
Photons, statistically
under-represented in this skim relative to the leptons, are
reproduced in good agreement with truth across all four
variables. The two jet types form the primary massive sector.
Both the four-momentum components and the mass marginal are
reproduced within statistical fluctuations, confirming that the
model preserves the on-shell relation
$m^{2} = E^{2} - |\mathbf{p}|^{2}$ at generation time. The corresponding collider-coordinate marginals for the
five non-muon types (electron, photon, $\tau_{\mathrm{had}}$,
small-$R$ jet, large-$R$ jet) are collected in
App.~\ref{app:collider}.

\subsection{Event-level transverse-momentum conservation}
\label{sec:results_conservation}

\begin{figure*}[!tbp]
  \includegraphics[width=\textwidth]{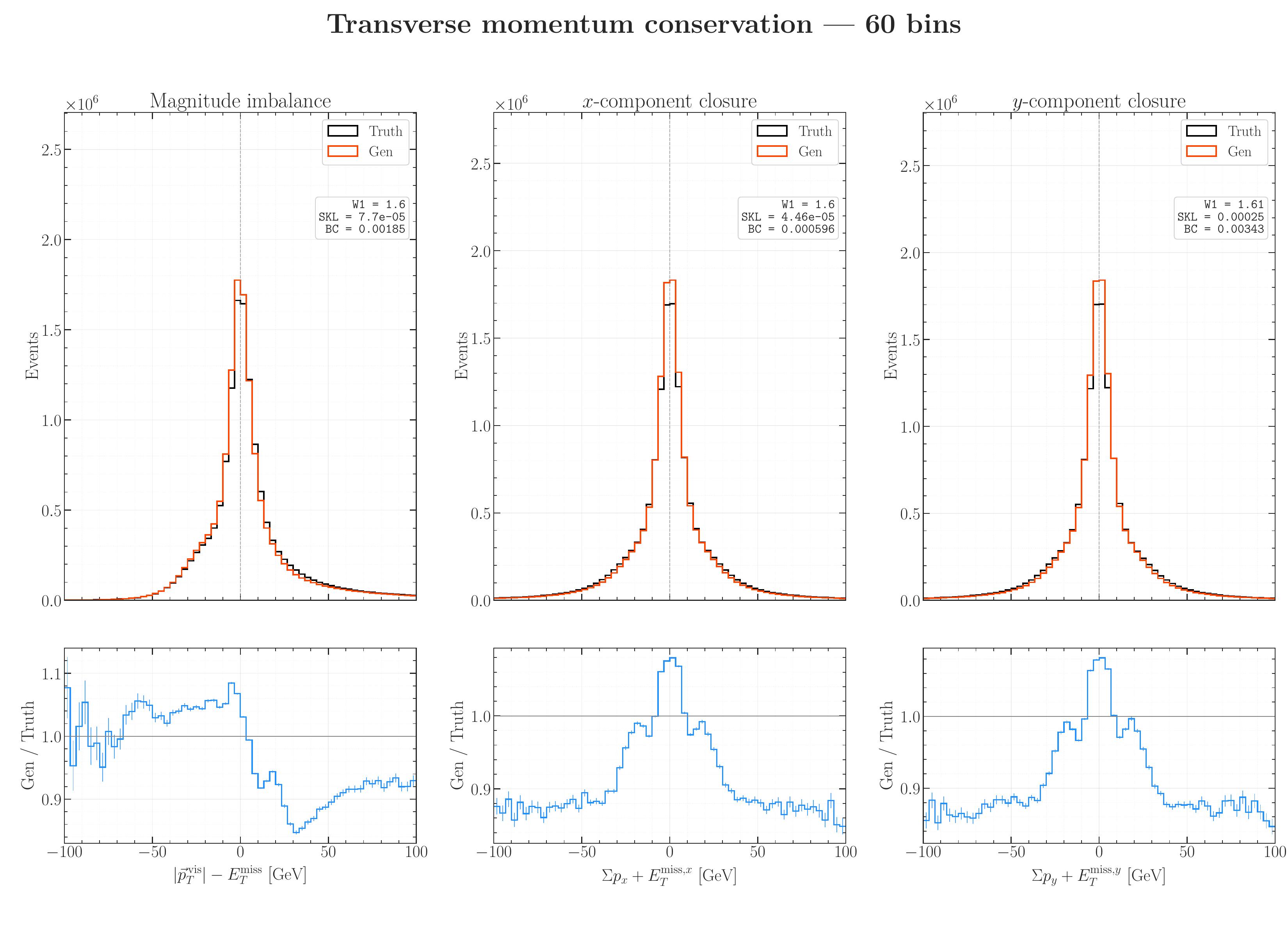}
  \caption{\label{fig:conservation}%
    \textbf{Transverse momentum conservation.}
    Event-level $p_T$ closure on the joint sample: signed components
    $\sum p_{T,x} + E_x^{\mathrm{miss}}$ and $\sum p_{T,y} +
    E_y^{\mathrm{miss}}$, plus the signed magnitude residual
    $|\mathbf{p}_T^{\mathrm{vis}}| - E_T^{\mathrm{miss}}$, truth vs.\
    generated. The signed components close at zero. The asymmetric
    magnitude bump on the negative side is carried by the 1LMET30
    component, whose $E_T^{\mathrm{miss}}$ selection and unstored
    soft activity make $E_T^{\mathrm{miss}}$ exceed
    $|\mathbf{p}_T^{\mathrm{vis}}|$ (see text). \emph{Selection:} all
    events in the validation sample.}
\end{figure*}

The simplest event-level closure tests the generator against the
kinematic relation $\sum_i \mathbf{p}_{T,i} + \mathbf{p}_T^{\mathrm{miss}}
\approx \mathbf{0}$. The proper definition of the missing transverse
momentum entering this relation is given in
Ref.~\cite{ATLAS:2024ivo}. The relation is not exact in
reconstructed data: the ATLAS $E_T^{\mathrm{miss}}$ includes a soft
term built from tracks not associated with any reconstructed
object, so both the recorded and the generated
events carry a residual, and the closure asks whether the generator
reproduces it. Figure~\ref{fig:conservation} shows the signed components
$\sum p_{T,x} + E_x^{\mathrm{miss}}$ and $\sum p_{T,y} +
E_y^{\mathrm{miss}}$ over the visible (non-MET) objects together
with the signed magnitude residual
$|\mathbf{p}_T^{\mathrm{vis}}| - E_T^{\mathrm{miss}}$, comparing truth
and generated. The signed-component distributions peak at zero and
the generated curves overlay truth within statistical fluctuations.

The signed-magnitude residual shows a visibly asymmetric shape with
a bump on the negative side. The bump is carried by the 1LMET30
component of the joint sample. Its events are selected with
$E_T^{\mathrm{miss}} > \SI{30}{GeV}$, and their recorded
$E_T^{\mathrm{miss}}$, which also receives the soft term and any
hadronic activity not stored as an object in the skim,
systematically exceeds the transverse momentum summed over the
stored objects, so the magnitude residual is negative. The 2-to-4
lepton component concentrates the residual around zero. The
generator reproduces this two-component shape, including its
asymmetry.

\subsection{Inter-particle observables without supervision}
\label{sec:results_unsup_kin}

\begin{figure*}[!tbp]
  \includegraphics[width=\textwidth]{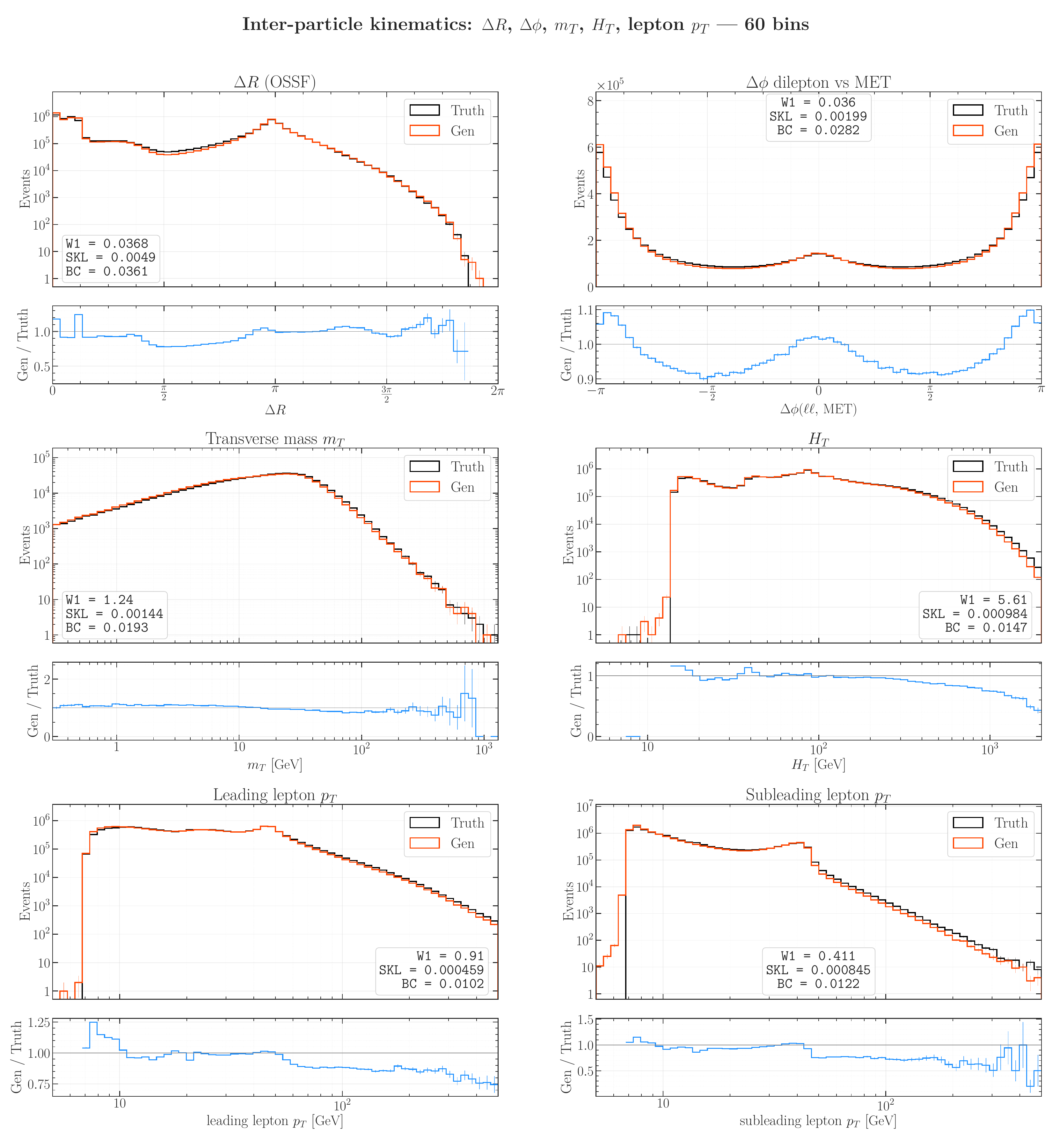}
  \caption{\label{fig:other_metrics_grid}%
    \textbf{Inter-particle kinematics.}
    Inter-particle observables that enter neither the primary
    flow-matching loss nor the auxiliary $K$-body loss and have no
    dedicated conditioning channel: the OSSF cone distance
    $\Delta R_{\ell\ell}$, the dilepton-to-MET azimuthal opening
    $\Delta\phi(\ell\ell,\,\mathrm{MET})$, the transverse mass
    $m_T$ of Eq.~\eqref{eq:mt_def}, the scalar transverse-momentum
    sum $H_T$, and the leading and sub-leading lepton $p_T$, all
    defined in the text. Truth (validation set) vs.\ generated. All
    six are computed after sampling, identically on truth and on
    generated. \emph{Selection:} all events for $H_T$; OSSF pairs
    for $\Delta R_{\ell\ell}$ and $\Delta\phi(\ell\ell,\,\mathrm{MET})$;
    three-lepton events for $m_T$; events with $\geq 2$ leptons for
    the lepton $p_T$ spectra.}
\end{figure*}

The transverse-momentum closure of the previous subsection follows
from the conditioning channel that feeds MET into the generator, and
so is in part an arithmetic check on the model. A stronger test is
whether the generator reproduces inter-particle observables that
enter neither the primary flow-matching loss nor the auxiliary
$K$-body loss
and have no dedicated conditioning channel. We collect six such
observables in Fig.~\ref{fig:other_metrics_grid}. The OSSF dilepton
cone distance
$\Delta R_{\ell\ell} = \sqrt{(\Delta\eta)^{2} + (\Delta\phi)^{2}}$
is computed for every OSSF lepton pair in
the event. The dilepton-to-MET azimuthal opening
$\Delta\phi(\ell\ell,\,\mathrm{MET})$ is the azimuthal angle
between $\mathbf{p}_T$ of the $Z$-candidate OSSF pair, chosen as
the pair whose invariant mass lies closest to $m_Z$, and
$\mathbf{p}_T^{\mathrm{miss}}$. The transverse mass
\begin{equation}
  m_T = \sqrt{2\, p_T^{\ell}\, E_T^{\mathrm{miss}}
        \bigl(1 - \cos\Delta\phi(\ell,\,
        \mathbf{p}_T^{\mathrm{miss}})\bigr)}
  \label{eq:mt_def}
\end{equation}
uses the highest-$p_T$ lepton $\ell$ not assigned to the
$Z$-candidate pair, making it the $W$-candidate transverse mass of
three-lepton topologies. The scalar transverse-momentum sum
$H_T = \sum_i p_{T,i}$ runs over all visible reconstructed objects
in the event. The leading and sub-leading lepton $p_T$ are ranked
among electrons and muons in events with at least two leptons. All
six are computed \emph{after} sampling, identically on truth and on
generated, and the generator has no objective that targets any of
them. Two indirect channels deserve note: the $K = 2$ auxiliary
density on $m_{\ell\ell}$ constrains a combination of
$\Delta R_{\ell\ell}$ and the lepton momenta, and the MET
conditioning supplies the reference vector for
$\Delta\phi(\ell\ell,\,\mathrm{MET})$ and $m_T$. Neither, however,
fixes the joint distributions examined here.

The generated and truth distributions agree to within the statistical
resolution of the validation sample across all six panels. The
$\Delta\phi(\ell\ell,\,\mathrm{MET})$ distribution closes across the
full $[-\pi, \pi]$ range, including the visible enhancement near
$|\Delta\phi| \approx \pi$. In events where the dilepton system
dominates the visible activity, the transverse-momentum balance of
Sec.~\ref{sec:results_conservation} forces
$\mathbf{p}_T^{\mathrm{miss}}$ to point opposite the pair. This
back-to-back configuration is the one selected on by
dilepton-plus-$E_T^{\mathrm{miss}}$
analyses~\cite{ATLAS:2017nyv}. The $\Delta R_{\ell\ell}$ distribution
reproduces the small-cone enhancement from collimated dilepton
resonances and the broad continuum at large $\Delta R$. The
$m_T$, $H_T$, and the leading and sub-leading lepton
$p_T$ marginals close in shape and normalisation across four decades
in events per bin. All six closures show that the model has
internalised multi-particle structure that no observable-level term
targets.

\subsection{Resonance recovery in the OSSF dilepton spectrum}
\label{sec:results_dilepton}

\begin{figure*}[!tbp]
  \includegraphics[width=\textwidth]{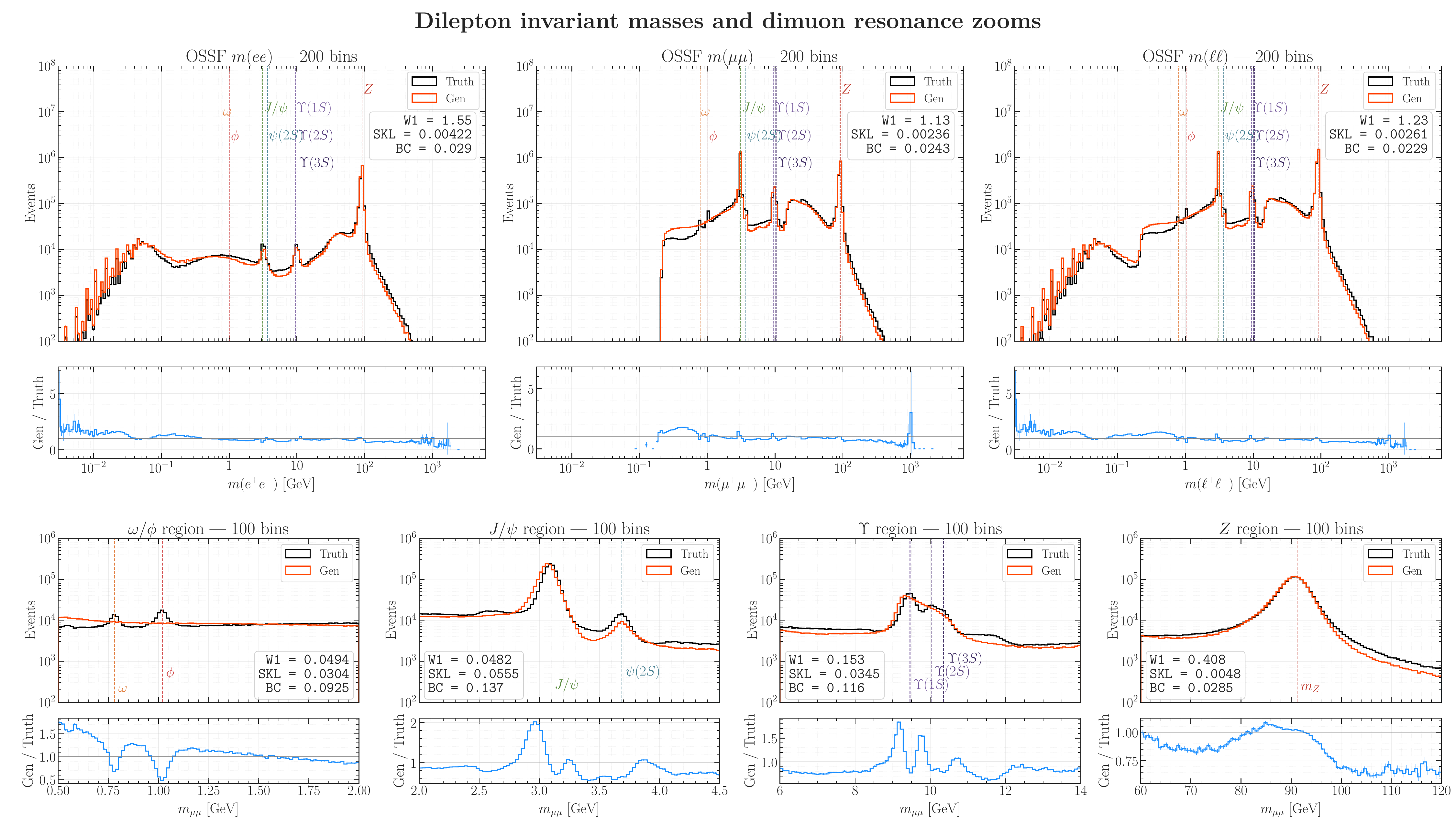}
  \caption{\label{fig:dilepton}% 
  \textbf{Main result:}
    OSSF dilepton invariant-mass spectrum, truth (validation set)
    vs.\ generated, across five decades in $m_{\ell\ell}$, with
    zooms on the four resonance regions. The $J/\psi(1S)$ and
    $\psi(2S)$ appear as two distinct peaks at their PDG positions,
    the bottomonium family $\Upsilon(1S, 2S, 3S)$ forms a single
    composite peak, and the $Z$ at $\approx \SI{91.2}{\GeV}$ rises
    above the falling Drell--Yan continuum. None of these masses
    was supplied to the model. In every panel the sub-panel below
    shows the bin-by-bin generated-to-truth ratio, and summary
    agreement metrics are annotated in the panel.
    \emph{Top-left:} dielectron channel $m(e^{+}e^{-})$.
    \emph{Top-middle:} dimuon channel $m(\mu^{+}\mu^{-})$.
    \emph{Top-right:} combined $m(\ell^{+}\ell^{-})$. The three
    top panels span the full mass range with $200$ log-spaced bins.
    \emph{Bottom row:} dimuon-channel zooms on each resonance
    region ($\omega/\phi$, $J/\psi$, $\Upsilon$, $Z$), $100$ bins
    each.
    \emph{Selection:} exactly two OSSF
    leptons (electrons or muons) with $p_{T} > \SI{10}{GeV}$.}
\end{figure*}

We turn next to the OSSF dilepton invariant-mass spectrum. The
$K = 2$ branch of the auxiliary loss
(Sec.~\ref{sec:method_architecture}) supervises $\log m_{\ell\ell}$
directly, so peak placement is a fidelity claim rather than an
emergence one. What is less obvious in advance is whether a single
model can simultaneously hold the narrow resonance peaks above the
Drell--Yan continuum across the five decades of $m_{\ell\ell}$ that
separate the dilepton kinematic threshold from the TeV continuum, in
both flavour channels, with the same hyperparameters. Figure
\ref{fig:dilepton} shows that it can.

Three distinct peak structures emerge above the Drell--Yan continuum.
The charmonium states $J/\psi(1S)$ and $\psi(2S)$ resolve as two
distinct peaks at their PDG positions. The bottomonium family
$\Upsilon(1S, 2S, 3S)$ forms a single composite peak. The individual
$\Upsilon$ states are not resolved at the present resolution and
training horizon. The $Z$ peak at $\approx \SI{91.2}{\GeV}$ rises
above the continuum at its expected position. The falling Drell--Yan
continuum underneath is captured across the full range in both
flavour channels. The agreement is the result of two physics priors
built into the construction: the on-shell condition, embedded in
the coordinates on which each particle is generated
(Sec.~\ref{sec:method_manifolds}), which removes the
single-particle mass smearing that would otherwise widen every pair
mass; and the auxiliary $K$-body loss
(Sec.~\ref{sec:method_architecture}), which supervises the
invariant mass of every particle tuple directly. The ablation of
Sec.~\ref{sec:ablation} (Fig.~\ref{fig:ablation},
Table~\ref{tab:ablation}) shows that both are necessary: retrained
variants lacking either ingredient, or both, fail to form the
quarkonium peaks and degrade the $Z$ peak.

\begin{figure*}[!tbp]
  \includegraphics[width=\textwidth]{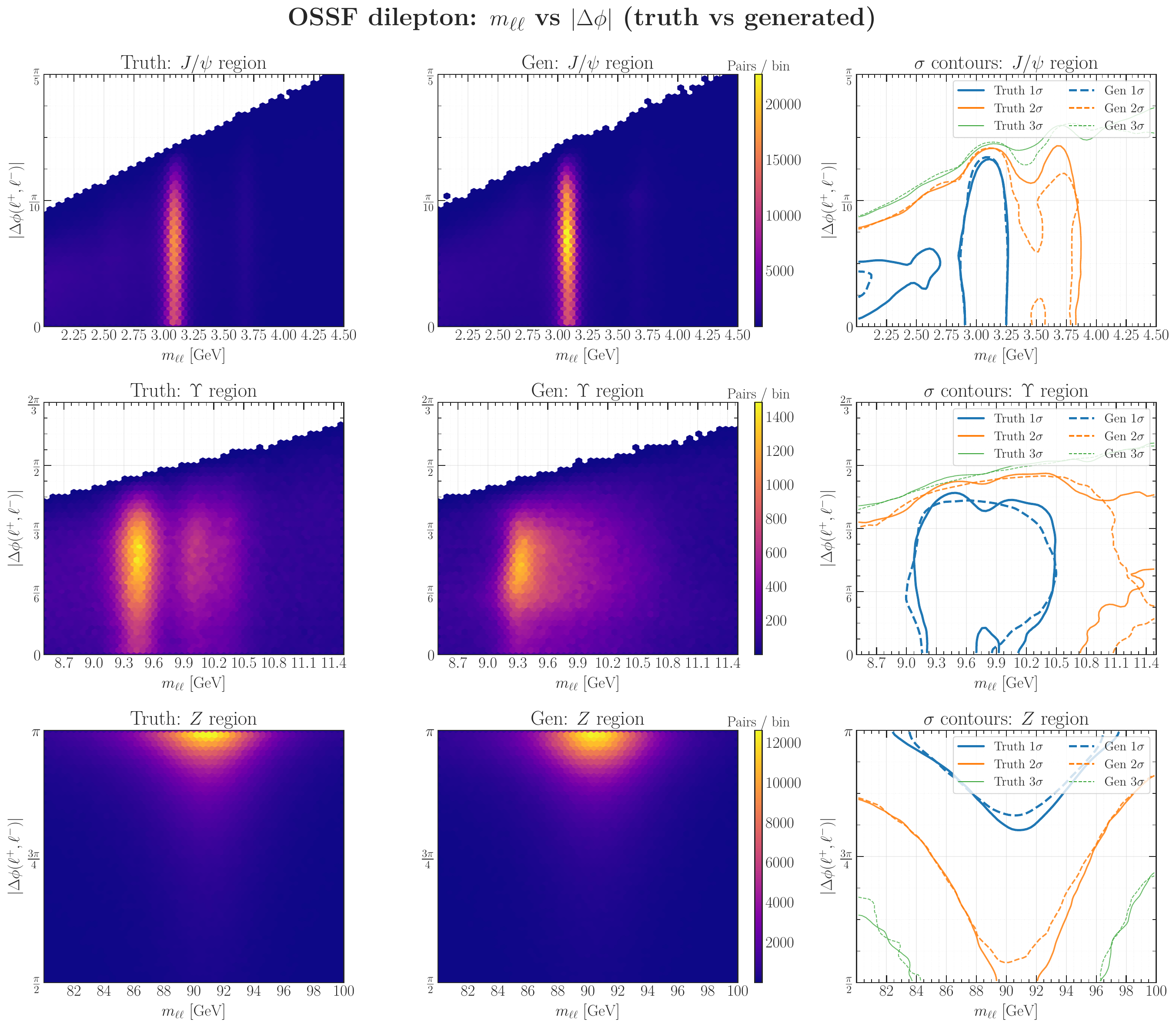}
  \caption{\label{fig:dilepton_resonance_and_angular_relation}%
    \textbf{Dilepton resonances and angular relations.}
    Joint distribution of the OSSF dilepton invariant mass and the
    azimuthal opening angle, displayed as a $3 \times 3$ grid (rows:
    $J/\psi$ family, $\Upsilon$ family, $Z$ peak, each restricted
    to a mass window centred on the resonance; columns: truth,
    generated, $\sigma$ contours). The third column overlays the
    $1\sigma$, $2\sigma$, and $3\sigma$ isocontours of the truth
    (solid) and generated (dashed) distributions.
    \emph{Selection:} OSSF dilepton, $p_{T} > \SI{10}{GeV}$
    per lepton.}
\end{figure*}

The marginal mass spectrum confirms that the model places events at
the correct invariant mass, but it does not by itself demonstrate
that the underlying decay kinematics have been learned. A sharper
test examines $m_{\ell\ell}$ jointly with the azimuthal opening angle
$|\Delta\phi|$ of the two leptons. Each resonance imposes a
characteristic correlation set by the boost kinematics of the parent
particle: the $J/\psi$ family is produced with high boost and decays
into a collimated lepton pair (small $|\Delta\phi|$), while the $Z$
is produced near rest in the transverse plane and decays into two
nearly back-to-back leptons ($|\Delta\phi| \to \pi$).
Figure~\ref{fig:dilepton_resonance_and_angular_relation} shows the joint distribution
$(m_{\ell\ell}, |\Delta\phi|)$ in a three-by-three grid, with rows
slicing the mass axis around the $J/\psi$, $\Upsilon$, and $Z$
resonances and columns showing truth, generated, and the
truth-vs-generated density contours. The
generated distributions reproduce the location and angular extent of
the truth densities in each window, including the back-to-back
configuration at the $Z$ pole and the high-boost collimated topology
of the $J/\psi$. In the $\Upsilon$ window the model has learned the
correct angular structure at every $m_{\ell\ell}$ (the angular
spread lies within the truth envelope throughout), even though it
resolves $\Upsilon(1S)$ cleanly while smearing $2S$ and $3S$
together.

The narrow resonance peaks of the dilepton spectrum are highly
sensitive to the forward noising that flow-matching training
applies to the data: each training event is corrupted along a path
parameterised by a flow time $t$, from the clean event at $t = 1$
to pure noise at $t = 0$ (Sec.~\ref{sec:method_rfm}), and the
model learns from the partially noised samples. Under this
noising the narrow light
quarkonia $\omega(782)$, $\phi(1020)$, and $\psi(2S)$ are
indistinguishable from the surrounding continuum well before the
flow time reaches $t \approx 0.99$, the $J/\psi(1S)$ and $\Upsilon$
peaks follow by $t \approx 0.9$, and only the broad $Z$ peak and
the Drell--Yan continuum survive past $t \approx 0.95$. We document
this in Appendix~\ref{app:noise_grid_dilepton}
(Fig.~\ref{fig:noise_grid_dilepton}). A loss that weights all
events equally cannot recover features that are buried in noise this
aggressively, which is the empirical motivation for the
distribution-matching weight $w^{\star}$
(Sec.~\ref{sec:method_architecture},
Eq.~\eqref{eq:dm_weight}): $w^{\star}$ tells the auxiliary loss that
certain mass regions are under-produced and need extra gradient
signal even when the forward-noised samples no longer carry their
structure. The $\omega$ and $\phi$ peaks remain absent from the
generated spectrum because they are erased so early in the forward
process that no amount of re-weighting on the auxiliary loss can
resurrect them at the present resolution and training horizon. A modest
event excess is also visible between the dimuon kinematic
threshold and the $\omega/\phi$ region, with small under-productions
at the immediate shoulders of the $J/\psi$ and $\Upsilon$ peaks
where $w^{\star}$ pulls events into the peaks at the expense of the
adjacent bins.

\subsection{Weinberg angle from the Drell--Yan asymmetry}
\label{sec:results_weinberg}

The dilepton spectrum carries more than just the mass distribution.
The Drell--Yan forward--backward asymmetry $A_{FB}(m_{\ell\ell})$
encodes the relative chiral couplings of the $Z$ and photon
intermediaries, and a fit to its shape yields the effective leptonic
weak mixing angle $\sin^{2}\theta_w$. Because the asymmetry is odd
in the decay angle, conventionally defined in the Collins--Soper
frame~\cite{Collins:1977iv}, and correlated with the lepton charge,
no inclusive invariant-mass density can encode it. Among the
observables studied in this paper it is the cleanest case of
structure the model can only have learned from the joint event
kinematics. We extract $\sin^{2}\theta_w$
from the trained model's generated events using a leading-order
Drell--Yan template fit applied identically to the truth validation
sample, separately in the dimuon and dielectron channels, in the
spirit of the CMS measurements at 8~TeV~\cite{CMS:2018ktx} and
13~TeV~\cite{CMS:2024ony} but at
the simpler leading-order level. The closure between truth and
generated is insensitive to the template refinements of the full
measurement, since both samples are fit with the same
template. The best-fit values are collected in
Table~\ref{tab:weinberg}. In each channel the generated and truth
fits are statistically compatible, with pulls on the
truth--generated difference of $0.3\sigma$ (dimuon) and
$1.2\sigma$ (dielectron), so the generator preserves the
Drell--Yan forward--backward asymmetry to the precision of the
truth sample itself. These pulls are stable under a more careful
error treatment: including the generation spread of
Table~\ref{tab:weinberg} and the truth--generated fit correlation
induced by the shared conditioning events, measured with a paired
event bootstrap to be $\rho \approx 0.3$, changes
each pull by less than $0.1\sigma$, since the two corrections act
in opposite directions. The truth fits sit $0.006$--$0.007$ above
the PDG effective leptonic mixing angle and the CMS Run-1 NLO
measurement~\cite{CMS:2018ktx, ParticleDataGroup:2024cfk}.
This offset reflects the choice of the LO fit template, not a
deficiency of the generator, since the generated and the truth fits
sit on the same side of PDG by approximately the same amount in
each channel.
The model therefore preserves the parity structure of the
Drell--Yan process in addition to its mass spectrum.

\begin{figure*}[!tbp]
  \includegraphics[width=\textwidth]{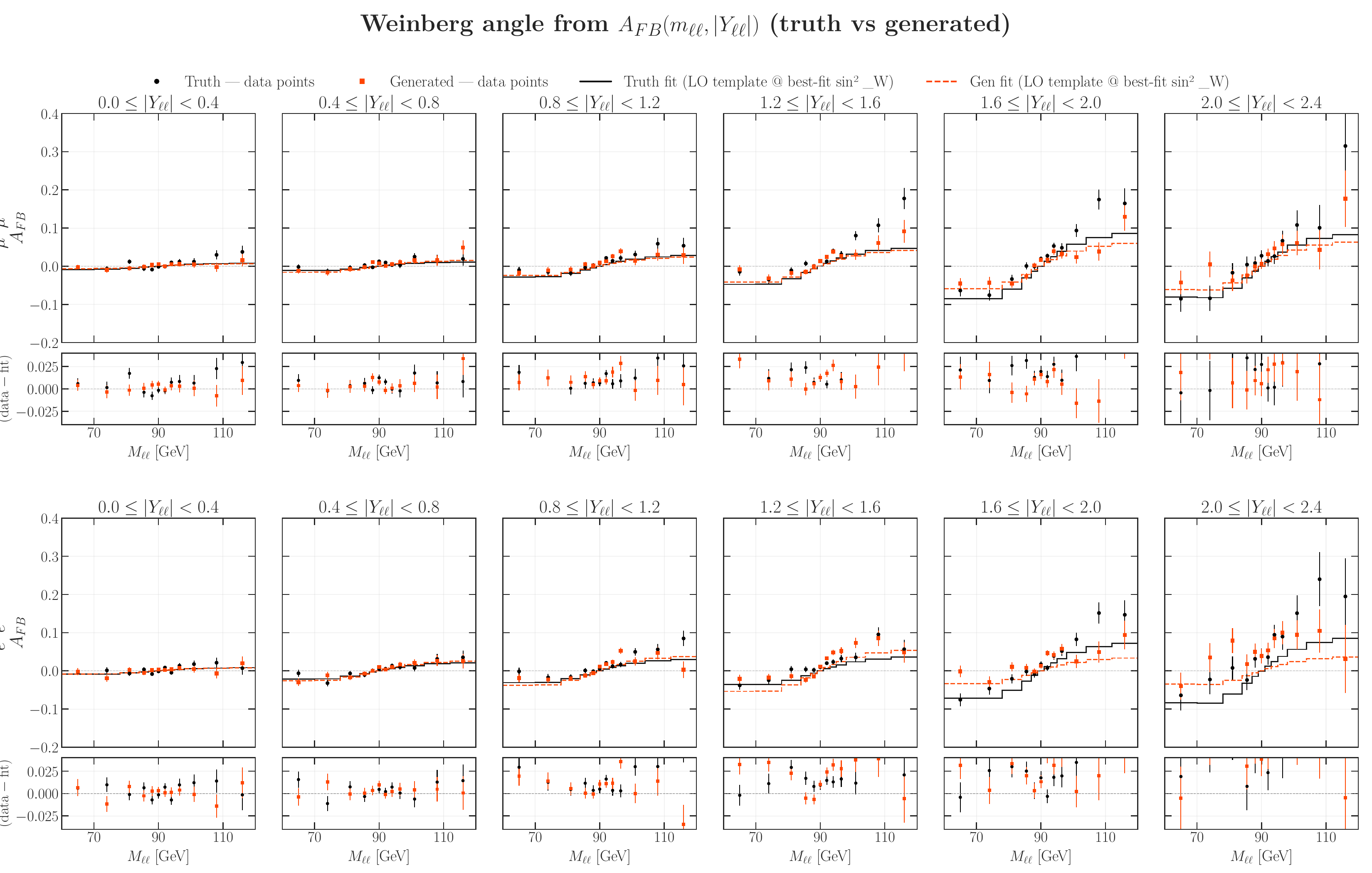}
  \caption{\label{fig:weinberg}%
    \textbf{Weinberg angle extraction.}
    Forward--backward asymmetry $A_{\mathrm{FB}}(m_{\ell\ell})$ in
    six bins of dilepton rapidity $|Y_{\ell\ell}|$, on truth
    (validation set) and on generated samples. In each panel the
    points are the measured asymmetry (truth in black, generated in
    orange), the curves are the LO Drell--Yan template evaluated at
    the best-fit $\sin^{2}\theta_w$ of the corresponding sample
    (truth solid, generated dashed),
    and the strip below each panel shows the data-minus-fit
    residuals. The fits to the generated sample and to the truth
    sample are statistically compatible ($0.3\sigma$ and
    $1.2\sigma$ pulls in the dimuon and dielectron channels), and both
    sit $\lesssim 0.007$ above the PDG effective leptonic mixing
    angle, an offset attributable to the LO fit template.
    \emph{Top row:} dimuon channel.
    \emph{Bottom row:} dielectron channel.
    \emph{Selection:} OSSF dilepton with $p_{T} > \SI{20}{GeV}$ on
    the leading lepton and $p_{T} > \SI{15}{GeV}$ on the
    sub-leading lepton, restricted to the Drell--Yan window
    $\SI{60}{GeV} < m_{\ell\ell} < \SI{120}{GeV}$.}
\end{figure*}

\begin{table}[!tbp]
  \caption{\label{tab:weinberg}%
    Fitted Weinberg angle $\sin^{2}\theta_w$ (LO $A_{\mathrm{FB}}$
    template). Best-fit values from the LO Drell--Yan
    $A_{\mathrm{FB}}(m_{\ell\ell})$ template fit on truth and
    generated samples, against the PDG effective leptonic mixing
    angle and the CMS Run-1 NLO value~\cite{CMS:2018ktx}. The fit
    uncertainty $\sigma_{\mathrm{fit}}$ and the generation spread
    $\sigma_{\mathrm{gen}}$, the standard deviation of the fitted
    value over five regenerations of the sample with independent
    prior seeds, are listed separately.}
  \begin{ruledtabular}
    \begin{tabular}{l c c c}
      \textbf{Source} & $\sin^{2}\theta_w$ & $\sigma_{\mathrm{fit}}$ & $\sigma_{\mathrm{gen}}$ \\
      \hline
      $\mu^{+}\mu^{-}$ truth        & $0.23690$ & $\pm 0.00084$ & --- \\
      $\mu^{+}\mu^{-}$ generated    & $0.23657$ & $\pm 0.00082$ & $\pm 0.00055$ \\
      \hline
      $e^{+}e^{-}$ truth            & $0.23792$ & $\pm 0.00103$ & --- \\
      $e^{+}e^{-}$ generated        & $0.23627$ & $\pm 0.00092$ & $\pm 0.00073$ \\
      \hline
      PDG eff.\ leptonic            & $0.23122$ & --- & --- \\
      CMS Run-1 NLO                 & $0.23140$ & --- & --- \\
    \end{tabular}
  \end{ruledtabular}
\end{table}

\subsection{Leading dijet kinematics}
\label{sec:results_dijet}

The leading dijet probes whether the model has captured the
longitudinal structure of the hadronic activity in these events,
predominantly jets recoiling against the leptonic system in the two
skims, through observables tied to the incoming parton momenta. The
average pseudorapidity of the two leading jets,
$\bar\eta = (\eta_1 + \eta_2)/2$, approximates the longitudinal
boost of the dijet centre-of-mass frame and tracks the asymmetry of
the parton momentum fractions $x_1, x_2$ of the incoming partons,
\begin{equation}
\begin{aligned}
  \bar\eta &\approx y_{\mathrm{boost}}
    = \tfrac{1}{2}\,\ln(x_1 / x_2) , \\
  x_{1,2} &= \tfrac{m_{jj}}{\sqrt{s}}\, e^{\pm\bar\eta} .
\end{aligned}
\label{eq:dijet_pdf}
\end{equation}
The kinematic constraints $x_1, x_2 \leq 1$ impose a triangular
envelope on the joint $(m_{jj}, \bar\eta)$ density:
$\ln m_{jj} \leq \ln\sqrt{s} - |\bar\eta|$, with the apex at
$m_{jj} = \sqrt{s} = 13$ TeV.

\begin{figure*}[!tbp]
  \includegraphics[width=\textwidth]{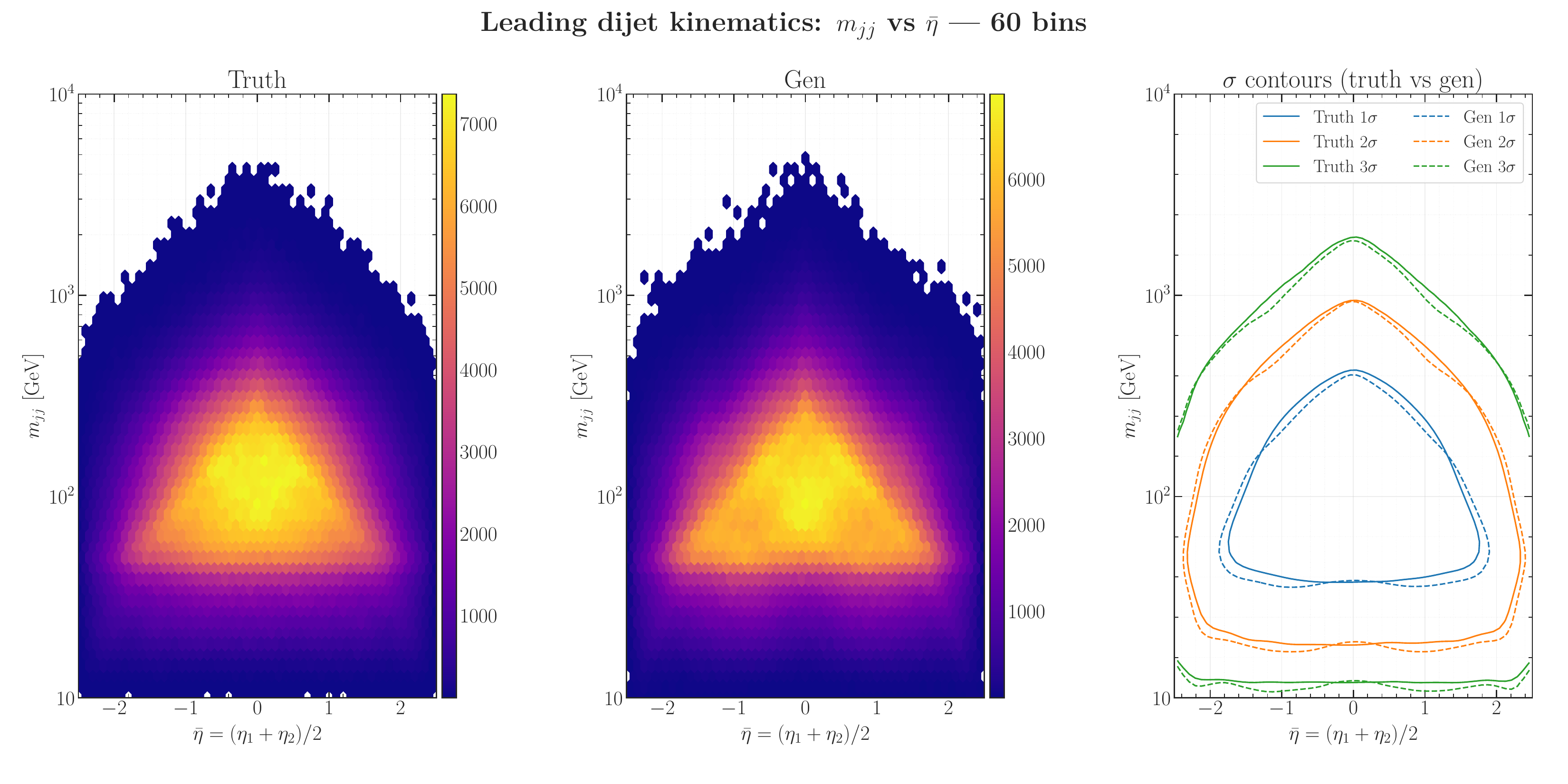}
  \caption{\label{fig:dijet_panel}%
    \textbf{Dijet kinematics.}
    Joint distribution of the leading-dijet invariant mass $m_{jj}$
    and average pseudorapidity $\bar\eta = (\eta_1 + \eta_2)/2$,
    truth (validation set) vs.\ generated. In all panels the two
    observables are built from the two leading jets of events with
    at least two reconstructed jets, and the triangular envelope is
    the kinematic constraint $x_1, x_2 \leq 1$ on the parton
    momentum fractions.
    \emph{Left:} truth.
    \emph{Middle:} generated.
    \emph{Right:} $1\sigma$, $2\sigma$, $3\sigma$ contours of truth
    (solid) and generated (dashed) overlaid for direct comparison.}
\end{figure*}

Figure~\ref{fig:dijet_panel} shows the joint distribution for the
truth and generated samples, with $1\sigma$, $2\sigma$, and
$3\sigma$ contours of the truth distribution overlaid. The generator
reproduces both the sharp kinematic edges of the triangular envelope
and the density inside. Beyond the supervised $m_{jj}$ marginal,
the joint $(m_{jj}, \bar\eta)$ correlation is reproduced with no
observable-level target.

\subsection{Three-particle invariant masses}
\label{sec:results_threebody}

Beyond the dilepton sector, each three-particle combination is a
distinct test of the model's correlation structure, because the
generated events must reproduce triple as well as pairwise
relationships between reconstructed objects. Most three-body spectra
are smooth continua or kinematic mixes from Standard-Model
processes. One combination, the all-jets triplet, carries a genuine
resonance from the hadronic decay $t \to W b \to q\bar q' b$.

Figure~\ref{fig:three_particle_grid}, collected in
Appendix~\ref{app:threebody}, compares truth and generated
distributions on eight three-particle combinations: $3\ell$
(multi-lepton continuum from $WZ$ and $ZZ$ topologies), $\gamma
\ell\ell$ (radiative $Z$ return at $m_Z$), $\ell\ell j$ ($Z +$ jet
with the dilepton at $m_Z$ and the jet carrying the QCD recoil),
$jjj$ (multijet QCD bulk plus the hadronic top peak at
$m \approx 173$ GeV), $\ell j j$ ($W +$ jets and semileptonic top
topologies), and the system transverse momenta of each triplet as
a 3-body momentum-balance check. The continua and kinematic mixes
are reproduced over the full mass range in each combination, and
the hadronic-top peak emerges above the multijet QCD continuum in
$m_{jjj}$.

\subsection{Heavy-particle mass closures: $W$, $Z$, top}
\label{sec:results_top}

The model is not told that the top quark exists, nor is it given the
top mass, the $W$ mass, or the $t\to Wb$ decay topology. The
$b$-tagging score of each jet is available to the model only as a
conditioning input, not as a generated quantity. Recovery of the top
peak under the standard semileptonic $t\bar t$ selection is
therefore a closure test on whether the generator has learned
the joint kinematics of the lepton, the neutrino, and the multi-jet
final state to the level a top decay requires, not a precision
measurement of the top mass.

The 1LMET30 component of the joint sample provides the standard
single-lepton plus jets topology of semileptonic $t\bar t$ events,
in which one top decays to $\ell\nu b$ and the other to
$q\bar q' b$. We apply the standard ATLAS-style $t\bar t$
selection~\cite{ATLAS:2018fwq} to the truth (validation) sample
and a generated sample of equal input size
($4.1 \times 10^{7}$ events each): exactly one
isolated lepton ($e$ or $\mu$) with $p_T > \SI{25}{GeV}$, missing
transverse energy $E_T^{\mathrm{miss}} > \SI{30}{GeV}$, at least
four jets with $p_T > \SI{25}{GeV}$, and exactly two of the four
leading jets $b$-tagged. The two non-$b$ jets are taken as the
hadronic-side $W$ candidate (unbiased by any $M_W$ selection), and
the $b$ jets are assigned to the hadronic and leptonic top
candidates by minimising the two-top mass difference. The
selection retains $44{,}763$ truth and $37{,}730$ generated
events. The neutrino
$p_z$ is solved from the $M_W$ constraint on the lepton and the
missing transverse energy, taking the smaller-$|p_z|$ root of the
resulting quadratic and its real part when the discriminant is
negative, as is standard in top-quark reconstruction. This follows
the spirit of KLFitter~\cite{Erdmann:2013rxa} without the full
likelihood machinery (hence the ``KLFitter-lite'' label used
below). The fit floats the top
mass with no $M_W$ constraint on the hadronic side, so the
reconstructed peak positions sit below the PDG values for both
truth and generated. The gen-vs-truth agreement is the operative
closure. The leptonic $Z$ panel uses a separate, simpler dilepton
selection: at least two leptons with $p_T > \SI{20}{GeV}$, the
leading OSSF pair, and the standard
$Z$ window $\SI{60}{GeV} < m_{\ell\ell} < \SI{120}{GeV}$, retaining
$2.6 \times 10^{6}$ truth and $2.5 \times 10^{6}$ generated events.

\begin{figure*}[!tbp]
  \includegraphics[width=\textwidth]{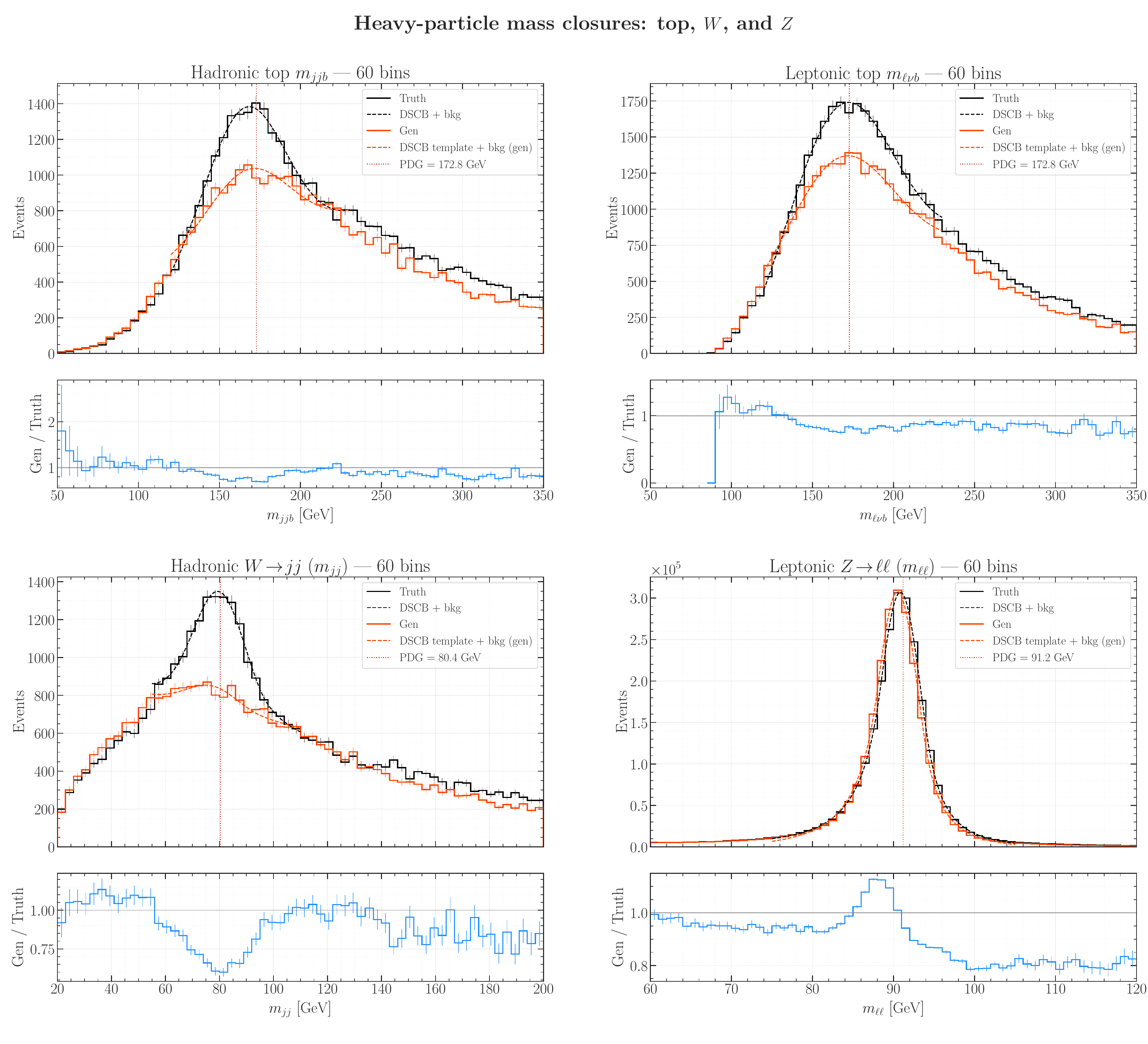}
  \caption{\label{fig:heavy_particle_mass}%
    \textbf{Heavy-particle masses.}
    Reconstructed-mass closures for the top quark, the $W$, and the
    $Z$, truth (validation) vs.\ generated. In every panel the
    truth peak is fit with a free-shape DSCB on a linear
    background, the generated peak with a truth-shape DSCB template
    in which only the normalisation, position, and background float
    (see text), and the lower sub-panel shows the bin-by-bin
    generated-to-truth ratio.
    \emph{Top-left:} hadronic top $m_{jjb}$.
    \emph{Top-right:} leptonic top $m_{\ell\nu b}$.
    \emph{Bottom-left:} hadronic $W$ $m_{jj}$. These three panels
    share the semileptonic $t\bar t$ selection and KLFitter-lite
    reconstruction described in the text.
    \emph{Bottom-right:} leptonic $Z$ $m_{\ell\ell}$, under the
    OSSF dilepton selection and the standard $Z$ window.}
\end{figure*}

Figure~\ref{fig:heavy_particle_mass} compares the truth and generated peaks
for four reconstructed heavy-particle observables, each fitted with
a Double-Sided Crystal Ball (DSCB) shape (a Gaussian core with two
independent power-law tails~\cite{ATLAS:2014jdv}, generalising the
standard Crystal Ball function of
Ref.~\cite{Skwarnicki:1986xj}) on a linear background.
The truth peaks are fitted with all shape parameters floating. The
generated peaks are fitted with a truth-shape template, in which the
resolution and tail parameters are fixed to the truth fit and only
the normalisation, position, and background float. The template is
required because a free-shape fit degenerates on the generated
hadronic-side distributions, whose resonant cores are diluted
relative to truth, and returns a width pinned at the fit bound with
no meaningful position. The fitted central values
and the residuals against PDG references are collected in
Table~\ref{tab:top_mass}.

The leptonic top $m_{\ell\nu b}$ closes
cleanly because the neutrino is reconstructed from the missing
transverse energy (a conditioning input) and the lepton via the W
constraint rather than being directly generated, so the closure
requires that the lepton kinematics, the b-jet kinematics, and the
missing transverse energy be simultaneously consistent with a parent
top decay. The hadronic top $m_{jjb}$ and the hadronic
$W$~$m_{jj}$ (the two non-$b$ jets, unbiased by any reference to
the W mass) close within the combined fit and regeneration
uncertainties quoted in Table~\ref{tab:top_mass}.
Fit-free cross-checks support the positional closure: the histogram
modes of the truth and generated distributions coincide within one
bin in every panel, and the interquartile ranges of the selected
distributions agree to $18\%$ or better (IQR-ratio column of
Table~\ref{tab:top_mass}). The tightest closure of the four sits
on the leptonic $Z \to \ell\ell$, where the truth and the
generated fits both land within $\SI{0.7}{GeV}$ of the PDG $m_Z$.
Lepton four-momenta are read directly off the generated
coordinates, with no intermediate reconstruction step, so the
dilepton invariant mass inherits the same sub-GeV precision as the
per-particle marginals.

What the generator has not yet
reproduced is the prominence of the hadronic resonant cores: at
matched sample size, the generated yield within one core width
($\mu \pm \sigma$) of the fitted peak sits $9\%$ below truth for
the hadronic top and $23\%$ for the hadronic $W$ ($3\%$ for the
leptonic top)\footnote{Computed from the histograms of
Fig.~\ref{fig:heavy_particle_mass} in three steps: the core window
is taken as $\mu \pm \sigma$ of the truth fit; the counts of the
bins whose centres fall inside the window are summed, separately
for the truth and the generated sample; and the generated-to-truth
ratio of the two sums is scaled by the truth-to-generated ratio of
selected-event counts ($44{,}763/37{,}730$), so that the shapes are
compared at matched sample size. The Poisson uncertainty on each
deficit is $1$--$1.5\%$.}, with the difference redistributed into the
surrounding continuum, and a free-shape fit consequently cannot
isolate the generated core width at all. We
attribute the core dilution to a combination of the resolution of
the coordinates in which jet four-momenta are generated
(Sec.~\ref{sec:method_manifolds}), the smoothing inherent in the
distribution-matching weight (Sec.~\ref{sec:method_architecture}),
and the present 30-epoch training
horizon. The bottom ratio panels of
Fig.~\ref{fig:heavy_particle_mass} sit at unity across the full
mass range in all four reconstructions: the closure is a
positional one, robust to the core dilution on the hadronic side,
but the dilution limits the precision to which the hadronic peaks
can be extracted from generated samples.

\begin{table*}[!tbp]
  \caption{\label{tab:top_mass}%
    Heavy-particle masses: fitted DSCB~\cite{Skwarnicki:1986xj} peak
    positions $\mu$ versus PDG~\cite{ParticleDataGroup:2024cfk}, on
    truth and generated samples, with residuals
    $\Delta = \mu - \mathrm{PDG}$, together with the truth core width
    $\sigma$ and a fit-free width comparison. The hadronic top
    reconstruction
    floats the top with no $M_W$ constraint on the hadronic side,
    which biases its peak position downward equally on truth and on
    generated. The gen-vs-truth agreement is the closure. The leptonic
    $Z$ row is the corresponding OSSF dilepton fit under the standard
    $Z$ window used by Fig.~\ref{fig:heavy_particle_mass}.}
  \begin{ruledtabular}
    \begin{tabular}{l c c c c c c c}
      \textbf{Observable} & Truth $\mu$ (GeV)\footnote{Peak position from a free-shape DSCB fit on a linear background. Uncertainties are fit uncertainties on the position, not peak widths.} & Gen $\mu$ (GeV)\footnote{Peak position from a truth-shape template fit: the DSCB shape is fixed from the truth fit; only normalisation, position, and background float. The first uncertainty is the fit uncertainty on the position; the second is the standard deviation of the fitted position over five regenerations of the sample with independent prior seeds.} & Truth $\sigma$ (GeV)\footnote{Width of the truth DSCB Gaussian core: the physical spread of the reconstructed mass around $\mu$.} & IQR ratio\footnote{Generated-to-truth ratio of the interquartile ranges within the fit window; a fit-free width comparison.} & PDG (GeV) & $\Delta$ Truth\footnote{$\Delta\,\mathrm{Truth} = \text{Truth } \mu - \mathrm{PDG}$.} & $\Delta$ Gen\footnote{$\Delta\,\mathrm{Gen} = \text{Gen } \mu - \mathrm{PDG}$.} \\
      \hline
      Hadronic top $m_{jjb}$        & $165.1 \pm 1.2$ & $167.5 \pm 2.3 \pm 3.6$ & $25.4 \pm 1.8$ & $1.10$ & $172.76$ & $-7.6$ & $-5.2$ \\
      Leptonic top $m_{\ell\nu b}$  & $166.9 \pm 1.3$ & $166.8 \pm 1.6 \pm 1.9$ & $32.9 \pm 2.4$ & $1.06$ & $172.76$ & $-5.8$ & $-5.9$ \\
      Hadronic $W$ $m_{jj}$         & $80.0 \pm 0.4$  & $77.3 \pm 1.8 \pm 3.5$  & $9.6 \pm 0.5$  & $1.18$ & $80.38$  & $-0.4$ & $-3.1$ \\
      Leptonic $Z$ $m_{\ell\ell}$   & $90.9 \pm 0.1$  & $90.5 \pm 0.1 \pm 0.05$  & $2.6 \pm 0.1$  & $0.96$ & $91.19$  & $-0.3$ & $-0.7$ \\
    \end{tabular}
  \end{ruledtabular}
\end{table*}

% Fig. 9 (higgs_m4l) is placed early in the source so the two-column
% float can share the page top with Table III instead of drifting to a
% page of its own after the Results text ends.
\begin{figure*}[!tbp]
  \includegraphics[width=\textwidth]{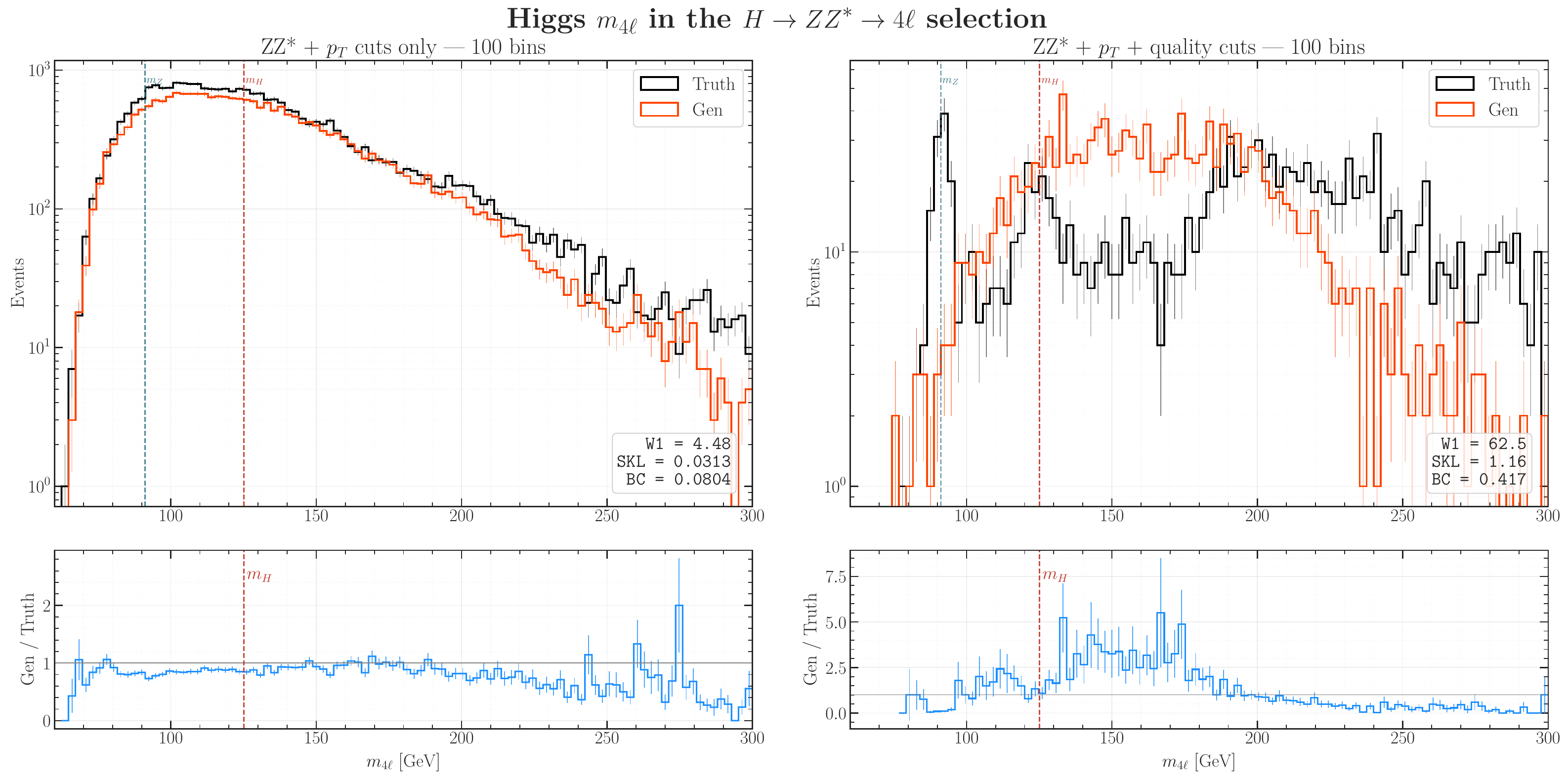}
  \caption{\label{fig:higgs_m4l}%
    \textbf{Higgs in the four-lepton invariant-mass spectrum.}
    Four-lepton invariant-mass spectrum $m_{4\ell}$ in the
    $H \to ZZ^{*} \to 4\ell$ golden channel, truth (validation)
    vs.\ generated, at the two selection stages defined in the
    text. The lower sub-panels show the bin-by-bin
    generated-to-truth ratio.
    \emph{Left:} kinematic stage ($\sim 2.3\times 10^{4}$ events
    per sample). The generator reproduces the continuum shape with
    a moderate excess in the $\SI{150}{}$--$\SI{200}{GeV}$ sideband.
    \emph{Right:} quality stage ($\sim 1.4\times 10^{3}$ events per
    sample). Truth resolves into a $Z \to 4\ell$ peak at
    $\SI{91}{GeV}$ and a Higgs peak at $\SI{125}{GeV}$, while the
    generated spectrum is consistent with a featureless continuum.}
\end{figure*}

\subsection{Higgs golden channel}
\label{sec:results_higgs}

The four-lepton system in the $H \to ZZ^{*} \to 4\ell$ golden channel
is the rarest and most demanding observable in the joint sample. The
branching fraction $\mathcal{B}(H \to 4\ell)$ is of order $10^{-4}$,
and the topology requires one on-shell $Z \to \ell\ell$ at
$m_Z \approx 91$ GeV plus one off-shell $Z^{*} \to \ell\ell$. The
analysis selection is patterned on the ATLAS
$H \to ZZ^{*} \to 4\ell$ measurement~\cite{ATLAS:2020rej} and
proceeds in two stages: a $ZZ^{*}$ pairing stage that selects
events with exactly four leptons satisfying
$p_T^{(1,2,3)} > 20, 15, 10$ GeV and OSSF pairs in
$m_Z \in [50, 106]$ GeV and $m_{Z^{*}} \in [12, 115]$ GeV, followed
by a per-lepton quality stage that imposes
$\texttt{ptvarcone30}/p_T < 0.15$,
$\texttt{topoetcone20}/p_T < 0.20$, and $|d_0/\sigma_{d_0}| < 5$ for
electrons and $< 3$ for muons. The kinematic thresholds, mass
windows, and impact-parameter cuts follow
Ref.~\cite{ATLAS:2020rej}, while the isolation cuts are adapted to the
isolation variables available in the Open Data format.

Figure~\ref{fig:higgs_m4l} compares the truth and generated
four-lepton invariant-mass spectra at the two selection stages. The
left panel applies only the $ZZ^{*}$ pairing and $p_T$ cuts and
retains $\sim 2.3\times 10^{4}$ events per sample. The right panel
applies the additional per-lepton quality cuts and reduces the
selected sample by an order of magnitude to $\sim 1.4\times 10^{3}$
events. With $p_T$ cuts only, the generator reproduces the bulk
$m_{4\ell}$ continuum shape and the location of the broad
$Z$-recoil shoulder, with a moderate excess in the $150$--$200$ GeV
sideband. Once the per-lepton quality cuts are imposed, the truth
spectrum resolves into the two-peak structure expected from
$Z \to 4\ell$ at $91$ GeV and $H \to 4\ell$ at $125$ GeV, while the
generated spectrum collapses to a featureless continuum that does
not reproduce either peak. The combination of the rare-process
branching fraction $\mathcal{B}(H \to 4\ell) \sim 10^{-4}$ and the
multi-cut topology makes this the hardest observable in the joint
sample to learn. The
correlated failure of the inner $Z$ shoulder and the Higgs peak
indicates that the model has not yet learned the joint quality-cut
acceptance: the per-lepton quality features that define this
selection are passed in as conditioning information but the
generator does not produce events whose lepton tracks satisfy all
four cuts simultaneously at the correct rate. We discuss this
failure mode further in Sec.~\ref{sec:limitations}.

\FloatBarrier% flush all Results floats before the Method section begins

%======================================================================
\section{Method}
\label{sec:method}

Flow matching~\cite{Lipman:CFM2023} casts generation as the solution
of an ordinary differential equation (ODE). A velocity field
$u_t^{\theta}(x)$ on $\mathbb{R}^{d}$, parameterised by a neural
network, defines the initial-value problem
\begin{equation}
  \frac{dx_t}{dt} = u_t^{\theta}(x_t),
  \qquad
  x_{t=0} = x_0 \sim p_{\mathrm{init}} ,
  \label{eq:fm_ode}
\end{equation}
whose solution carries a sample $x_0$ of a simple base distribution
$p_{\mathrm{init}}$ (here a standard Gaussian) to a generated sample
$x_{t=1}$. Training adjusts $\theta$ so that the distribution of
$x_{t=1}$ matches the data distribution $p_{\mathrm{data}}$. The
link between a velocity field and the distribution it transports is
the continuity equation: a field $u_t^{\mathrm{target}}$ generates
the probability path
$p_t : [0, 1] \to \mathcal{P}(\mathbb{R}^{d})$ with endpoints
$p_{t=0} = p_{\mathrm{init}}$ and $p_{t=1} = p_{\mathrm{data}}$ when
\begin{equation}
  \partial_t\, p_t(x)
  + \nabla_x \!\cdot\! \bigl(p_t(x)\, u_t^{\mathrm{target}}(x)\bigr)
  = 0 .
  \label{eq:continuity}
\end{equation}
The marginal field $u_t^{\mathrm{target}}(x)$ is not tractable, but
it does not need to be. For a data point $z \sim p_{\mathrm{data}}$
and a noise sample $x_0 \sim p_{\mathrm{init}}$, the straight-line
path $x_t = (1 - t)\, x_0 + t\, z$ is generated by the conditional
field $u_t^{\mathrm{target}}(x | z) = (z - x)/(1 - t)$, which
along the path equals the constant velocity $z - x_0$. Regressing
the network on it,
\begin{equation}
  \mathcal{L}_{\mathrm{CFM}}(\theta)
  = \mathbb{E}_{t,\, x_0,\, z}\,
    \bigl\| u_t^{\theta}(x_t)
          - u_t^{\mathrm{target}}(x_t | z) \bigr\|^{2} ,
  \label{eq:cfm_euclidean}
\end{equation}
with $t \sim \mathcal{U}(0, 1)$, has the same $\theta$-gradient as
the intractable regression on the marginal
field~\cite{Lipman:CFM2023}. Each training step therefore only draws
a time, a data event, and a noise sample. No simulation or
likelihood evaluation is required. This is Conditional Flow Matching (CFM). The
Euclidean construction, however, places no constraint on where the
generated samples live, yet collider objects satisfy on-shell relations
that a free vector in $\mathbb{R}^{d}$ does not respect. We
therefore train the Riemannian extension of CFM, summarised next.

\subsection{Conditional flow matching on Riemannian manifolds}
\label{sec:method_rfm}

Riemannian flow matching~\cite{Chen:RiemannianFM2024} transfers the
construction above onto a curved space $\mathcal{M}$ with three
replacements: straight lines become geodesics, vector differences
become logarithm maps, and the Euclidean norm becomes the norm
induced by the metric.

Let $\mathcal{M}$ carry at every point $x$ a metric $g_x$, an inner
product on the tangent space $T_x \mathcal{M}$ (the local
linearisation of $\mathcal{M}$ at $x$). On every manifold used in
this work, $g_x$ is simply the Euclidean dot product of an ambient
embedding space restricted to the tangent space, so
$g_x(v, w) = \langle v, w \rangle$ for tangent vectors $v, w$. Two
maps replace vector addition and subtraction. The exponential map
$\exp_x : T_x \mathcal{M} \to \mathcal{M}$ takes a tangent vector
$v$ to the endpoint of the geodesic that leaves $x$ with initial
velocity $v$ and runs for unit time; the logarithm map
$\log_x : \mathcal{M} \to T_x \mathcal{M}$ is its inverse where
defined, returning the initial velocity of the unit-time geodesic
from $x$ to its argument. Their argument types differ because the
two maps invert one another: $\exp_x(\log_x y) = y$.

The conditional path anchored on a data point $z$ is the geodesic
interpolant
\begin{equation}
  x_t = \exp_{x_0}\!\bigl(t\, \log_{x_0} z\bigr) ,
  \label{eq:geodesic_interp}
\end{equation}
the curved-space analogue of the straight line, and the conditional
field that generates it takes the closed
form~\cite{Chen:RiemannianFM2024}
$u_t^{\mathrm{target}}(x | z) = \log_x(z)/(1 - t)
\in T_x \mathcal{M}$, of which the Euclidean field above is the
flat-space case. The objective becomes the Riemannian conditional
flow matching loss
\begin{equation}
  \mathcal{L}_{\mathrm{RCFM}}(\theta)
  = \mathbb{E}_{t,\, x_0,\, z}\,
    \bigl\| u_t^{\theta}(x_t) - u_t^{\mathrm{target}}(x_t | z)
    \bigr\|_{g(x_t)}^{2} ,
  \label{eq:cfm}
\end{equation}
with $\|w\|_{g(x)}^{2} \equiv g_x(w, w)$ and the network output
projected onto the tangent space at $x_t$. The full derivation is
given in~\cite{Chen:RiemannianFM2024}. When
$\mathcal{M} = \mathbb{R}^{d}$ and $g_x \equiv \mathbf{I}$, every
element above reduces to its Euclidean counterpart from
Eq.~\eqref{eq:cfm_euclidean}. Generation solves the initial-value
problem of Eq.~\eqref{eq:fm_ode} on the manifold, stepping with
$\exp_{x_t}$ so that the state never leaves $\mathcal{M}$. What
remains is the physics input, the choice of $\mathcal{M}$ and its
charts, which is the subject of the next subsection.

\subsection{On-shell manifolds for collider objects}
\label{sec:method_manifolds}

A physical particle satisfies the on-shell relation
$E^{2} = m^{2} + |\mathbf{p}|^{2}$, and a generator should not be
able to produce anything else. Modelling each reconstructed particle
as a free four-vector in $\mathbb{R}^{4}$ offers no such guarantee:
the components $(E, p_x, p_y, p_z)$ fluctuate independently, the
generated mass $m^{2} = E^{2} - |\mathbf{p}|^{2}$ can land anywhere
(including at unphysical $m^{2} < 0$), and the invariant mass of a
$K$-body system,
\begin{equation}
\begin{aligned}
  m_{K}^{2}
  &= \sum_i (E_i^{2} - |\mathbf{p}_i|^{2})
     + 2 \!\sum_{i < j}\! (E_i E_j - \mathbf{p}_i \cdot \mathbf{p}_j) ,
\end{aligned}
\label{eq:mK_decomp}
\end{equation}
inherits independent fluctuations from every diagonal and angular
component, broadening any sharp resonance into a continuum. We
therefore place each particle on its on-shell manifold from the
outset, so that the mass-shell condition holds by construction for
every generated sample.

Doing so raises two practical difficulties. First, the mass shell of
a particle of mass $m$ is a hyperboloid whose curvature scales as
$1/m^{2}$. Its $\exp_x$ and $\log_x$ maps are built from $\cosh$ and
$\sinh$ of the boost rapidity $\chi = \operatorname{arccosh}(E/m)$,
and for an electron at LHC energies $\chi \approx 12$, deep in the
regime where these functions overflow and lose precision in
single-precision arithmetic, a problem several orders of
magnitude harsher for electrons and muons than for the heaviest
jets. Second, reconstructed composites such as hadronic taus and
jets do not carry a single rest mass. Their reconstructed $m$ is a
distribution broadened by the jet algorithm and the underlying QCD
shower. We solve the first problem by factorising the per-particle
manifold so that all curvature is confined to a unit two-sphere
$S^{2}$, with the remaining degrees of freedom on flat charts, and
the second by promoting mass to a generated coordinate in a log-mass
chart for the composite-mass sector.

\paragraph{Charts.}
Table~\ref{tab:charts} collects the four charts and their encode and
decode maps: encode takes a reconstructed four-momentum to the chart
coordinates the flow operates on, decode inverts it at generation
time. Throughout, $k$ indexes the six reconstructed object types,
and $E_{\mathrm{ref},k}$ and $m_{\mathrm{floor},k}$ are per-type
reference constants estimated on the training set.

\begin{table}[!htbp]
  \caption{\label{tab:charts}%
    Per-coordinate charts. Each reconstructed four-momentum is
    encoded into chart coordinates before the flow acts on it and
    decoded back at generation time. The on-shell row is not a
    generated coordinate: it is the relation the decode enforces,
    which ties $E$, $|\mathbf{p}|$, and $m$ together for every
    generated particle. Flat (non-$S^{2}$) coordinates are normalized
    per type after encoding, Eq.~\eqref{eq:whitening}.}
  \begin{ruledtabular}
    \begin{tabular}{lll}
      \textbf{Chart} & \textbf{Encode} & \textbf{Decode} \\
      \hline
      Direction ($S^{2}$)
        & $\hat{\bm n} = \mathbf{p}/|\mathbf{p}|$
        & $\mathbf{p} = |\mathbf{p}|\, \hat{\bm n}$ \\
      Log energy
        & $y_E = \log(E / E_{\mathrm{ref},k})$
        & $E = E_{\mathrm{ref},k}\, e^{y_E}$ \\
      Boost rapidity
        & $\chi = \operatorname{arccosh}(E/m)$
        & $E = m \cosh\chi$ \\
      Log mass
        & $z_m = \log(m / m_{\mathrm{floor},k})$
        & $m = m_{\mathrm{floor},k}\, e^{z_m}$ \\
      On-shell (imposed)
        & $m = \sqrt{E^{2} - |\mathbf{p}|^{2}}$
        & $|\mathbf{p}| = \sqrt{E^{2} - m^{2}}$ \\
    \end{tabular}
  \end{ruledtabular}
\end{table}

The \emph{direction chart} on $S^{2}$ parameterises the
three-momentum direction $\hat{\bm n} = \mathbf{p}/|\mathbf{p}|$,
with the metric obtained by restricting the Euclidean dot product
$\langle \cdot, \cdot \rangle$ of the ambient $\mathbb{R}^{3}$ to
the sphere. Geodesics on the unit sphere are great circles with
closed-form expressions, so the three maps the training loop needs
(the logarithm, the exponential, and their composition along the
conditional path) require no numerical integration. For points
$x, y \in S^{2}$ separated by the angle
$\theta = \arccos\langle x, y\rangle$ and a tangent vector
$v \in T_x S^{2}$, they read~\cite{Shoemake:slerp1985}:
\begin{equation}
\begin{aligned}
  \log_x(y)
    &= \frac{\theta}{\sin\theta}\,
       \bigl(y - \langle x, y\rangle\,x\bigr) , \\[2pt]
  \exp_x(v)
    &= \cos\lVert v\rVert\,x
     + \frac{\sin\lVert v\rVert}{\lVert v\rVert}\,v , \\[2pt]
  \mathrm{slerp}(\hat n_0, \hat n_1; t)
    &= \frac{\sin\!\bigl((1 - t)\,\Omega\bigr)}{\sin\Omega}\,\hat n_0
     + \frac{\sin\!\bigl(t\,\Omega\bigr)}{\sin\Omega}\,\hat n_1 .
\end{aligned}
\label{eq:slerp}
\end{equation}
Here $\hat n_0$ and $\hat n_1$ are the direction coordinates of the
noise sample and of the data sample,
$\cos\Omega = \hat n_0 \cdot \hat n_1$, and the slerp
(spherical linear interpolation) expression
on the third line is the closed-form value of
$\exp_{\hat n_0}\!\bigl(t\, \log_{\hat n_0}\hat n_1\bigr)$,
i.e.\ the geodesic interpolant of
Eq.~\eqref{eq:geodesic_interp} on $S^{2}$. We work with the slerp
form directly because the $\theta / \sin\theta$ ratio in $\log_x$
loses precision when the two directions approach back-to-back
($\theta \to \pi$), while the slerp expression stays stable over
almost the full range of angular separations. In the opposite limit
of nearly parallel directions ($\Omega \to 0$), where the
$\sin\Omega$ denominator itself degenerates, we fall back to linear
interpolation followed by re-projection onto the sphere.

The three flat charts of Table~\ref{tab:charts} carry the remaining
degrees of freedom. The \emph{log-energy chart} $y_E$ parameterises
the energy of the fixed-mass sector, where the rest mass is a known
constant and energy is the only magnitude to generate. The
\emph{boost-rapidity chart} $\chi \in [0, \infty)$ parameterises the
momentum magnitude of a massive object so that the on-shell relation
$m^{2} = E^{2} - |\mathbf{p}|^{2}$ reads
\begin{equation}
  E = m \cosh\chi, \qquad |\mathbf{p}| = m \sinh\chi ,
  \label{eq:on_shell}
\end{equation}
exact by construction. The \emph{log-mass chart} $z_m$ promotes the
reconstructed mass of the composite sector to a generated
coordinate, with the per-type floor $m_{\mathrm{floor},k}$ keeping
$\chi$ well-defined at small reconstructed masses.

\paragraph{Per-particle manifolds.}
The massless sector\footnote{The photon is exactly massless and the
electron is treated as such ($m_e \simeq 0$). The muon keeps its PDG
mass, imposed at decode through
$|\mathbf{p}| = \sqrt{E^{2} - m_\mu^{2}}$. Its manifold is
nevertheless that of the massless sector.}
($e,\, \mu,\, \gamma$) is parameterised on
\begin{equation}
  \mathcal{M}_{\mathrm{massless}}
  = \mathbb{R} \times S^{2},
  \qquad
  \psi := (y_E,\, \hat n) ,
  \label{eq:massless_manifold}
\end{equation}
with the rest mass fixed to its PDG value and the energy recovered as
$E = E_{\mathrm{ref},k} \exp(y_E)$. The massive sector
($\tau_{\mathrm{had}}$, small-$R$ jet, large-$R$ jet) lives on
\begin{equation}
  \mathcal{M}_{\mathrm{massive}}
  = \mathbb{R} \times S^{2} \times \mathbb{R},
  \qquad
  \psi := (\chi,\, \hat n,\, z_m) ,
  \label{eq:massive_manifold}
\end{equation}
with the reconstructed mass generated by the flow. In both cases the
full four-momentum at decode time satisfies
$E^{2} - |\mathbf{p}|^{2} = m^{2}$ automatically by construction.
Table~\ref{tab:manifolds} collects the per-type manifold assignment
across the six reconstructed object types. The per-event manifold is
the product of per-particle manifolds over the objects present in the
event,
$\mathcal{M} = \prod_{p \in \text{event}} \mathcal{M}_{\mathrm{type}(p)}$,
and the conditional flow factorises across particles. Missing
transverse energy enters the model as a conditioning input via
cross-attention, not as a generated coordinate.

\begin{table}[!h]
  \caption{\label{tab:manifolds}%
    Per-type on-shell manifold assignment and generated chart
    coordinates. The massless sector
    ($\mathcal{M}_{\mathrm{massless}}$, Eq.~\eqref{eq:massless_manifold})
    fixes the rest mass to its PDG value and only the energy magnitude
    is generated. The massive sector
    ($\mathcal{M}_{\mathrm{massive}}$, Eq.~\eqref{eq:massive_manifold})
    also generates the reconstructed mass. MET conditions the model
    via cross-attention and is not generated. ``Dim'' is the
    chart-embedding dimension on which the flow operates, with the
    unit direction $\hat n \in S^{2}$ counted by its three $\mathbb{R}^{3}$
    components. The on-shell condition is enforced by the manifold
    geometry rather than by the chart dimension.}
  \begin{ruledtabular}
    \begin{tabular}{l l c l}
      \textbf{Type} & \textbf{Manifold} & \textbf{Dim} & \textbf{Chart} \\
      \hline
      Electron      & $\mathbb{R} \times S^{2}$              & 4 & $(y_E, \hat n)$ \\
      Muon          & $\mathbb{R} \times S^{2}$              & 4 & $(y_E, \hat n)$ \\
      Photon        & $\mathbb{R} \times S^{2}$              & 4 & $(y_E, \hat n)$ \\
      $\tau_{\mathrm{had}}$
                    & $\mathbb{R} \times S^{2} \times \mathbb{R}$ & 5 & $(\chi, \hat n, z_m)$ \\
      Small-$R$ jet & $\mathbb{R} \times S^{2} \times \mathbb{R}$ & 5 & $(\chi, \hat n, z_m)$ \\
      Large-$R$ jet & $\mathbb{R} \times S^{2} \times \mathbb{R}$ & 5 & $(\chi, \hat n, z_m)$ \\
    \end{tabular}
  \end{ruledtabular}
\end{table}

The Euclidean chart coordinates are normalized per type, so that
geodesic distances sit on a comparable scale across object types,
\begin{equation}
  \widetilde y_E      = \frac{y_E - \mu_{y_E,k}}{\sigma_{y_E,k}}, \quad
  \widetilde \chi     = \frac{\chi - \mu_{\chi,k}}{\sigma_{\chi,k}}, \quad
  \widetilde z_m      = \frac{z_m - \mu_{z_m,k}}{\sigma_{z_m,k}} ,
  \label{eq:whitening}
\end{equation}
with per-type means $\mu_{\bullet,k}$ and standard deviations
$\sigma_{\bullet,k}$ measured on the training set and inverted at
decode time. The $S^{2}$ factor is passed through unchanged, since
its round metric already has a fixed unit curvature shared by every
particle type.

\subsection{Architecture and dual-head loss}
\label{sec:method_architecture}

\begin{figure*}[!tbp]
  \centering
  \includegraphics[scale=0.85]{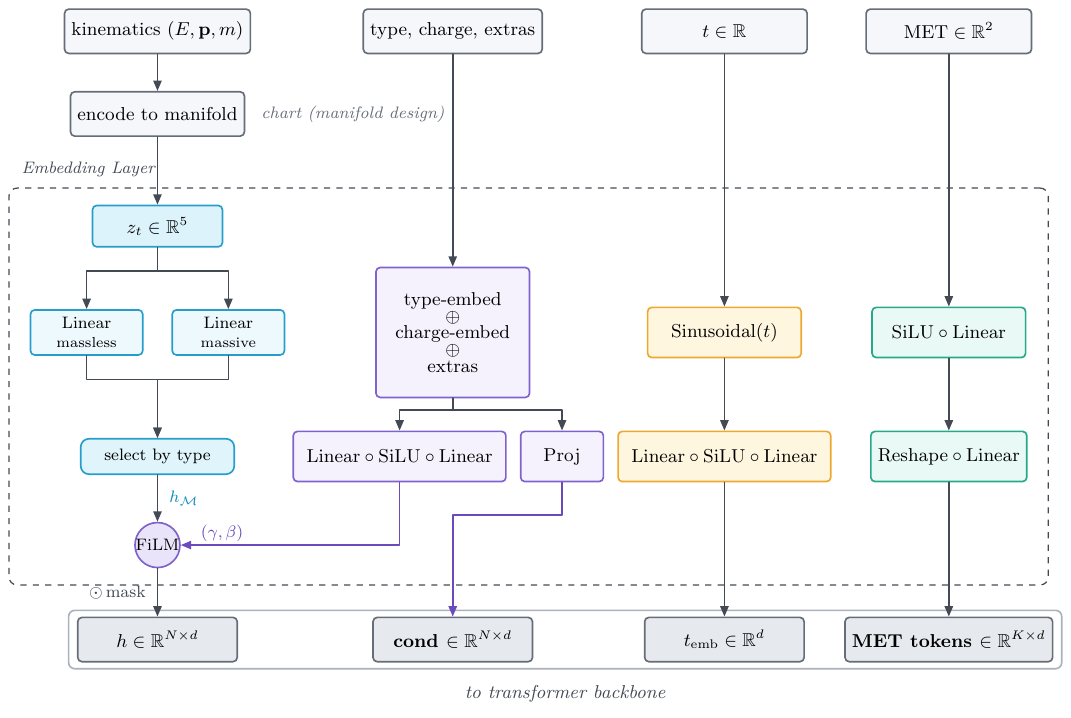}
  \caption{\label{fig:embedding_layer}%
    \textbf{Embedding-layer architecture.} Raw inputs (top row, grey, outside
    the model) are lifted into the four conditioning streams of the
    transformer backbone: a per-particle token $h$, a per-particle
    adaLN-Zero modulation signal $\mathrm{cond}$, a global time token
    $t_{\mathrm{emb}}$, and $K$ MET cross-attention tokens. The
    diffusion state $z_t$ on the chart of
    Sec.~\ref{sec:method_manifolds} is the only stream that is fed
    into the type-conditional two-path linear projection (massless or
    massive branch).}
\end{figure*}

The model is a transformer~\cite{vaswani2017attention} preceded by a
type-aware embedding layer and followed by two parallel output heads.
The embedding layer (Fig.~\ref{fig:embedding_layer}) lifts the raw
per-particle inputs into the four conditioning streams of the
backbone: a per-particle token $h$, which carries the diffusion
state; a per-particle modulation signal $\mathrm{cond}$, built from
the particle type, charge, and per-type detector extras, which
drives the adaptive layer-norm modulation inside each block; a
global time token $t_{\mathrm{emb}}$, the sinusoidal encoding of the
flow time $t$; and a small set of MET tokens. The chart-coordinate
path is the only one that carries the diffusion state
$\psi_t \equiv z_t$, the per-particle chart coordinates of the state
at flow time $t$: the interpolant $x_t$ of
Eq.~\eqref{eq:geodesic_interp} during training, and the integrated
flow state during generation. It is processed through two parallel linear
projections, one for the massless and one for the massive
per-particle manifolds of Sec.~\ref{sec:method_manifolds}, and a
type-conditional selector routes each particle through the
projection that matches its on-shell manifold. The event-level
missing transverse energy $\mathrm{MET} \in \mathbb{R}^{2}$ is
projected into $K$ cross-attention tokens consumed once per block by
the backbone, not generated by the flow.

\begin{figure*}[!tbp]
  \centering
  \includegraphics[scale=0.85]{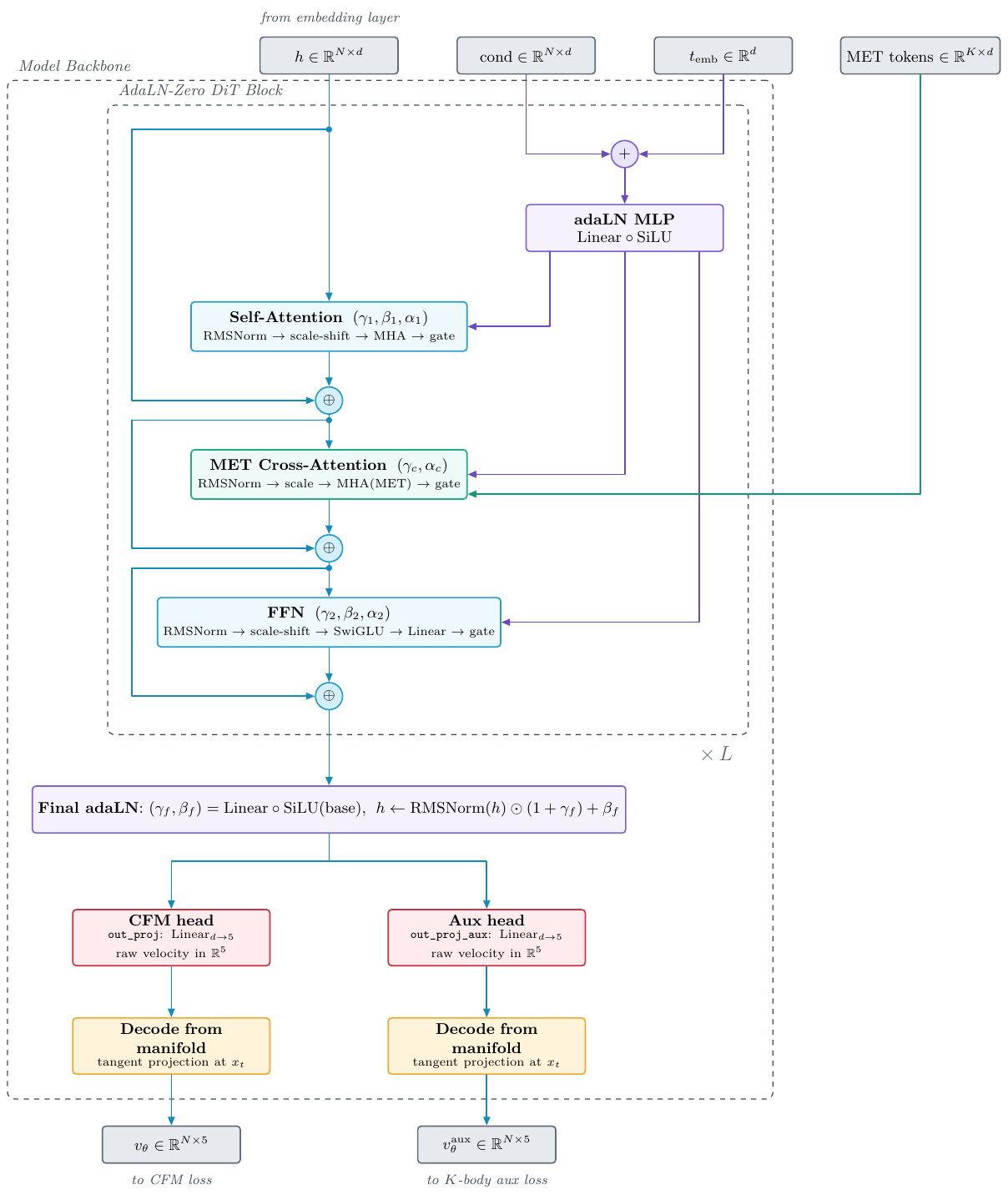}
  \caption{\label{fig:backbone}%
    \textbf{Model backbone.} A stack of $L$ adaLN-Zero DiT blocks operates on
    the per-particle token stream $h$ produced by the embedding layer
    of Fig.~\ref{fig:embedding_layer}. Each block routes $h$ through
    multi-head self-attention (MHA), multi-head cross-attention to the
    MET tokens (MHA(MET)), and a feed-forward sub-layer, each preceded
    by RMSNorm~\cite{zhang2019rmsnorm}\footnote{$\mathrm{RMSNorm}(x)
    = x \,/\, \sqrt{\frac{1}{d}\sum_{i=1}^{d} x_i^{2} + \epsilon}$,
    applied per token over the feature dimension $d$.} and gated by
    adaLN modulation parameters derived from
    $t_{\mathrm{emb}} + \mathrm{cond}$. A final adaLN modulation feeds
    two parallel output heads: the primary head predicts the
    per-particle velocity $v_\theta$ consumed by the flow-matching ODE
    integrator, and the auxiliary head predicts per-particle quantities
    that are combined into the $K$-body invariant-mass targets of
    Eq.~\eqref{eq:mK_targets}.}
\end{figure*}

The backbone (Fig.~\ref{fig:backbone}) is a stack of $L$ Diffusion
Transformer (DiT) blocks with adaLN-Zero
modulation~\cite{peebles2023scalable}, followed by a final
adaLN modulation and two parallel output heads. The trained
configuration uses $L = 6$ blocks of width
$d_{\mathrm{model}} = 128$ with four attention heads, for a total of
$3.0 \times 10^{6}$ trainable parameters
(Table~\ref{tab:hyperparams}). The physics closures of
Sec.~\ref{sec:results} are obtained already at this compact model
size. The residual stream
$h$ runs vertically through each block. The three sub-layers
(self-attention, MET cross-attention, and a position-wise
feed-forward network) branch off,
each preceded by RMSNorm~\cite{zhang2019rmsnorm} and modulated by the
adaLN parameters
$(\gamma_\bullet, \beta_\bullet, \alpha_\bullet)$ produced once per
block from $t_{\mathrm{emb}} + \mathrm{cond}$, and join back into the
residual stream. Both heads predict a per-particle velocity. The
\emph{primary head} produces the velocity
$v_\theta(\psi_t, t) \equiv u_t^{\theta}(\psi_t)$
consumed by the flow-matching ODE integrator at sampling time, and it
carries the loss described next. The \emph{auxiliary head} produces
a parallel velocity $v_\theta^{\mathrm{aux}}$ that never touches the
integrator: it exists solely to carry $K$-body invariant-mass
supervision back to the shared backbone, and is described together
with its loss below. Splitting
the supervision in this way is essential. In single-head ablations
the $K$-body Huber added directly to the primary loss dominated the
optimisation, the resonance peaks sharpened at the expense of the
underlying continuum, and the intra-particle CFM marginals of
Sec.~\ref{sec:results_singleparticle} degraded measurably. Isolating
the $K$-body signal on its own head breaks this coupling and allows
the resonance-sharpening and the continuum-modelling losses to be
tuned independently.

The primary loss is the RCFM objective of Eq.~\eqref{eq:cfm} with the
quadratic norm replaced by a Huber loss~\cite{Huber:1964robust} on
the velocity residual,
\begin{equation}
  H_\delta(r) =
  \begin{cases}
    \tfrac{1}{2}\, r^{2} & |r| \leq \delta , \\[2pt]
    \delta \bigl(|r| - \tfrac{\delta}{2}\bigr) & |r| > \delta ,
  \end{cases}
  \label{eq:huber}
\end{equation}
with $\delta = 1$, quadratic near zero and linear in the tails so
that rare hard events cannot dominate the gradient. The residual is
modulated by a soft per-particle quality weight
$q_i^{\mathrm{eff}}$:
\begin{equation}
  \mathcal{L}_{\mathrm{RCFM}}
  = \mathbb{E}\!\left[\, q_i^{\mathrm{eff}}\,
    H_\delta\!\bigl(v_\theta(\psi_t, t)_i
                  - u_t^{\mathrm{target}}(\psi_t)_i\bigr)
  \,\right],
  \label{eq:cfm_primary}
\end{equation}
where the expectation is taken \emph{per particle type, per event},
so that each event contributes equally to the gradient regardless of
its reconstructed multiplicity. The effective weight $q_i^{\mathrm{eff}}$
acts as a per-particle \emph{particleness}, the likelihood
that the reconstructed object is a genuine particle rather than a
misreconstructed one, and is a linear floor of the raw quality
score $q_i \in [0, 1]$,
\begin{equation}
  q_i^{\mathrm{eff}}
  = q_{\mathrm{floor}}^{\mathrm{cfm}}
    + \bigl(1 - q_{\mathrm{floor}}^{\mathrm{cfm}}\bigr)\, q_i
  \;\in\; [\,q_{\mathrm{floor}}^{\mathrm{cfm}},\, 1\,],
  \label{eq:q_eff}
\end{equation}
with floor $q_{\mathrm{floor}}^{\mathrm{cfm}} = 0.5$. The floor keeps
the model exposed to the bulk continuum while steering it towards
well-reconstructed objects: high-particleness objects pull on the
gradient with full weight, low-quality objects only half. The raw
quality score $q_i$ is a non-learned type-conditional sigmoid product
of detector quality features chosen to match standard ATLAS analysis
cuts. The explicit per-type definition and the corresponding cuts and
sigmoid slopes are collected in Appendix~\ref{app:quality_score}.

The auxiliary loss exposes the $K$-body invariant-mass identity
directly to the gradient. The auxiliary velocity
$v_\theta^{\mathrm{aux}}$ is decoded to per-particle four-momenta
together with their time derivatives, from which pair sums
$(E_{ij}, \mathbf{p}_{ij}, \dot E_{ij}, \dot{\mathbf{p}}_{ij})$ are
constructed once and reused at $K = 3$ (pair plus single) and
$K = 4$ (pair of disjoint pairs). Writing
$\mu_K(\mathcal{I}) \equiv \log m_K^{2}(\mathcal{I})$ for the log
squared invariant mass of a $K$-tuple $\mathcal{I}$ of active
particles, the per-tuple residuals on the position and velocity of
$\mu_K$ are passed through a Huber loss with tuple weight
$q_\mathcal{I} = \prod_{i \in \mathcal{I}} q_i$ and
distribution-matching weight $w^{\star}_\mathcal{I}$:
\begin{equation}
\begin{aligned}
  \mathcal{L}_{\mathrm{aux}}
  &= \sum_{K=2}^{K_{\max}}
     \bigl\langle\, q_\mathcal{I}\, w^{\star}_\mathcal{I}\,
     \rho_\mathcal{I}(t) \,\bigr\rangle_{\mathcal{I} \in \mathcal{C}_K}, \\
  \rho_\mathcal{I}(t)
  &= \mathbf{1}_{t > t_\mathrm{gate}}\,
     H_\delta(\Delta^{\mathrm{pos}}_\mathcal{I})
   + H_\delta(\Delta^{\mathrm{vel}}_\mathcal{I}),
\end{aligned}
\label{eq:aux_loss}
\end{equation}
with the per-tuple residuals
\begin{equation}
\begin{aligned}
  \Delta^{\mathrm{pos}}_\mathcal{I}
    &= \log m_K^{\mathrm{pred}}(\mathcal{I})
       - \log m_K^{\mathrm{truth}}(\mathcal{I}), \\
  \Delta^{\mathrm{vel}}_\mathcal{I}
    &= \dot\mu_K^{\mathrm{pred}}(\mathcal{I})
     - \dot\mu_K^{\mathrm{truth}}(\mathcal{I}).
\end{aligned}
\label{eq:mK_targets}
\end{equation}
Here $\mathcal{C}_K$ is the set of active $K$-tuples per event and
$\langle \cdot \rangle$ is the
$q_\mathcal{I} w^{\star}_\mathcal{I}$-weighted average over tuples.
The loss runs over \emph{all} $K$-tuples of active particles for
$K \in \{2, 3, 4\}$, with no restriction on particle type or charge,
and no specific channel is targeted. The ceiling $K_{\max} = 4$ covers
the invariant masses that appear in two-, three-, and four-body
final states, of which the dilepton resonances and the four-lepton
Higgs channel are examples rather than targets. The position
term is gated to late flow times ($t > t_\mathrm{gate} = 0.99$) so
the auxiliary signal does not dominate the flow at early diffusion
times where the destination mass is still buried in noise.

The distribution-matching weight $w^{\star}$ is the central
ingredient that lets the auxiliary loss resolve narrow peaks above
the surrounding continuum. We estimate Gaussian kernel densities of
the $K$-body log-mass distribution on both the truth and the
generator sides,
\begin{equation}
\begin{aligned}
  \hat p(\log m) &= \frac{1}{n h} \sum_{j=1}^{n}
    \mathcal{K}\!\Bigl(\tfrac{\log m - \log m_j}{h}\Bigr) , \\
  \mathcal{K}(u) &= \tfrac{1}{\sqrt{2\pi}}\, e^{-u^{2}/2} ,
\end{aligned}
\label{eq:kde}
\end{equation}
with bandwidth $h = 0.01$ in log-mass units, and combine them into
the smoothed importance-weight
\begin{equation}
  w^{\star}(r) = \bigl((1 - \eta)\, r^{k} + \eta\bigr)^{-\alpha} ,
  \quad
  r = \frac{\hat p_{\mathrm{gen}}(\log m)}{\hat p_{\mathrm{truth}}(\log m)} ,
  \label{eq:dm_weight}
\end{equation}
parameterised by an exponent $\alpha$, a sharpness $k$, and a
truth-prior mix $\eta$ that sets the ceiling
$w^{\star}_{\max} = \eta^{-\alpha}$. At a deficit ($r \to 0$) the
weight saturates at the ceiling; at a match ($r = 1$) the weight is
unity; at an excess ($r \to \infty$) the weight tends to zero. The
$\eta$-mix removes the naive-ratio singularity at
$\hat p_{\mathrm{gen}} \to 0$ that would otherwise produce a
discontinuous loss landscape.

The total training loss is the sum of the primary and auxiliary
contributions,
\begin{equation}
  \mathcal{L}_{\mathrm{total}}
  = \mathcal{L}_{\mathrm{RCFM}} + \mathcal{L}_{\mathrm{aux}} ,
  \label{eq:total_loss}
\end{equation}
with the internal weights of Eq.~\eqref{eq:aux_loss} chosen so that
the two contributions are of comparable magnitude after a short
warm-up.

\subsection{Dataset and training}
\label{sec:method_dataset}

The model is trained on the de-duplicated union of the 2-to-4 lepton
and 1LMET30 skims of the ATLAS Open Data 13~TeV $pp$
release~\cite{ATLAS:opendata:1LMET30, ATLAS:opendata:2to4lep},
filtered to events whose total reconstructed particle multiplicity
satisfies $N_{\mathrm{particles}} \leq 8$ to keep the per-batch
padding fraction low. The union is constructed by retaining each
ATLAS event exactly once: events that appear in both source skims
under the same $(\mathrm{run}, \mathrm{event})$ identifier are kept
in a single copy. The retained sample is split $95/5$ with seed $42$
into roughly $7.8 \times 10^{8}$ training events and $4.1 \times 10^{7}$
held-out validation events. Each event is
preprocessed into an unordered set of reconstructed objects across
the six supported types
($e,\, \mu,\, \tau_\mathrm{had},\, \gamma,\, \text{jet},\, \text{large-}R\,\text{jet}$),
together with the event-level missing transverse energy used as
conditioning. We retain the standard ATLAS analysis fields used to
construct the per-type quality score and the kinematic charts.
Further details of the preprocessing pipeline and the per-type
reference scales are deferred to the Supplemental Material. The
full list of selected features is given in
Appendix~\ref{app:selected_fields}.

The model is optimised end-to-end with
AdamW~\cite{loshchilov2019decoupled}, with decoupled weight decay and
gradients clipped to a global norm of $1.0$ before each update. The
learning rate follows a cosine schedule with a peak of
$6 \times 10^{-4}$, a $10{,}000$-step linear warm-up,
and a floor of $10^{-6}$. An exponential moving average~(EMA) of the
parameters $\theta^{\mathrm{EMA}}_t = \rho\, \theta^{\mathrm{EMA}}_{t-1} + (1-\rho)\, \theta_t$
with decay $\rho = 0.9995$ is maintained throughout training, and all
evaluation, sample generation, and reported metrics use
$\theta^{\mathrm{EMA}}$ rather than the live training weights.
Training runs in distributed data-parallel on eight GPUs at a
per-device batch size of $4{,}096$, giving an effective global batch
of $32{,}768$ events. Mixed-precision \texttt{bf16} is used
throughout, and the model is compiled with \texttt{torch.compile}.
The full schedule runs for $30$ epochs on the joint sample
($7.2 \times 10^{5}$ optimiser steps). The complete set of training
hyperparameters, together with the hardware and software environment
of the training run, is collected in
Appendix~\ref{app:training_details}
(Tables~\ref{tab:hyperparams} and~\ref{tab:system}). The CFM
forward process is the geodesic interpolant of
Eq.~\eqref{eq:geodesic_interp} on the per-type chart, with the flow
time drawn uniformly on $[10^{-2},\, 1 - 10^{-6}]$, clipped at both
ends to keep the interpolant away from its endpoint degeneracies. At
inference time generation reduces to numerical integration of
$v_\theta$ from $t = 0$ to $t = 1$ with a fixed-step solver of $200$
steps, each step applied through the exponential map on the
spherical factor so the state remains on the manifold.
The model generates only the particle kinematics ($p_T$, $\eta$,
$\phi$, $E$, $m$). The particle type, charge, MET, and the remaining
reconstruction features are conditioning inputs, taken at sampling
time directly from the validation events. Every truth-vs.\ generated
comparison in Sec.~\ref{sec:results} is therefore paired on
identical composition, and any distributional difference is
attributable to the generated kinematics alone.

%======================================================================
\section{Discussion}
\label{sec:discussion}

\subsection{Design choices that mattered}
\label{sec:design}

The headline result, that resonance peaks emerge at PDG positions
without the model ever being given the masses, depends on two physics
priors built into the model rather than learned from data. The first
is the on-shell condition, embedded in the chart geometry of
Sec.~\ref{sec:method_manifolds}. Without it, the resonance peaks are
smeared away. The second is the
hierarchical $K$-body auxiliary loss of
Sec.~\ref{sec:method_architecture}, which supervises the
invariant mass of every $K$-tuple of particles directly. Without
it, peaks below the high-statistics $Z$
mass are either absent or broad enough to merge into the surrounding
continuum. Both statements are demonstrated by the ablation of
Sec.~\ref{sec:ablation}.

Two soft re-weighting mechanisms shape where the gradient signal
lands. The per-particle quality score $q_i$ multiplies every
per-particle contribution, so well-reconstructed objects pull harder
on the gradient than fakes. The distribution-matching weight
$w^{\star}$ along the $K$-body mass axis boosts the loss where the
generator under-produces and damps it where it overshoots. Without
$w^{\star}$, a bulk of excess events would be generated in the
high-statistics middle of the mass spectrum, with compensating
deficits in the adjacent continuum. The dual-head architecture
prevents the strong gradient signal from peaks like $J/\psi$ and
$\Upsilon$ from dragging neighbouring events into the peak via shared
output weights. Encoding the three-momentum direction on $S^{2}$
removes the need for the model to learn
$\phi$ periodicity and produces a uniform $\phi$ marginal by
construction.

The distribution-matching weight $w^{\star}$ is therefore a form of
mass-density supervision. It tells the auxiliary head where on the
log-mass axis the model is under-producing relative to truth, and
corrects it. It is not peak supervision. The weight carries a
single marginal density, the truth distribution of $\log m_K^{2}$.
Nothing in it specifies the angular separation of the decay
products, the flavour rule that selects OSSF
leptons, the boost of the parent particle that places a peak above
the continuum, or the joint correlation structure between any two of
these. The closure tests on
$\Delta\phi(\ell\ell,\,\mathrm{MET})$, $\Delta R_{\ell\ell}$, and
the Drell--Yan forward--backward asymmetry, none of which has a
corresponding mass-density target, are recovered by the primary head
from the underlying kinematics rather than supplied by $w^{\star}$.

Despite these designs, \textsc{ShellFlow} fails to reconstruct three known
peaks: the $\omega(782)$, the $\phi(1020)$, and the Higgs. For the
two sub-GeV mesons, the forward noising washes the peaks into the
continuum earlier than any other feature of the dilepton spectrum
(App.~\ref{app:noise_grid_dilepton}). The Higgs peak emerges only
after the joint per-lepton quality selection, which leaves a sample
too rare for the model to learn at the present training statistics.
Section~\ref{sec:limitations} examines both cases.

\subsection{Ablation of the two priors}
\label{sec:ablation}

\begin{figure*}[!tbp]
  \includegraphics[width=\textwidth]{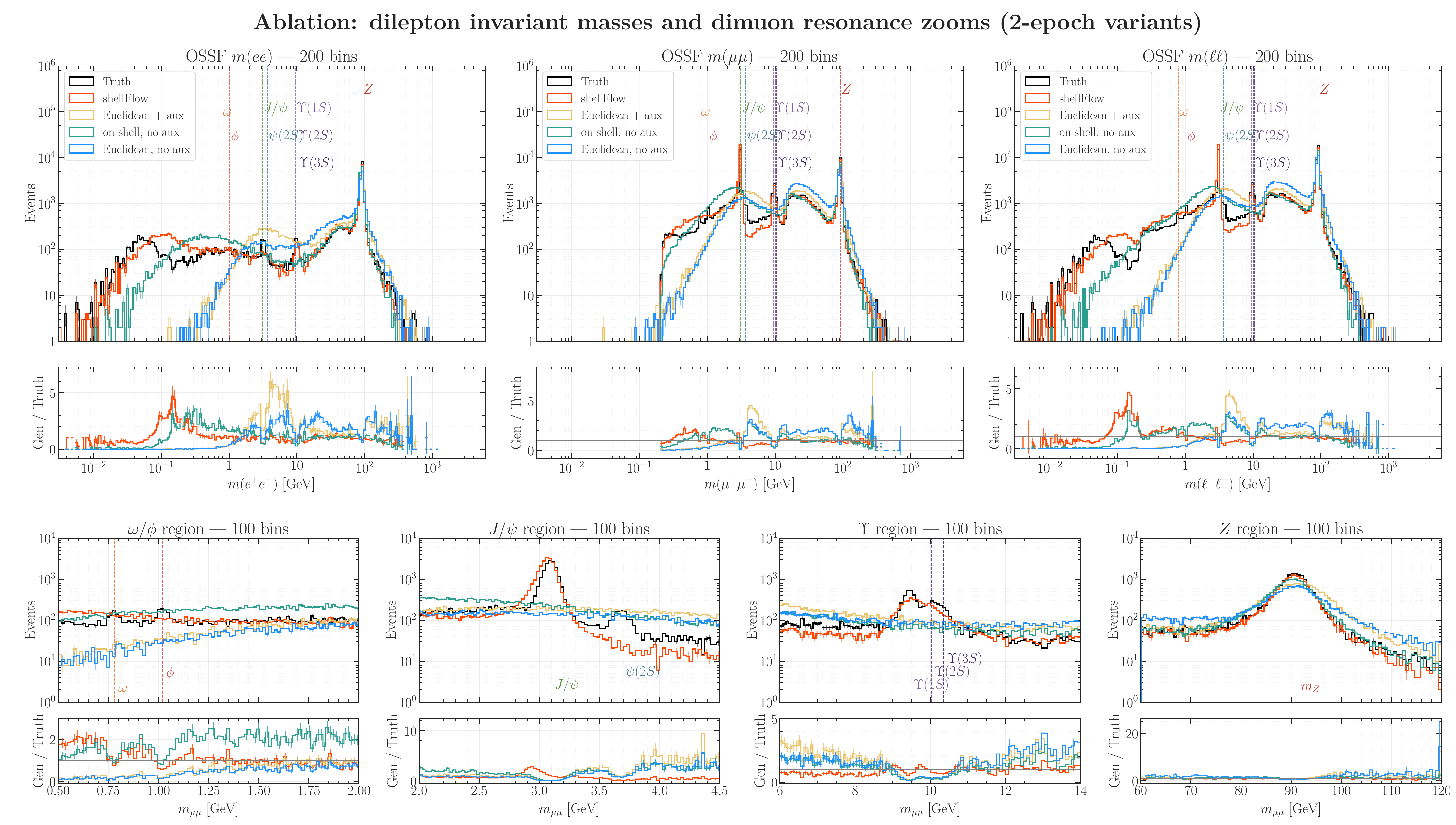}
  \caption{\label{fig:ablation}%
  \textbf{Ablation:} OSSF dilepton spectra of the four variants in
    the format of Fig.~\ref{fig:dilepton}, with the truth in black.
    \textsc{ShellFlow} is the full configuration, with both the
    on-shell chart and the $K$-body auxiliary loss; \textit{on
    shell, no aux} keeps the chart but drops the auxiliary loss;
    \textit{Euclidean + aux} keeps the auxiliary loss but generates
    energy, direction, and momentum as independent Euclidean
    coordinates; \textit{Euclidean, no aux} has neither. Without the chart the spectrum below
    $\approx \SI{0.5}{\GeV}$ is empty and every peak washes out;
    without the auxiliary loss only a degraded $Z$ survives; only
    \textsc{ShellFlow} recovers peaks and continuum simultaneously. All
    variants are trained for two epochs, so the \textsc{ShellFlow} variant
    shown here is a shorter training of the model of
    Fig.~\ref{fig:dilepton}.}
\end{figure*}

In order to demonstrate that the on-shell chart of
Sec.~\ref{sec:method_manifolds} and the $K$-body auxiliary loss of
Sec.~\ref{sec:method_architecture} are responsible for the
resonance recovery, we retrain the model from scratch in four
configurations: (i) with both (\textsc{ShellFlow}), (ii) with
the auxiliary loss but without the chart, (iii) with the chart but
without the auxiliary loss, and (iv) with neither. The two
configurations without the chart generate $\log E$, the
three-momentum direction $\hat{\bm n} = \mathbf{p}/|\mathbf{p}|$,
and $\log|\mathbf{p}|$ as independent Euclidean coordinates, so the
energy and the momentum of a generated particle are not tied to
each other. The architecture, the dataset, and all hyperparameters
are kept identical, and each variant is trained for two epochs. The
\textsc{ShellFlow} variant is therefore a shorter training of the model of
Sec.~\ref{sec:results}. All four variants are sampled with
identical conditioning on $5 \times 10^{5}$ validation events.
Figure~\ref{fig:ablation} shows the resulting dilepton spectra, and
Table~\ref{tab:ablation} collects the metrics.

\begin{table}[!tbp]
  \caption{\label{tab:ablation}%
    Ablation metrics on the OSSF dimuon spectrum, windows as in
    Fig.~\ref{fig:ablation}, identical conditioning for all
    variants. $W_{1}$ is in \si{\GeV}, while SKL and BC are
    dimensionless. The last two rows quantify the muon on-shell
    condition, counting muons off shell at
    $|m^{2} - m_{\mu}^{2}| > \SI{0.01}{\GeV\squared}$.}
  \begin{ruledtabular}
    \begin{tabular}{lcccc}
      On-shell chart & \checkmark & --- & \checkmark & --- \\
      $K$-body auxiliary loss & \checkmark & \checkmark & --- & --- \\
      \hline
      \multicolumn{5}{l}{\textit{Full dimuon spectrum}} \\
      $W_{1}$ [\si{\GeV}] & 1.78 & 3.19 & \textbf{1.59} & 6.39 \\
      SKL & \textbf{0.013} & 0.158 & 0.043 & 0.234 \\
      BC & \textbf{0.052} & 0.242 & 0.086 & 0.323 \\
      \multicolumn{5}{l}{\textit{$\omega/\phi$ window, 0.5--2\,\si{\GeV}}} \\
      $W_{1}$ [\si{\GeV}] & 0.098 & 0.217 & \textbf{0.067} & 0.208 \\
      SKL & 0.059 & 0.186 & \textbf{0.036} & 0.183 \\
      BC & \textbf{0.137} & 0.303 & 0.291 & 0.398 \\
      \multicolumn{5}{l}{\textit{$J/\psi$ window, 2--4.5\,\si{\GeV}}} \\
      $W_{1}$ [\si{\GeV}] & \textbf{0.094} & 0.344 & 0.326 & 0.347 \\
      SKL & \textbf{0.129} & 0.770 & 0.694 & 0.752 \\
      BC & \textbf{0.192} & 0.448 & 0.454 & 0.475 \\
      \multicolumn{5}{l}{\textit{$\Upsilon$ window, 6--14\,\si{\GeV}}} \\
      $W_{1}$ [\si{\GeV}] & \textbf{0.233} & 0.887 & 0.850 & 0.859 \\
      SKL & \textbf{0.047} & 0.384 & 0.370 & 0.363 \\
      BC & \textbf{0.149} & 0.384 & 0.342 & 0.353 \\
      \multicolumn{5}{l}{\textit{$Z$ window, 60--120\,\si{\GeV}}} \\
      $W_{1}$ [\si{\GeV}] & \textbf{0.186} & 1.69 & 0.852 & 3.41 \\
      SKL & \textbf{0.008} & 0.126 & 0.035 & 0.162 \\
      BC & \textbf{0.051} & 0.208 & 0.117 & 0.248 \\
      \hline
      $\Delta R_{\ell\ell}$ $W_{1}$ & 0.073 & 0.189 & \textbf{0.056} & 0.401 \\
      median $|m^{2} - m_{\mu}^{2}|$ [\si{\GeV\squared}]
        & $\bm{7{\times}10^{-5}}$ & 1.45 & $\bm{7{\times}10^{-5}}$ & 1.12 \\
      off-shell muon fraction
        & \textbf{0.7\%} & 99.4\% & \textbf{0.7\%} & 99.2\% \\
    \end{tabular}
  \end{ruledtabular}
  \par\smallskip
  {\footnotesize The lowest (best) value of each row is shown in
  bold, and ties are both bold.}
\end{table}

Only \textsc{ShellFlow} produces the $J/\psi$ and $\Upsilon$ peaks. In
the other three variants the corresponding windows contain only
continuum. The $Z$ peak is produced by all four variants, but
\textsc{ShellFlow} matches the truth most closely, with
$\mathrm{SKL} = 0.008$ in the $Z$ window against $0.035$--$0.162$
for the ablated variants. The two variants without the chart
generate heavily off-shell muons (median
$|m^{2} - m_{\mu}^{2}| > \SI{1}{\GeV\squared}$, with more than
99\% of the muons off shell), so plain Euclidean flow matching
does not learn the on-shell condition from the data. One visible
consequence is the empty spectrum below $\approx \SI{0.5}{\GeV}$,
where nearly collinear on-shell muon pairs would lie.
Table~\ref{tab:ablation} agrees with these observations: \textsc{ShellFlow}
gives the lowest, and hence best, value of every metric
in the $J/\psi$, $\Upsilon$, and $Z$ windows. The chart geometry
is thus required for physical four-momenta, and the auxiliary loss
for resolving the narrow resonances.

\subsection{Limitations and open issues}
\label{sec:limitations}

We group the limitations of \textsc{ShellFlow} into physics issues, where a
known structure of the data is not yet reproduced by the generator,
and systematic limitations of the setup that
constrain what the generator can in principle say about real
collider events.

\paragraph{Physics issues.}
Beyond the muon $\eta$ marginal, where the muon-spectrometer
acceptance gap at $|\eta| < 0.1$~\cite{ATLAS:2016lqx} is a real
feature of the data, a consistent shallow deficit at
$\eta \approx 0$ appears across particle types that have no
physical gap there. We
attribute this to the parameterisation of the $S^{2}$ direction
chart, whose density distorts mildly at the equator of the sphere.

The Higgs peak at $m_{4\ell} \approx 125$ GeV and the
$Z \to 4\ell$ peak at $\SI{91}{GeV}$ are absent from the generated
four-lepton spectrum after the simultaneous per-lepton quality cut
(Sec.~\ref{sec:results_higgs}). The model has learned part of the
selection structure but did not converge on the combination of
per-lepton isolation, impact-parameter, and calorimeter cuts that
the analysis requires at this rare-process branching fraction.
Whether longer training closes this gap has not been tested and is
the natural next step.

The narrow sub-GeV vector mesons $\omega(782)$ and $\phi(1020)$ are
absent from the generated dilepton spectrum. The CFM forward
noising washes these peaks into the continuum substantially earlier
than any other resonance (Sec.~\ref{sec:results_dilepton}). On top
of that, the log-energy chart is normalised against per-type
reference energies of tens of GeV, where the model performs well,
which compresses the sub-GeV regime into a narrow interval of the
chart and lowers the model's effective resolution there.

The resonant cores of the generated hadronic-side mass peaks are
shallower and broader than in truth (Sec.~\ref{sec:results_top}):
at matched sample size the generated yield within one core width of
the peak falls $9\%$ below truth on the hadronic top and $23\%$ on
the hadronic $W$, the selected distributions are $10$--$18\%$
broader in their fit-free interquartile range, and a free-shape
DSCB fit cannot isolate the generated core width. The template-fit positions of these peaks
nevertheless agree with truth within the combined fit and
regeneration uncertainties of Table~\ref{tab:top_mass}, so the
closure tests are robust to the dilution, but sharpening the cores to truth prominence requires longer
training and possibly finer chart resolution on jet four-momenta.
We stress that the core dilution is not a limitation of the
demonstration the present paper is making. The model is given
no top-mass information, and the recovery of the peak at the correct
location and within the correct selection acceptance already implies
that an internal feature corresponding to top-like events must
exist. Closing the residual prominence gap is a precision question, to be
revisited once dedicated precision-measurement methodology is built
on top of the framework demonstrated here.

\paragraph{Systematic limitations.}
Three properties of the present setup constrain what the generator
can say about real collider events.

\emph{Particle identification and charge are conditioning, not
inferred.} The particle type and charge of each object are passed to
the model as conditioning inputs rather than predicted from
kinematics. The generator produces kinematics conditioned on a
given event composition and does not itself assign identities.
The conditioning is also taken as truth, so a particle mislabelled by
the reconstruction is generated as its assigned type. Once the
model is upgraded to generate the particle type and charge, the
identity can instead be treated probabilistically.

\emph{Missing transverse energy is conditioning, not generated.}
MET enters the model via cross-attention rather than being sampled
alongside the visible particles, so generated events are not
self-contained. Downstream analyses that require a self-consistent
MET, in particular those involving neutrinos, are not yet
supported.

\emph{Fixed particle multiplicity with padding.} The training set is
restricted to events with reconstructed particle multiplicity
$N_{\mathrm{particles}} \leq 8$ to keep the per-batch padding
fraction low. The high-multiplicity tail of the joint sample is
therefore discarded, and any event-level completeness claim is
implicitly conditioned on $N_{\mathrm{particles}} \leq 8$.

\subsection{Outlook}
\label{sec:outlook}

\paragraph{Physics or statistics?}
The model reproduces much of the Standard Model from the joint
sample: the resonance peaks at PDG positions, the Weinberg angle,
the top-quark mass closures, the leading-dijet kinematic envelope.
Whether the network has \emph{learned} this physics or has only
reproduced the marginals through correlations in the training
statistics is not settled by the distributions reported here. The
answer lies in the internal representation of the trained network,
and reaching it requires mechanistic interpretability of the model,
which is an ongoing line of work in our group.

\paragraph{Mechanistic interpretability.}
If the network has internalised the structures of
Sec.~\ref{sec:results}, internal features in the backbone should
activate selectively on events carrying the corresponding physics.
Locating the feature subspace that distinguishes, for example, a
$Z \to \ell\ell$ event from the Drell--Yan continuum would expose
the network's own representation of the resonance and would in
principle allow the relevant feature to be amplified at inference
time, sharpening the peak on demand.

\paragraph{Multi-signal BSM search.}
Training the same architecture on Monte-Carlo samples that contain
the Standard Model together with BSM processes turns
the generator into an event-level anomaly detector that is not tied
to any single signal hypothesis. Different BSM scenarios excite
different internal features, and the activation pattern itself
becomes the discriminator. A single trained generator then covers
many signal hypotheses at once, extending the data-driven
anomaly-detection programme~\cite{Kasieczka:2021xcg,
Karagiorgi:2021ngt} from low-dimensional background templates to
complete events.

\paragraph{Generating the unobserved.}
A further upgrade is to generate the particles that the detector
cannot see, starting with the neutrinos behind the missing
transverse energy. This step requires Monte Carlo. Trained on
simulated events in which the neutrinos are known, the model learns
to predict them from the observed particles. Applied to recorded
events, it then generates a neutrino prediction for each one, and
the resulting sample presents the full scope of the collision, the
observed part measured and the unobserved part supplied by the
model.

%======================================================================
\section{Conclusion}
\label{sec:conclusion}

\paragraph{What was learned.}
The \textsc{ShellFlow} generator, a single Riemannian Conditional Flow
Matching model with a
transformer backbone, trained on the union of the 1LMET30 and
2-to-4 lepton skims of the ATLAS Open Data 13~TeV $pp$ release with
no supervision tied to any particular channel, captures a
substantial fraction of the Standard Model directly from real
collider data. The
$J/\psi$, $\Upsilon$, and $Z$ resonances appear at their PDG
positions in both flavour channels, and the Weinberg angle is
preserved at the precision of the truth-sample leading-order fit.
The hadronic and leptonic top-quark masses, the hadronic $W$ mass,
and the leptonic $Z$ mass close against the truth under the
standard semileptonic $t\bar t$ and OSSF dilepton reconstructions.
The leading-dijet joint distribution reproduces the
$(m_{jj}, \bar\eta)$ correlation beyond its supervised $m_{jj}$
marginal, and the
event-level $p_T$ closure reproduces the two-component structure of
the joint sample. The only physics priors built into
the model are the on-shell condition and the invariant-mass
formula, both Lorentz-invariant statements that impose no bias
toward any specific channel or specific resonance. The model is
therefore independent of process and mass by construction.

\paragraph{What remains.}
The Standard Model is not yet fully captured. The narrow sub-GeV
vector mesons $\omega(782)$ and $\phi(1020)$, the sharpness of the
reconstructed top and $W$ peaks, and the Higgs $H \to 4\ell$
resonance under the simultaneous per-lepton quality cut remain
beyond the present training horizon. These gaps call for further
training and for the model extensions outlined above, in particular
the addition of a generated MET field that carries implicit
neutrinos, and the relaxation of the fixed-multiplicity assumption.

\paragraph{Why it matters.}
A generator that learns physics directly from data, rather than
from the chain of simulation and analysis cuts, opens a class of
downstream tasks that the present paper does not itself address.
Mechanistic interpretability of the trained network can ask whether
the model has truly learned the resonances or only their statistical
shadows. Searches for physics beyond the Standard Model that are
not tied to a single signal hypothesis can use the same
architecture, trained on simulated samples that contain new physics
processes, as an event-level anomaly detector. Generation of the
unobserved at the event level can extend the conditioning to the
MET field and reconstruct the full Standard Model interaction event
by event, including the particles that the detector cannot see. The
model reported here is the natural starting point for each of these.

\section*{Companion website}
\label{sec:website}

The complete set of results presented in this paper, together with
the full collection of truth-vs.\ generated validation figures, can
be found on the companion website at
\url{https://hep-ssl-webapp.pages.dev/}.

\begin{acknowledgments}
We thank Jean-Loup Tastet and Craig Wiglesworth for their comments
on the manuscript, and Arnau Morancho Tardà and Troels
C.~Petersen for helpful discussions and advice.
We acknowledge the work of the ATLAS Collaboration to record or simulate, reconstruct, and distribute the Open Data used in this paper, and to develop and support the software with which it was analysed. The GPU compute used to train the generative models was
provided by the NVIDIA Academic Grant Program. This work was
supported by a research grant (VIL57416) from VILLUM FONDEN.
\end{acknowledgments}

\section*{AI usage}
\label{sec:ai_usage}

Large language models were used in the preparation of this
manuscript to revise prose, correct grammar, and draft portions
of the text. Parts of the model implementation and the majority
of the visualization code were also generated with the assistance
of large language models, primarily through the Claude Code
agentic coding environment (Anthropic). All output was reviewed
and validated by the authors.

\clearpage% flush all main-text floats before the appendices
\appendix

\newpage
\section{Per-particle collider marginals}
\label{app:collider}

Figure~\ref{fig:single_particle_muon} of
Sec.~\ref{sec:results_singleparticle} shows the muon
intra-particle marginals $(p_T, \eta, \phi, E)$ in the main
text. The remaining five reconstructed object types are reported
here: Figs.~\ref{fig:collider_electron}--\ref{fig:collider_photon}
for the two other light types and
Figs.~\ref{fig:collider_tau}--\ref{fig:collider_largerjet} for the
three hadronic types ($\tau_{\mathrm{had}}$, small-$R$ jet,
large-$R$ jet). The massive types add the reconstructed mass $m$ to
the four collider variables.

\onecolumngrid
\begin{center}
  \includegraphics[width=\textwidth]{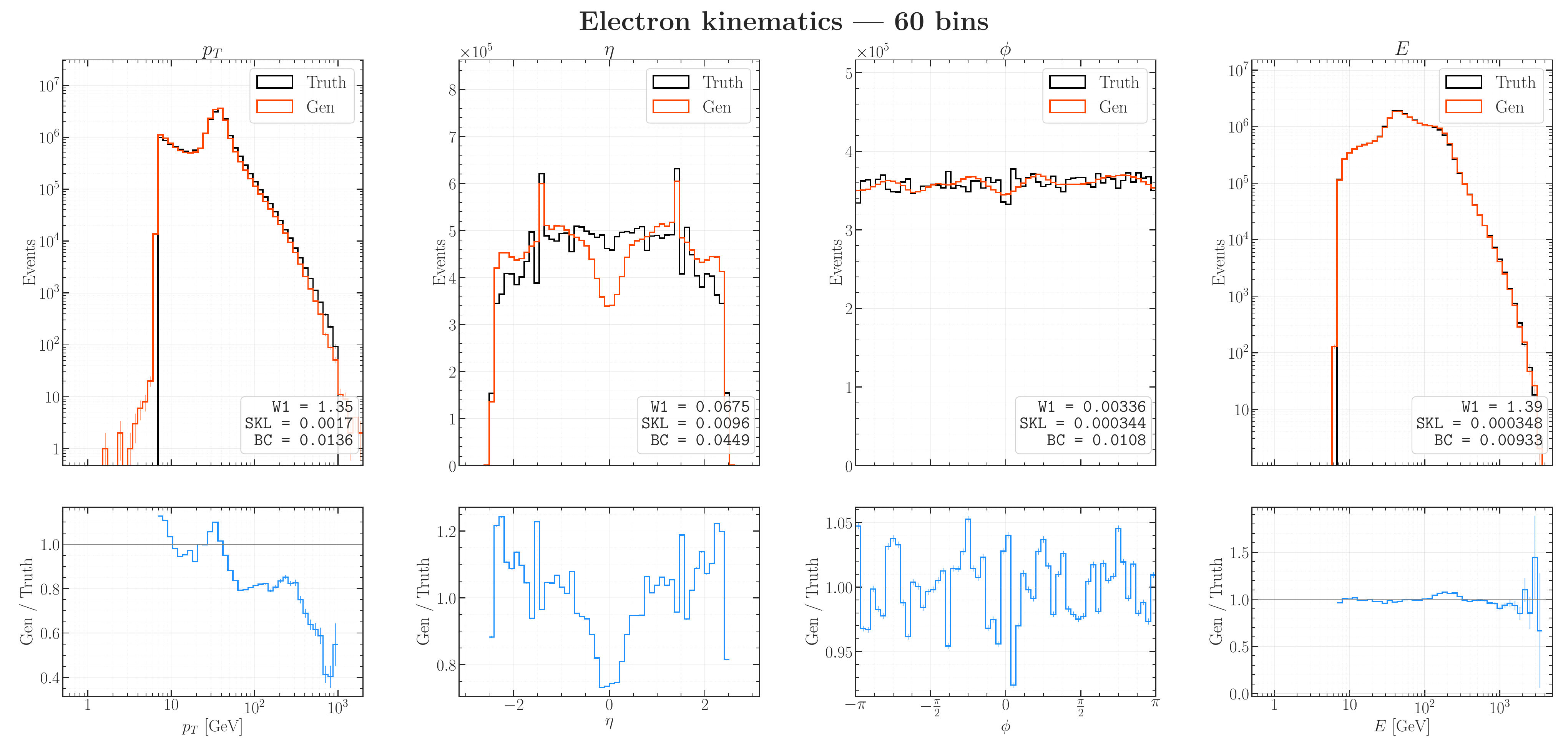}
  \captionof{figure}{\label{fig:collider_electron}%
    \textbf{Electron kinematics.}
    Electron intra-particle marginals $(p_T, \eta, \phi, E)$.
    Truth (validation set) vs.\ generated. Same conventions as
    Fig.~\ref{fig:single_particle_muon}.}
\end{center}
\twocolumngrid

\onecolumngrid
\begin{center}
  \includegraphics[width=\textwidth]{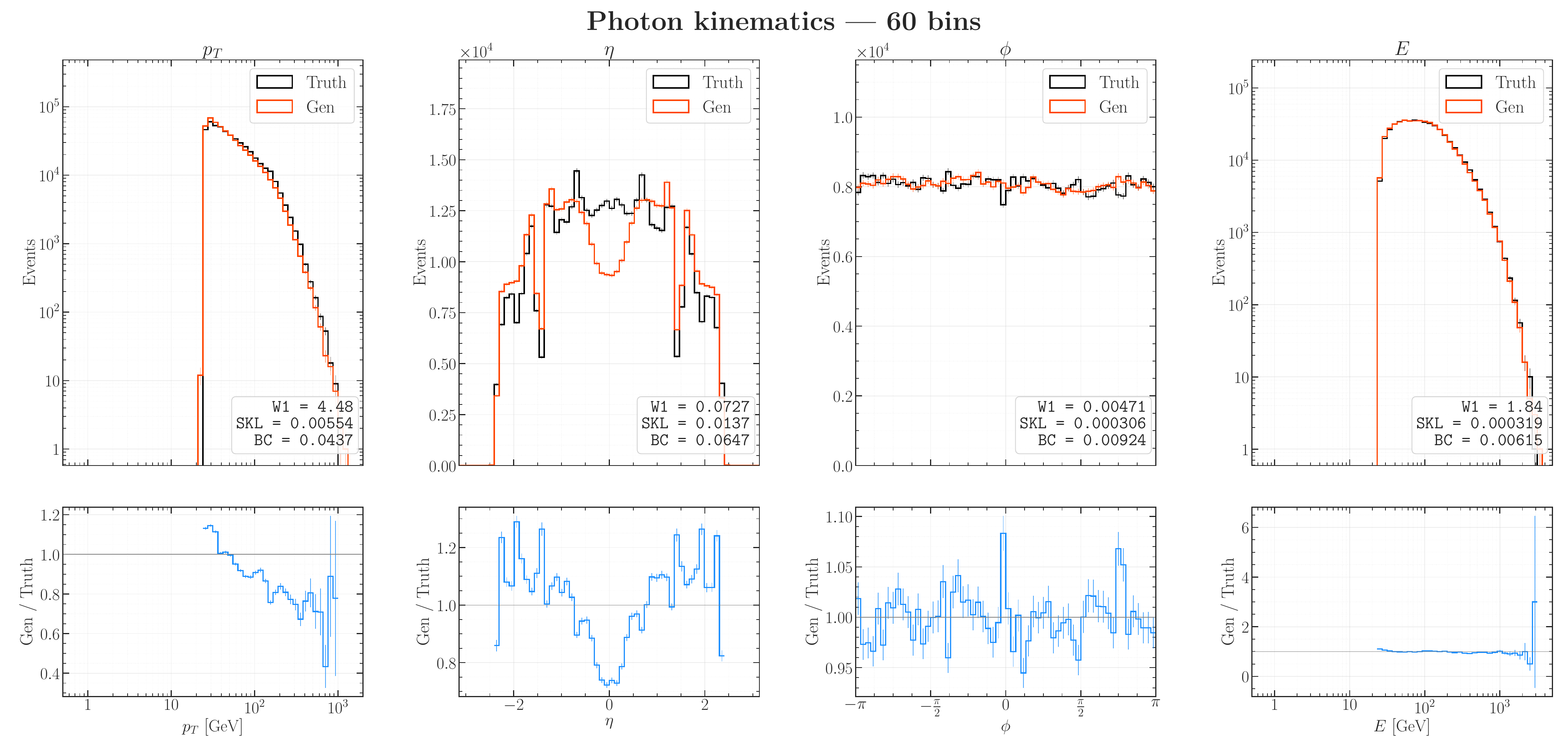}
  \captionof{figure}{\label{fig:collider_photon}%
    \textbf{Photon kinematics.}
    Photon intra-particle marginals $(p_T, \eta, \phi, E)$.
    Truth (validation set) vs.\ generated. Same conventions as
    Fig.~\ref{fig:single_particle_muon}.}
\end{center}
\twocolumngrid

\onecolumngrid
\begin{center}
  \includegraphics[width=\textwidth]{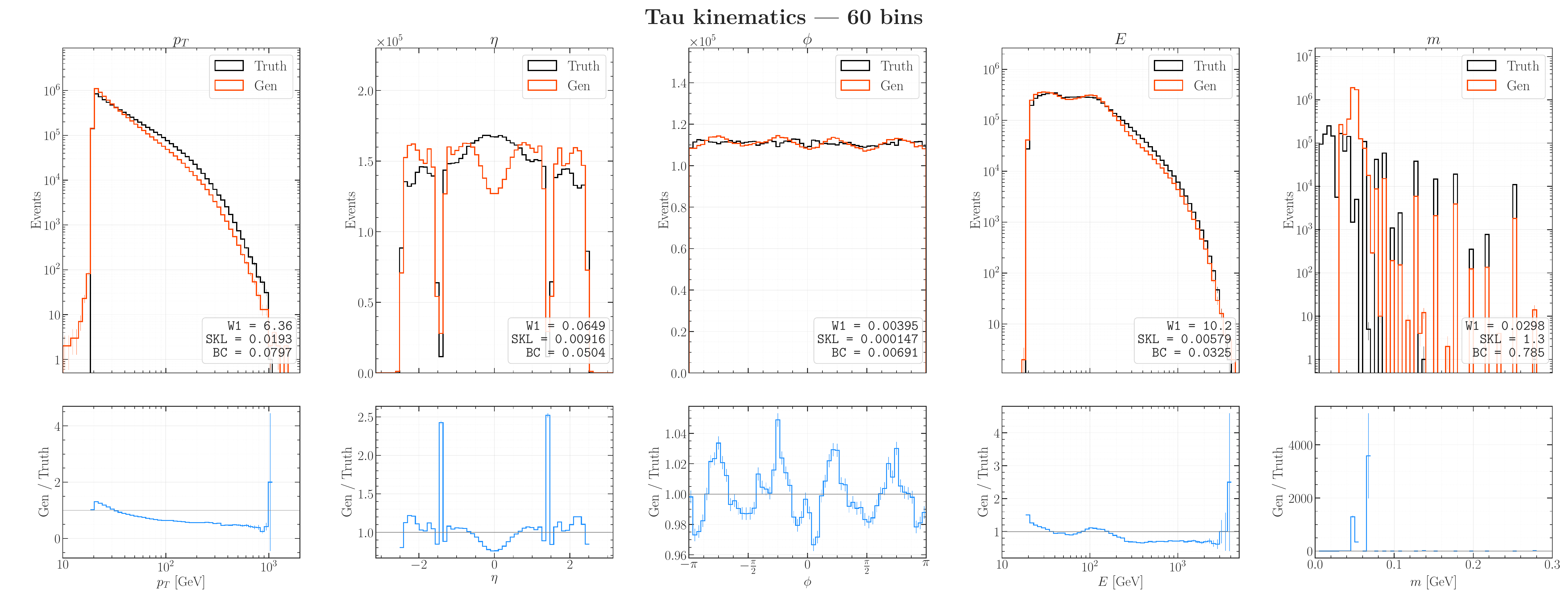}
  \captionof{figure}{\label{fig:collider_tau}%
    \textbf{Hadronic-$\tau$ kinematics.}
    Hadronic-$\tau$ intra-particle marginals
    $(p_T, \eta, \phi, E, m)$. The reconstructed mass $m$ is included
    because $\tau_{\mathrm{had}}$ uses the massive chart
    $\mathbb{R} \times S^{2} \times \mathbb{R}$
    (Table~\ref{tab:manifolds}). Truth vs.\ generated; same
    conventions as Fig.~\ref{fig:single_particle_muon}.}
\end{center}
\twocolumngrid

\onecolumngrid
\begin{center}
  \includegraphics[width=\textwidth]{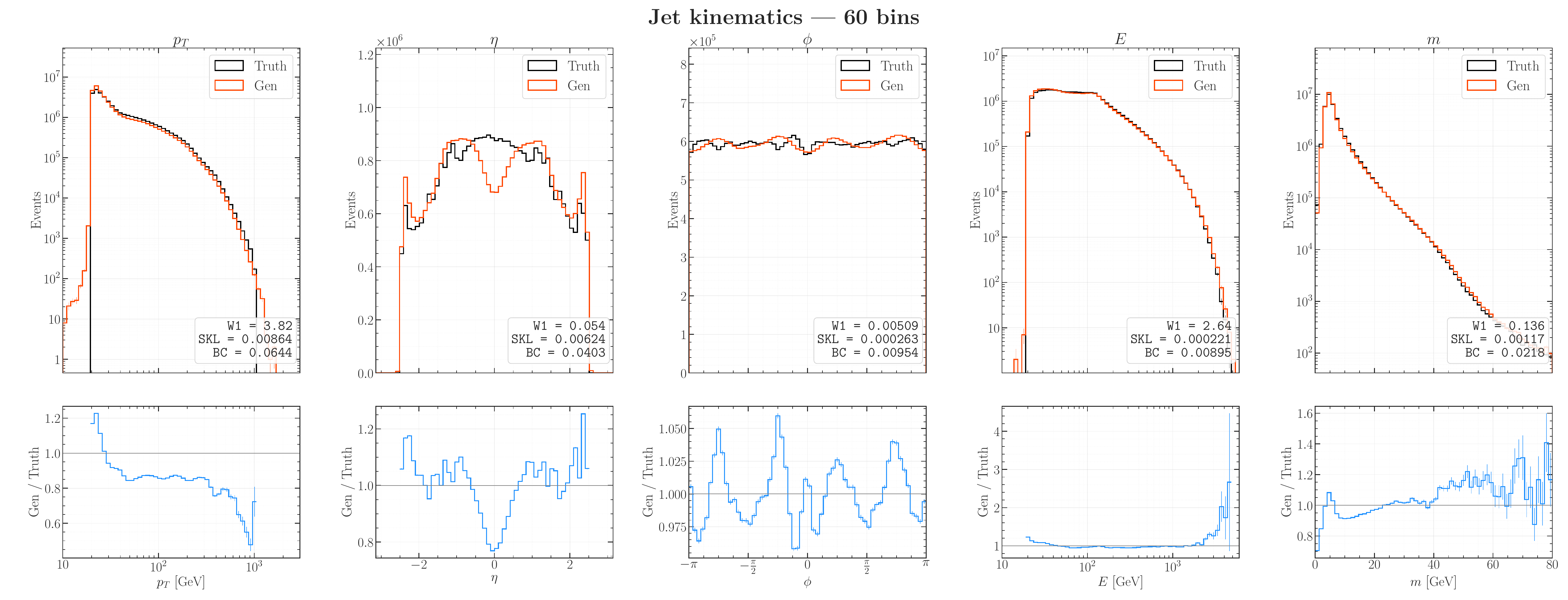}
  \captionof{figure}{\label{fig:collider_jet}%
    \textbf{Small-$R$ jet kinematics.}
    Small-$R$ jet intra-particle marginals
    $(p_T, \eta, \phi, E, m)$. The reconstructed mass $m$ is
    included because jets use the massive chart
    $\mathbb{R} \times S^{2} \times \mathbb{R}$
    (Table~\ref{tab:manifolds}). Truth vs.\ generated; same
    conventions as Fig.~\ref{fig:single_particle_muon}.}
\end{center}
\twocolumngrid

\onecolumngrid
\begin{center}
  \includegraphics[width=\textwidth]{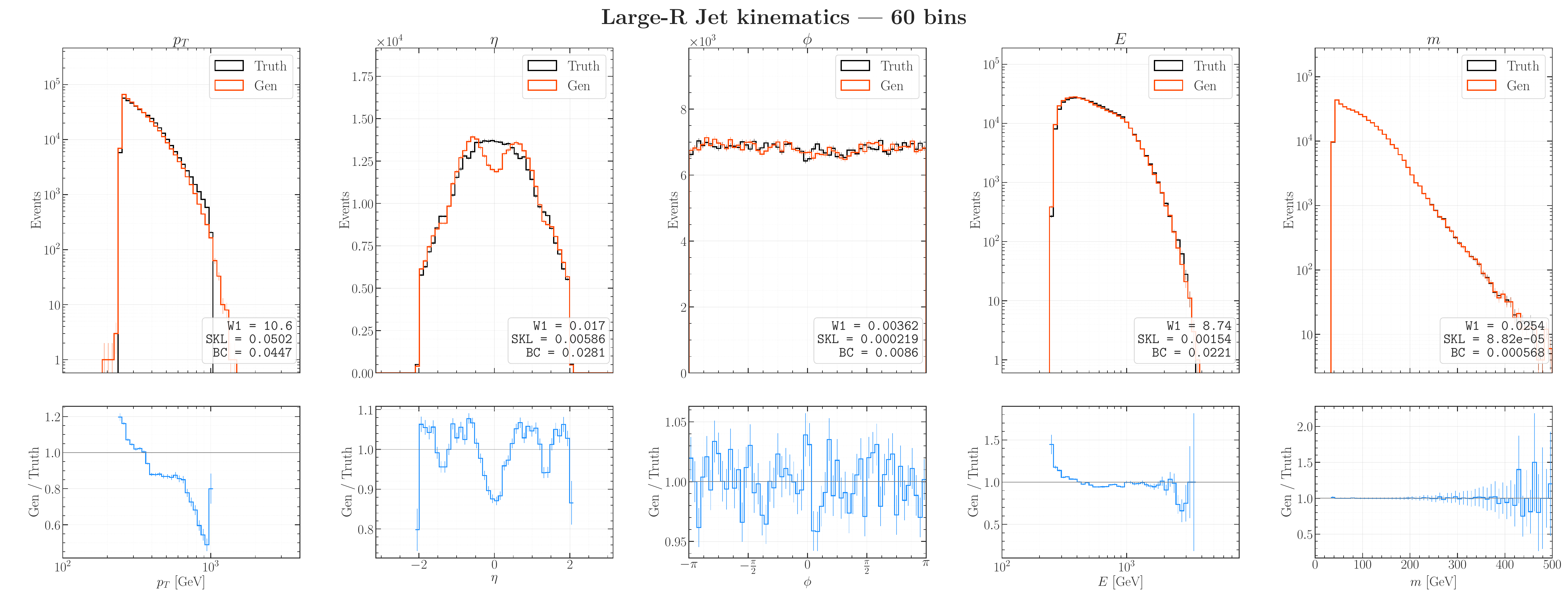}
  \captionof{figure}{\label{fig:collider_largerjet}%
    \textbf{Large-$R$ jet kinematics.}
    Large-$R$ jet intra-particle marginals
    $(p_T, \eta, \phi, E, m)$. The reconstructed mass $m$ is
    included because large-$R$ jets use the massive chart
    $\mathbb{R} \times S^{2} \times \mathbb{R}$
    (Table~\ref{tab:manifolds}). Truth vs.\ generated; same
    conventions as Fig.~\ref{fig:single_particle_muon}.}
\end{center}
\twocolumngrid

\section{Per-particle Cartesian marginals}
\label{app:cartesian}

Figures~\ref{fig:cartesian_electron}--\ref{fig:cartesian_largerjet}
report the Cartesian momentum marginals $(p_x, p_y, p_z)$ for all
six reconstructed object types. The agreement matches the collider
marginals of Sec.~\ref{sec:results_singleparticle} and
App.~\ref{app:collider}.

\onecolumngrid
\begin{center}
  \includegraphics[width=\textwidth]{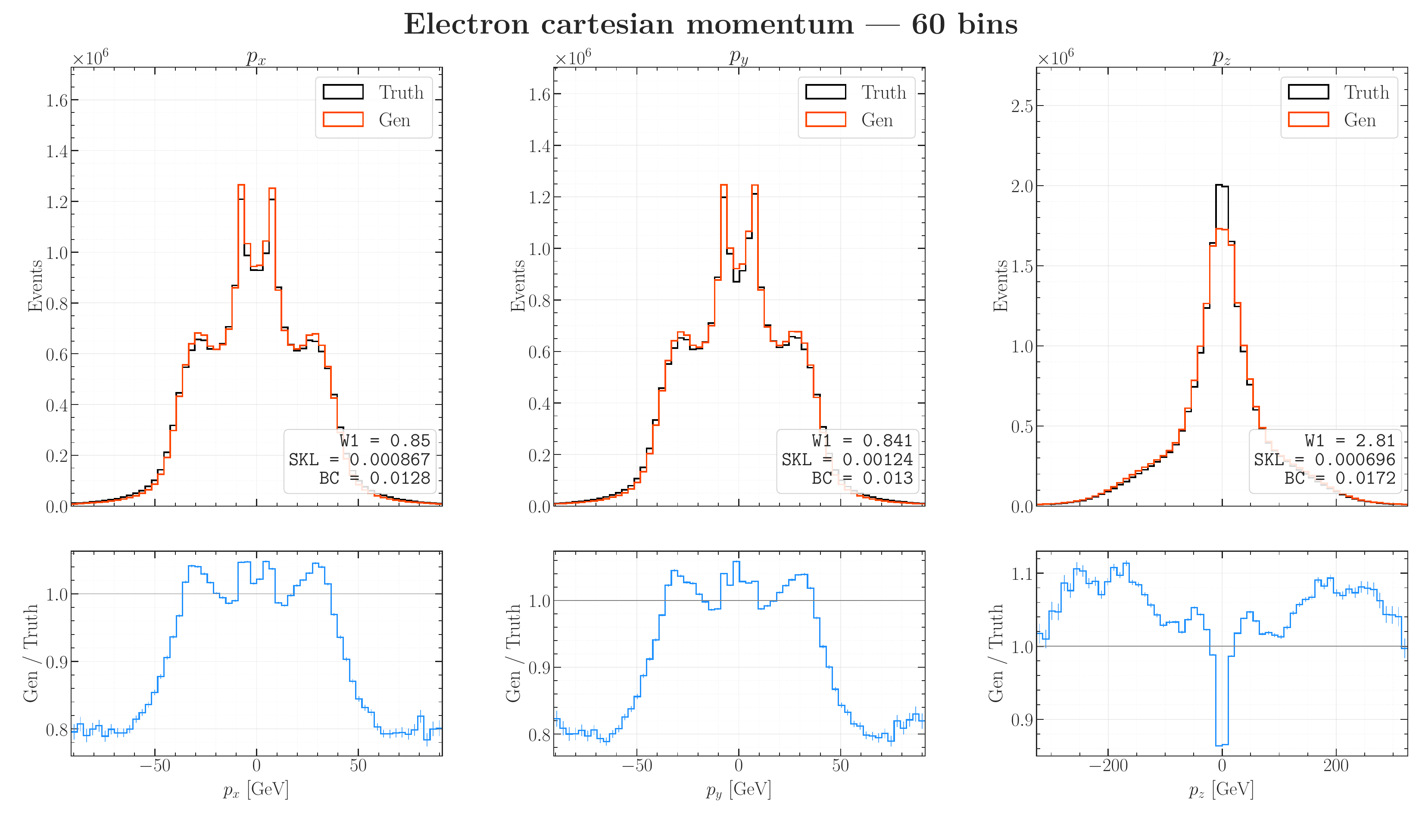}
  \captionof{figure}{\label{fig:cartesian_electron}%
    \textbf{Electron Cartesian components.}
    Electron Cartesian marginals $(p_x, p_y, p_z)$.
    Truth vs.\ generated.}
\end{center}
\twocolumngrid

\onecolumngrid
\begin{center}
  \includegraphics[width=\textwidth]{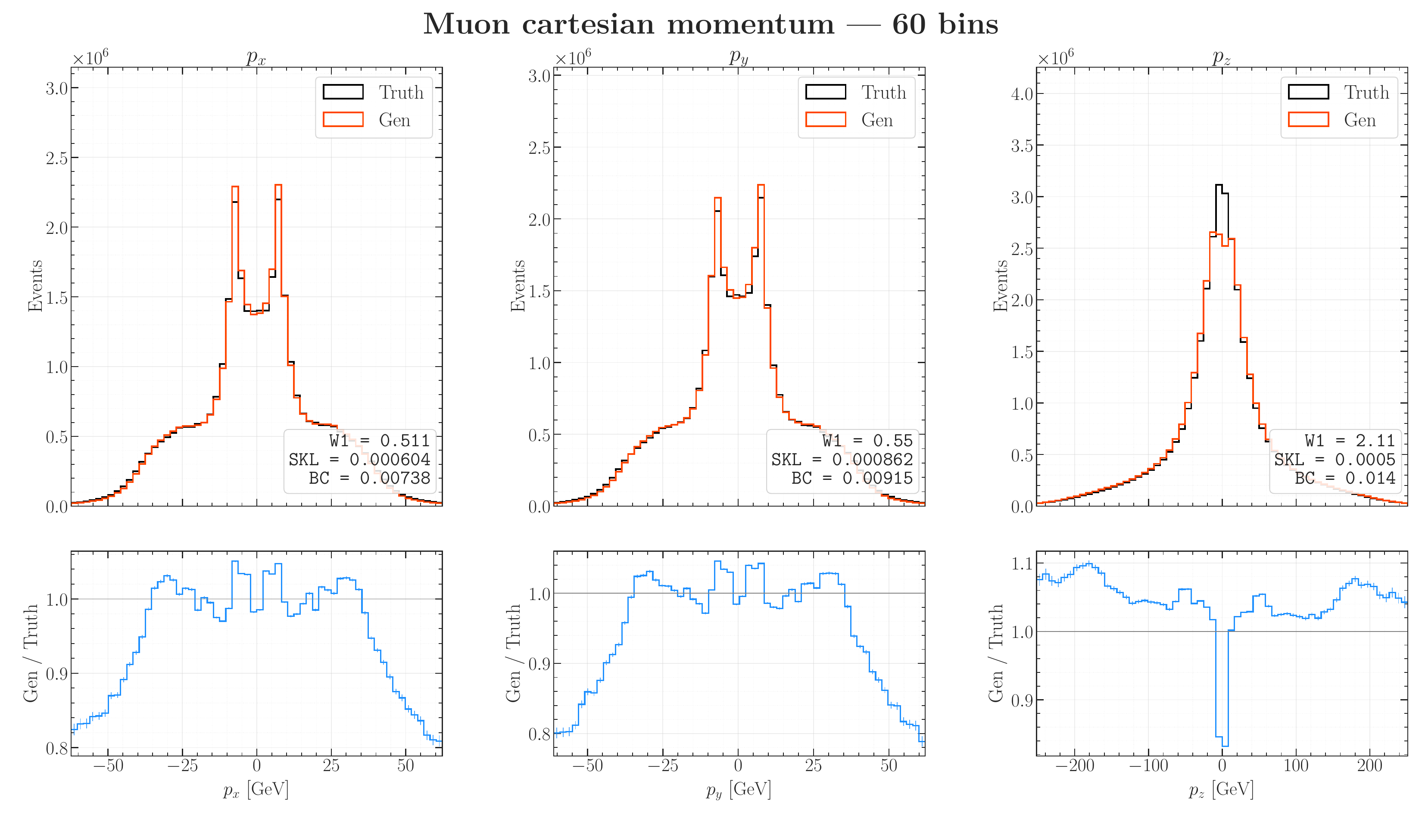}
  \captionof{figure}{\label{fig:cartesian_muon}%
    \textbf{Muon Cartesian components.}
    Muon Cartesian marginals $(p_x, p_y, p_z)$.
    Truth vs.\ generated.}
\end{center}
\twocolumngrid

\onecolumngrid
\begin{center}
  \includegraphics[width=\textwidth]{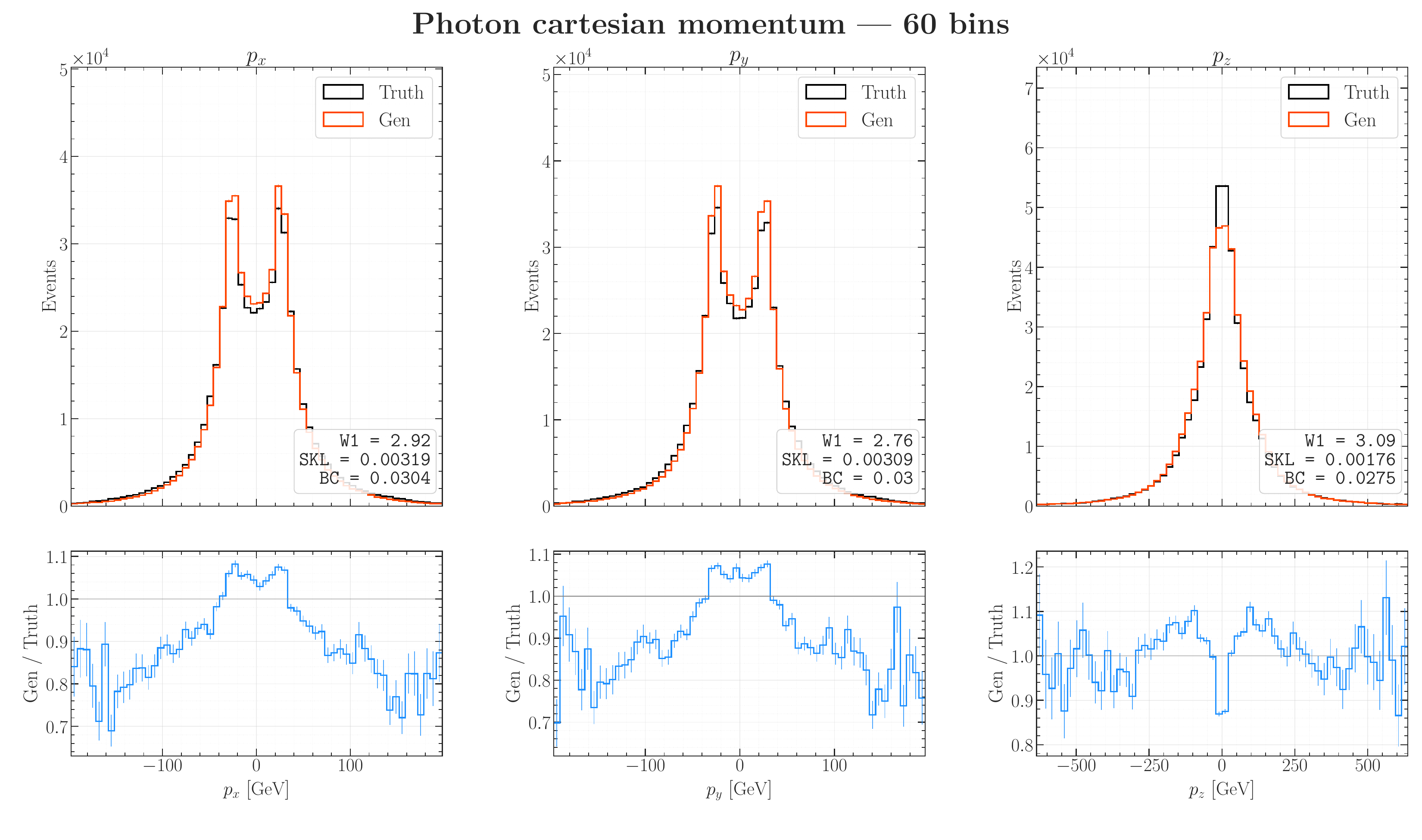}
  \captionof{figure}{\label{fig:cartesian_photon}%
    \textbf{Photon Cartesian components.}
    Photon Cartesian marginals $(p_x, p_y, p_z)$.
    Truth vs.\ generated.}
\end{center}
\twocolumngrid

\onecolumngrid
\begin{center}
  \includegraphics[width=\textwidth]{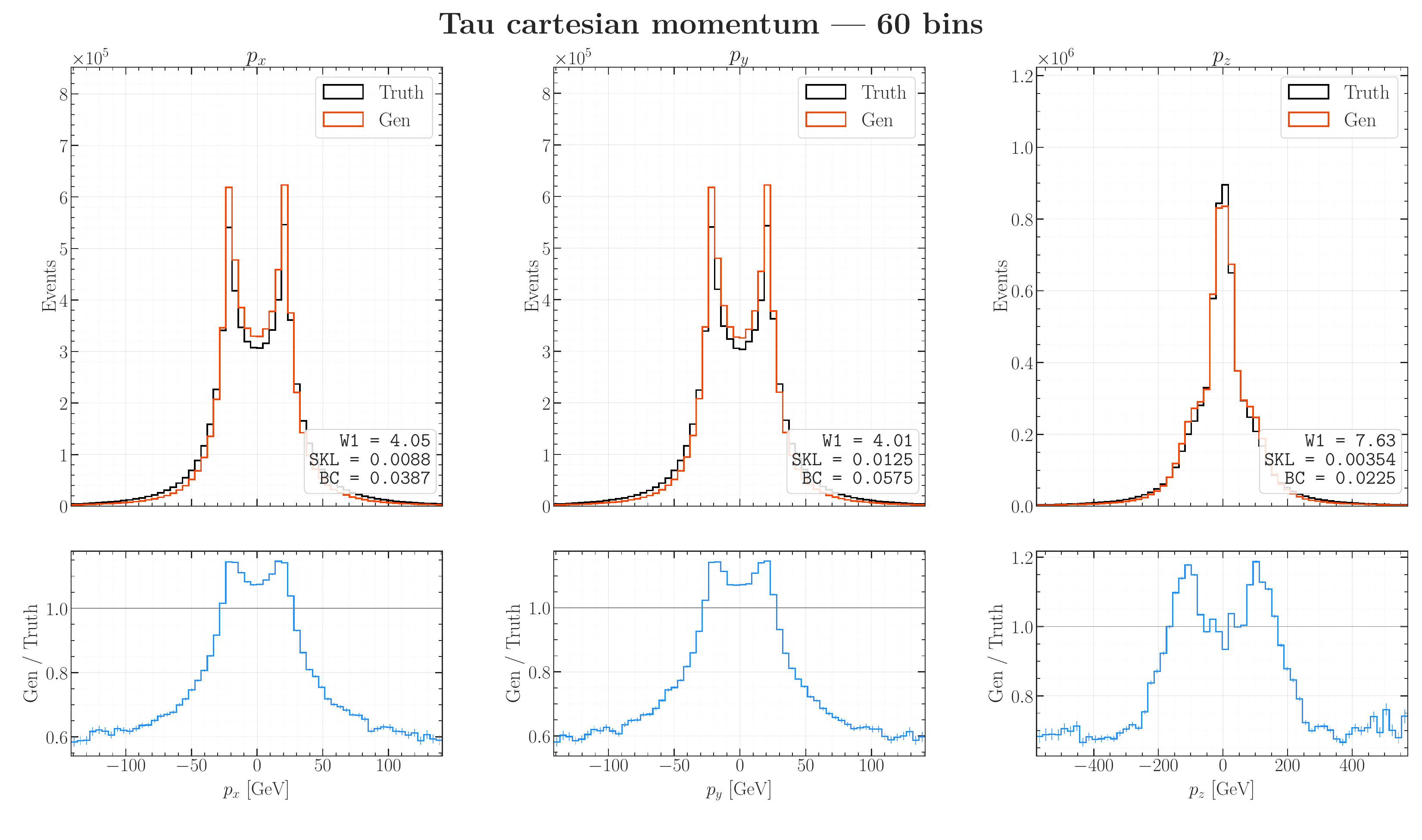}
  \captionof{figure}{\label{fig:cartesian_tau}%
    \textbf{Hadronic-$\tau$ Cartesian components.}
    Hadronic-$\tau$ Cartesian marginals $(p_x, p_y, p_z)$.
    Truth vs.\ generated.}
\end{center}
\twocolumngrid

\onecolumngrid
\begin{center}
  \includegraphics[width=\textwidth]{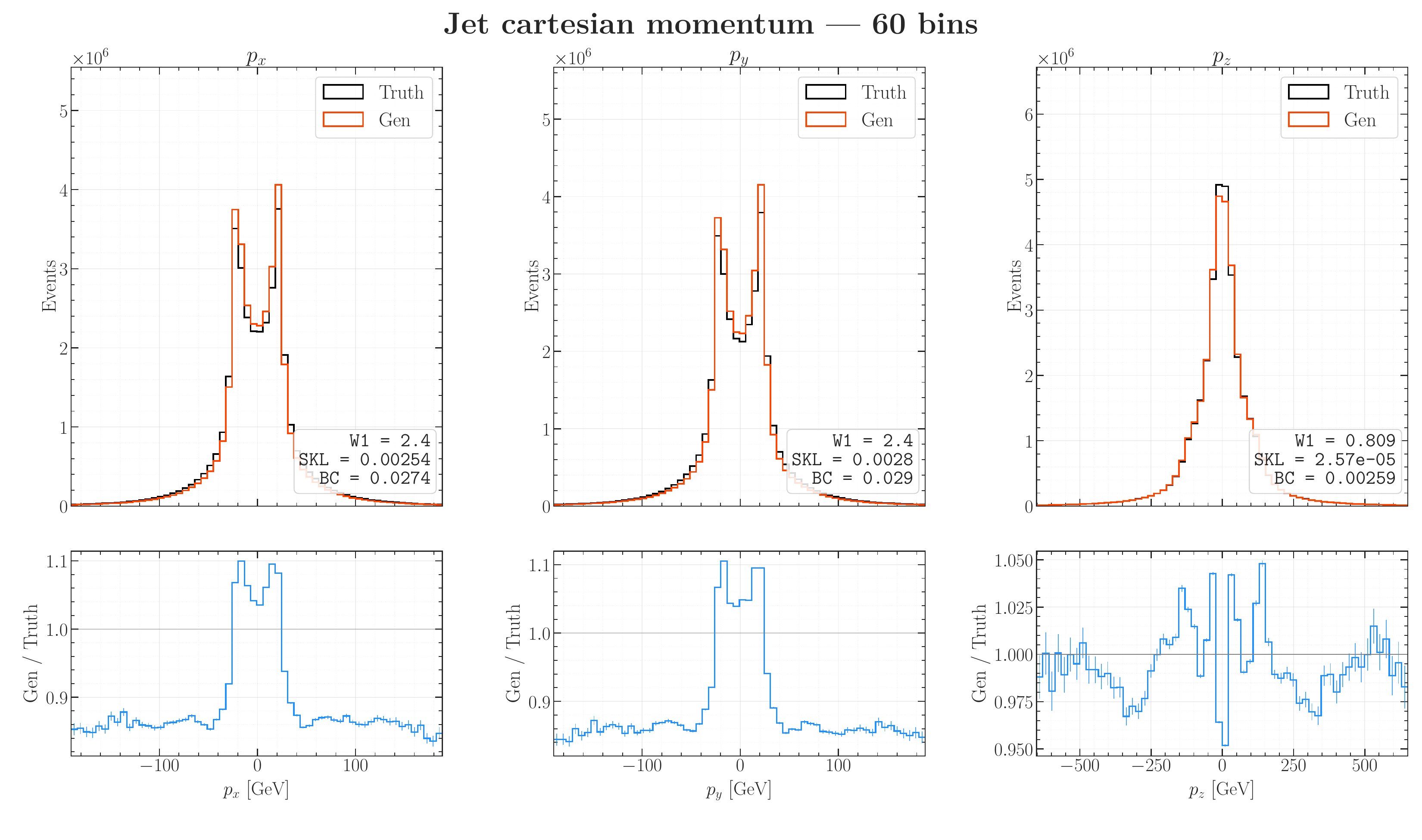}
  \captionof{figure}{\label{fig:cartesian_jet}%
    \textbf{Small-$R$ jet Cartesian components.}
    Small-$R$ jet Cartesian marginals $(p_x, p_y, p_z)$.
    Truth vs.\ generated.}
\end{center}
\twocolumngrid

\onecolumngrid
\begin{center}
  \includegraphics[width=\textwidth]{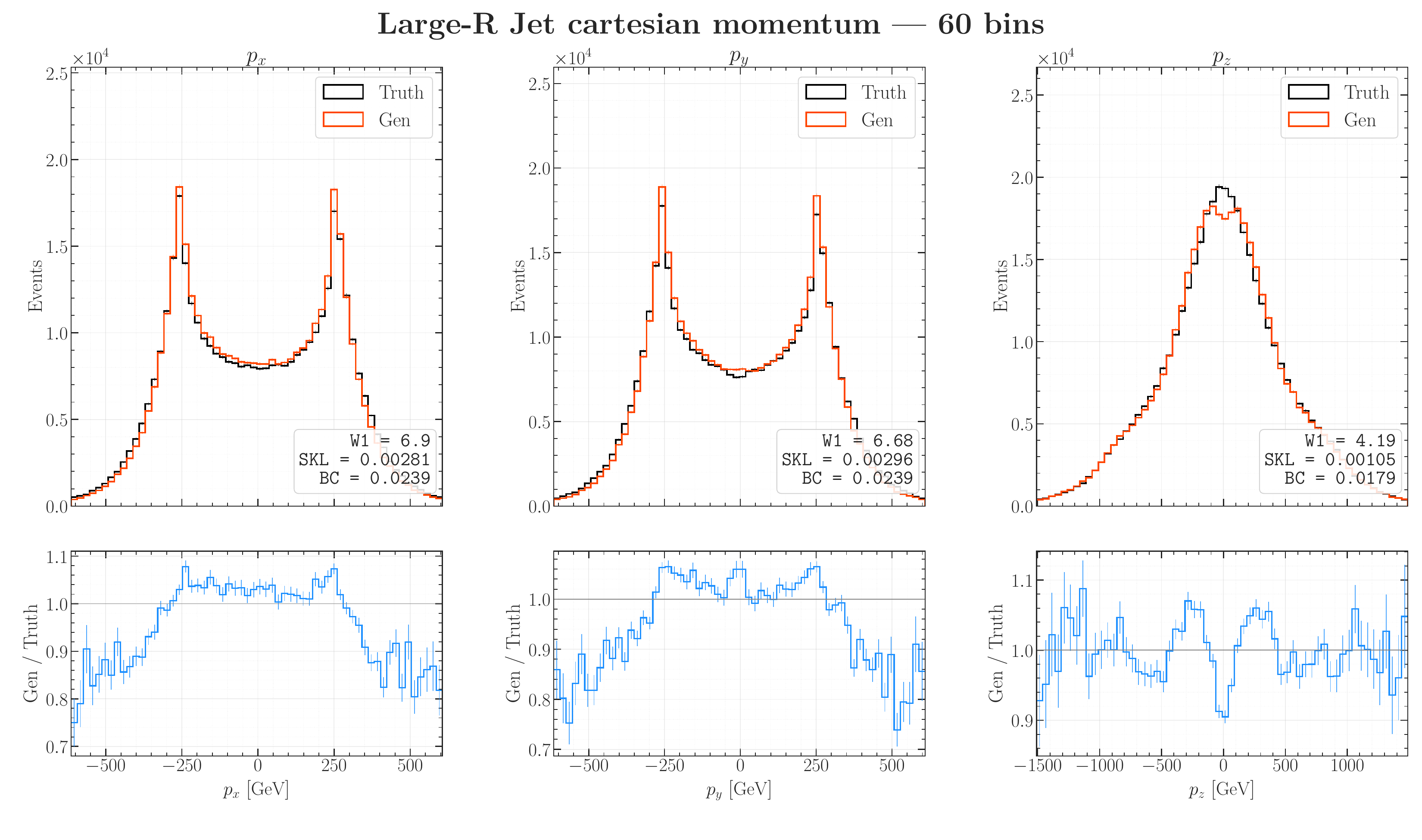}
  \captionof{figure}{\label{fig:cartesian_largerjet}%
    \textbf{Large-$R$ jet Cartesian components.}
    Large-$R$ jet Cartesian marginals $(p_x, p_y, p_z)$.
    Truth vs.\ generated.}
\end{center}
\twocolumngrid

\section{Forward-process noising of the dilepton spectrum}
\label{app:noise_grid_dilepton}

This appendix documents how the OSSF dilepton invariant-mass
distribution degrades under the CFM forward noising process at a
sequence of flow times
$t \in \{1,\, 0.99,\, 0.95,\, 0.9,\, 0.8,\, 0.5,\, 0.3,\, 0.1\}$, where
the noising follows
$x_t = t\,x_{\mathrm{data}} + (1 - t)\,z$ with $z \sim \mathcal{N}(0,\,I)$
on the chart, so $t = 1$ is the clean truth distribution and $t = 0$ is
the noise prior. The progression explains why the narrow resonance peaks
of Sec.~\ref{sec:results_dilepton} are the hardest part of the dilepton
spectrum to learn and motivates the KDE-based distribution-matching
weight $w^{\star}$ of Eq.~\eqref{eq:dm_weight}: already at
$t \approx 0.99$ the narrow light quarkonia $\omega(782)$, $\phi(1020)$,
and $\psi(2S)$ are indistinguishable from the surrounding continuum;
by $t \approx 0.9$ the $J/\psi(1S)$ and $\Upsilon$ peaks have followed;
and below $t \approx 0.8$ only the broad $Z$ shoulder remains before it
too dissolves into the Drell--Yan continuum. Once these features are
buried in noise, the only signal the primary CFM loss can use to recover
them at sampling time is their small residual weight in the conditional
velocity field, and the auxiliary $w^{\star}$ weight is what amplifies
that residual at the under-produced mass regions.

\begin{figure*}[!tbp]
  \includegraphics[width=\textwidth]{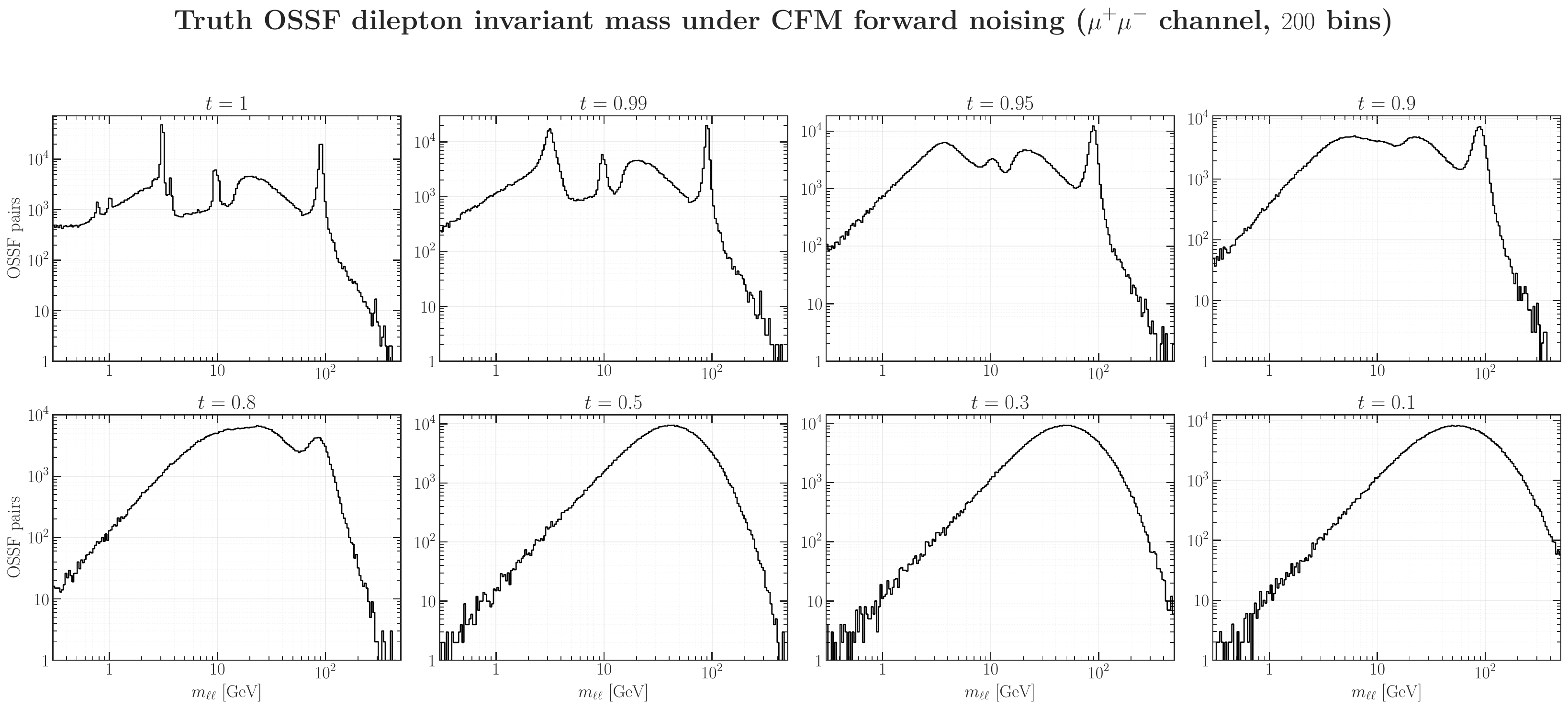}
  \caption{\label{fig:noise_grid_dilepton}%
    \textbf{CFM forward noising of the dimuon spectrum.}
    Truth OSSF dimuon invariant-mass spectrum after application of the
    CFM forward noising
    $x_t = t\,x_{\mathrm{data}} + (1 - t)\,z$ at flow times
    $t \in \{1,\, 0.99,\, 0.95,\, 0.9,\, 0.8,\, 0.5,\, 0.3,\, 0.1\}$
    (clean $\to$ noisy). The narrow light-quarkonium peaks $\omega$ and
    $\phi$ and the $\psi(2S)$ are washed into the continuum already at
    $t \approx 0.99$; the $J/\psi(1S)$ and $\Upsilon$ families follow
    by $t \approx 0.9$; below $t \approx 0.8$ only the broad $Z$
    shoulder remains before it too dissolves into the Drell--Yan
    continuum. \emph{Selection:} same OSSF dilepton selection as
    Fig.~\ref{fig:dilepton}, restricted to the $\mu^{+}\mu^{-}$
    channel (the cleanest channel for resonance visibility); $\sim 4.2\times 10^{5}$
    dimuon OSSF pairs from the validation sample,
    histogrammed with $200$ log-spaced bins between $0.3$ and $500$~GeV.}
\end{figure*}

\section{Three-particle invariant masses}
\label{app:threebody}

\begin{figure*}[!tbp]
  \includegraphics[width=\textwidth]{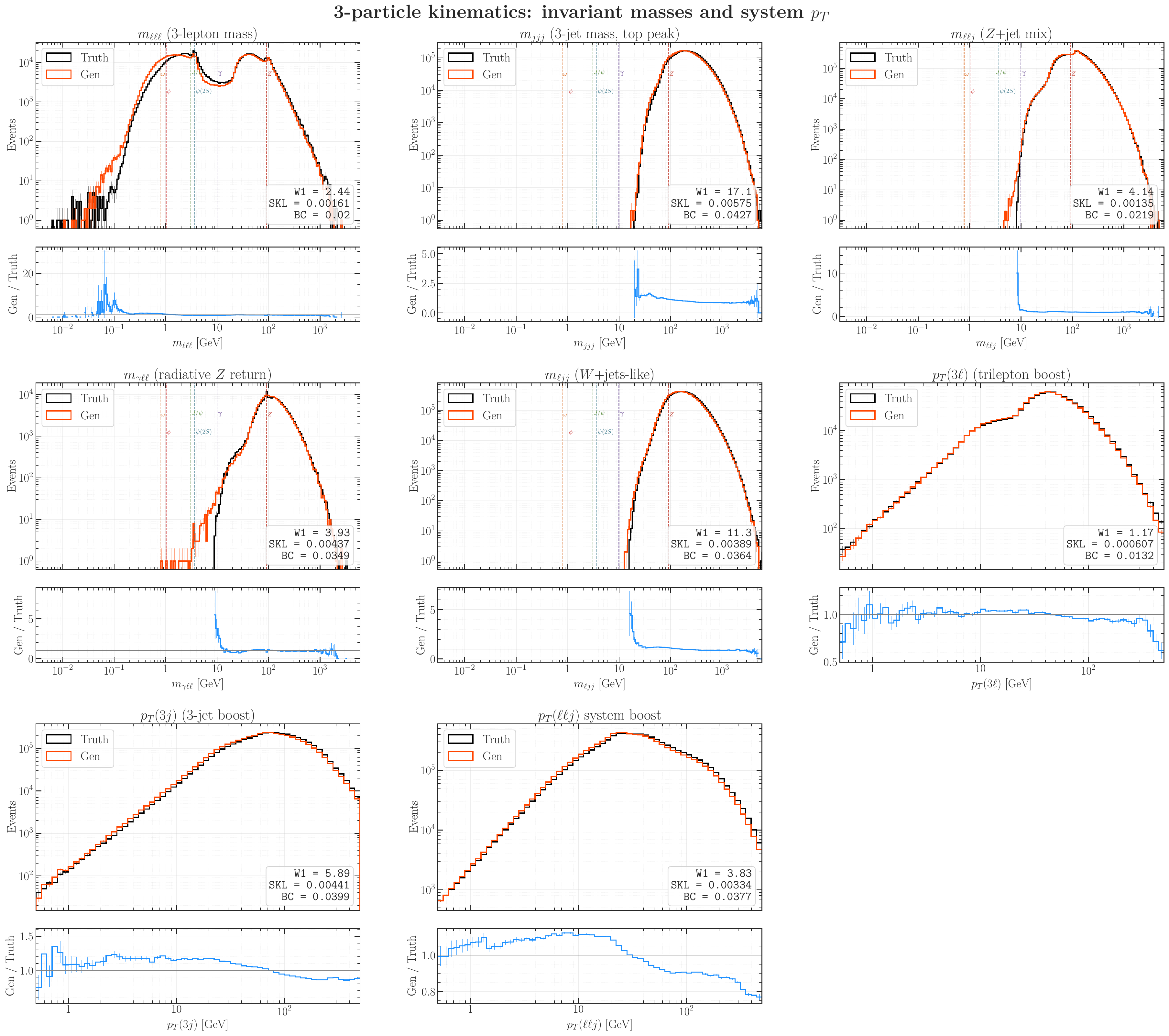}
  \caption{\label{fig:three_particle_grid}%
    \textbf{Three-particle kinematics.}
    Three-particle invariant masses and system $p_T$ spectra,
    truth vs.\ generated (log-log). The lower sub-panel of each pair
    shows the bin-by-bin generated-to-truth ratio. Eight combinations
    are shown: $m_{\ell\ell\ell}$, $m_{\gamma\ell\ell}$,
    $m_{\ell\ell j}$, $m_{jjj}$, $m_{\ell jj}$, and the corresponding
    system $p_T$ distributions. The hadronic top peak at
    $m_{jjj} \approx 173$~GeV emerges above the multijet QCD
    continuum.}
\end{figure*}

This appendix collects the truth-vs.-generated comparison on eight
three-particle combinations summarised in
Sec.~\ref{sec:results_threebody}, shown in
Fig.~\ref{fig:three_particle_grid}.

\section{Per-particle quality score}
\label{app:quality_score}

The raw quality score $q_i$ entering the effective weight
$q_i^{\mathrm{eff}}$ of Eq.~\eqref{eq:q_eff} is a type-conditional
sigmoid product of detector quality features. For each feature
$f$ with cut $c_f$ and slope $\alpha_f$ we write the shorthand
$\sigma_f \equiv \sigma\!\bigl(\alpha_f\,(c_f - f)\bigr)$ (with the sign
flipped when the cut sets a lower bound), so that
$\sigma_f \in (0, 1)$ smoothly tracks the corresponding hard
analysis cut. With this shorthand the per-type score reads
\begin{equation}
\begin{aligned}
  q_e &= q_\mu = \sigma_{\mathrm{iso}}^{\ell}\,
                \sigma_{\mathrm{ip}}^{\ell}, \\
  q_\gamma &= \sigma_{\mathrm{iso}}^{\gamma}, \\
  q_{\mathrm{jet}} &= \sigma_{\mathrm{JVT}}, \\
  q_{\tau_{\mathrm{had}}} &= \sigma_{\mathrm{RNN}}, \\
  q_{\text{large-}R\text{ jet}} &= \sigma_{D_2},
\end{aligned}
\label{eq:quality_score}
\end{equation}
with MET and padding tokens carrying $q = 1$ by convention. The
isolation variables are
$r_{\mathrm{iso}} = \min(\texttt{ptvarcone30},\, \texttt{topoetcone20})/p_T$
for leptons and
$r_{\mathrm{iso}}^{\gamma} = \min(\texttt{topoetcone40},\, \texttt{ptcone20})/p_T$
for photons. The score is non-learned: no gradient flows through
$q_k$, and the cuts $c$ and slopes $\alpha$ collected in
Table~\ref{tab:quality_score} are fixed per-type detector constants
chosen to track the standard ATLAS analysis cuts but with a soft
sigmoid edge rather than a hard step, so that borderline objects are
neither fully accepted nor fully rejected.

\begin{table}[!h]
  \caption{\label{tab:quality_score}%
    Per-type detector quality cuts $c$ and sigmoid slopes $\alpha$
    used in Eq.~\eqref{eq:quality_score}. Lepton isolation
    $r_{\mathrm{iso}}$ and photon isolation $r_{\mathrm{iso}}^{\gamma}$
    are defined in the surrounding text. MET and padding tokens carry
    $q = 1$ by convention and are not listed.}
  \begin{ruledtabular}
    \begin{tabular}{l l c c}
      \textbf{Type} & \textbf{Feature} & $c$ & $\alpha$ \\
      \hline
      Leptons $(e, \mu)$
        & $r_{\mathrm{iso}}$
        & $0.15$ & $10.0$ \\
        & $|d_0/\sigma_{d_0}|$
        & $3.00$ & $1.00$ \\
      Photon
        & $r_{\mathrm{iso}}^{\gamma}$
        & $0.15$ & $10.0$ \\
      Jet & \texttt{jvt} & $0.50$ & $5.00$ \\
      $\tau_{\mathrm{had}}$ & \texttt{RNNJetScore} & $0.50$ & $5.00$ \\
      Large-$R$ jet & $D_2$ & $1.50$ & $5.00$ \\
    \end{tabular}
  \end{ruledtabular}
\end{table}

\section{Training configuration and computing environment}
\label{app:training_details}

Table~\ref{tab:hyperparams} collects the architecture and training
hyperparameters of the single training run analysed throughout this
paper. All values are quoted from the logged configuration of that
run. The loss constants refer to the symbols of
Sec.~\ref{sec:method_architecture}: the Huber scale $\delta$ and the
gate $t_{\mathrm{gate}}$ of Eq.~\eqref{eq:aux_loss}, the
distribution-matching exponent $\alpha$, sharpness $k$, and
truth-prior mix $\eta$ of Eq.~\eqref{eq:dm_weight}, and the quality
floor $q_{\mathrm{floor}}^{\mathrm{cfm}}$ of Eq.~\eqref{eq:q_eff}.

\begingroup
\definecolor{groupbg}{RGB}{226,232,240}%   slate-100, group header band
\newcommand{\hpgroup}[1]{\rowcolor{groupbg}\multicolumn{2}{l}{\textit{#1}}}%
\begin{table}[!h]
  \caption{\label{tab:hyperparams}%
    Architecture and training hyperparameters of the reported run.}
  \begin{ruledtabular}
    \begin{tabular}{l r}
      \textbf{Hyperparameter} & \textbf{Value} \\
      \hline
      \hpgroup{Architecture} \\
      DiT blocks $L$ & $6$ \\
      model width $d_{\mathrm{model}}$ & $128$ \\
      attention heads & $4$ \\
      feed-forward expansion ratio & $4$ \\
      MET cross-attention tokens & $4$ \\
      dropout & $0$ \\
      trainable parameters & $3.0 \times 10^{6}$ \\
      \hpgroup{Optimiser (AdamW)} \\
      peak learning rate & $6 \times 10^{-4}$ \\
      weight decay & $0.01$ \\
      $(\beta_1,\, \beta_2)$ & $(0.9,\, 0.999)$ \\
      $\varepsilon$ & $10^{-8}$ \\
      gradient clip (global norm) & $1.0$ \\
      EMA decay $\rho$ & $0.9995$ \\
      \hpgroup{Learning-rate schedule} \\
      type & cosine \\
      warm-up & $10^{4}$ steps, linear \\
      minimum learning rate & $10^{-6}$ \\
      epochs & $30$ \\
      total optimiser steps & $717{,}180$ \\
      \hpgroup{Data and batching} \\
      batch size per GPU & $4{,}096$ \\
      global batch size & $32{,}768$ \\
      train\,/\,validation split & $0.95\,/\,0.05$ \\
      split seed & $42$ \\
      \hpgroup{Loss} \\
      Huber scale $\delta$ & $1.0$ \\
      $K$-body orders & $K \in \{2, 3, 4\}$, equal weight \\
      position-term gate $t_{\mathrm{gate}}$ & $0.99$ \\
      exponent $\alpha$ & $1.0$ \\
      sharpness $k$ & $5$ \\
      truth-prior mix $\eta$ & $1/3$ ($w^{\star}_{\max} = 3$) \\
      quality floor $q_{\mathrm{floor}}^{\mathrm{cfm}}$ & $0.5$ \\
      \hpgroup{Flow-time sampling} \\
      distribution & $t \sim \mathcal{U}[t_{\min},\, t_{\max}]$ \\
      $(t_{\min},\, t_{\max})$ & $(0.01,\, 1 - 10^{-6})$ \\
      \hpgroup{Compute} \\
      GPUs & $8$, data-parallel \\
      numerical precision & \texttt{bf16-mixed} \\
      compilation & \texttt{torch.compile} \\
    \end{tabular}
  \end{ruledtabular}
\end{table}
\endgroup

The run executes on a single dedicated node whose hardware and
software environment is summarised in Table~\ref{tab:system}. The
full $30$-epoch schedule completes in approximately $61$~h of
wall-clock time.

\begin{table}[!h]
  \caption{\label{tab:system}%
    Hardware and software environment of the training node.}
  \begin{ruledtabular}
    \begin{tabular}{l l}
      \textbf{Component} & \textbf{Specification} \\
      \hline
      GPU & $8\times$ NVIDIA A100 ($80$~GB), NVLink \\
      CPU & $2\times$ AMD EPYC 7J13 ($128$ cores) \\
      System memory & $2$~TB \\
      Storage & $19$~TB NVMe RAID array \\
      Software & PyTorch $2.10$ (CUDA $12.8$), Linux \\
    \end{tabular}
  \end{ruledtabular}
\end{table}

\section{Selected input fields}
\label{app:selected_fields}

Table~\ref{tab:selected_fields} lists the ATLAS Open Data fields
retained by the preprocessing pipeline, grouped by reconstructed
object type and named as in the source ntuples. The four kinematic
fields ($p_T$, $\eta$, $\phi$, $E$) are common to all types and
carry the generated state. The remaining fields are type-specific
detector quantities that enter the model as conditioning, through
the conditioning streams of Sec.~\ref{sec:method_architecture} and
the per-particle quality score of Appendix~\ref{app:quality_score}.

\begingroup
\definecolor{groupbg}{RGB}{226,232,240}%   slate-100, group header band
\definecolor{condblue}{HTML}{3776CC}%      conditioning field names
\newcommand{\fgroup}[1]{\rowcolor{groupbg}\multicolumn{2}{l}{\textit{#1}}}%
\newcommand{\cfield}[1]{\textcolor{condblue}{\texttt{#1}}}%
\newcommand{\fdesc}[1]{\parbox[t]{0.52\columnwidth}{\raggedright #1\strut}}%
\renewcommand{\arraystretch}{0.95}%
\begin{table}[!h]
  \caption{\label{tab:selected_fields}%
    Selected ATLAS Open Data fields per particle type. Fields in
    blue enter the model as conditioning inputs, while the remaining
    kinematic fields are generated on the per-type charts of
    Sec.~\ref{sec:method_manifolds}. The event metadata does not
    enter the model.}
  \footnotesize
  \begin{ruledtabular}
    \begin{tabular}{l l}
      \textbf{Field} & \textbf{Description} \\
      \hline
      \fgroup{Event metadata} \\
      \texttt{runNumber}   & \fdesc{ATLAS run number} \\
      \texttt{eventNumber} & \fdesc{Event number within the run; with the
                             run number it uniquely identifies the
                             event} \\
      \fgroup{Missing transverse energy} \\
      \cfield{met}         & \fdesc{Magnitude of the missing transverse
                             momentum} \\
      \cfield{met\_phi}    & \fdesc{Azimuthal angle of the missing
                             transverse momentum} \\
      \fgroup{Leptons ($e$, $\mu$)} \\
      \texttt{lep\_pt}     & \fdesc{Transverse momentum $p_T$} \\
      \texttt{lep\_eta}    & \fdesc{Pseudorapidity $\eta$} \\
      \texttt{lep\_phi}    & \fdesc{Azimuthal angle $\phi$} \\
      \texttt{lep\_e}      & \fdesc{Energy $E$} \\
      \cfield{lep\_charge} & \fdesc{Electric charge} \\
      \cfield{lep\_z0}     & \fdesc{Longitudinal track coordinate $z_0$ at
                             the primary vertex} \\
      \cfield{lep\_d0}     & \fdesc{Transverse impact parameter $d_0$} \\
      \cfield{lep\_ptvarcone30}  & \fdesc{Variable-cone track isolation,
                             $\Delta R = 0.3$~\cite{ATLAS:2024vdo}} \\
      \cfield{lep\_topoetcone20} & \fdesc{Topological calorimeter isolation,
                             $\Delta R = 0.2$~\cite{ATLAS:2024vdo}} \\
      \cfield{lep\_d0sig}  & \fdesc{Significance of $d_0$} \\
      \fgroup{Photon} \\
      \texttt{photon\_pt}  & \fdesc{Transverse momentum $p_T$} \\
      \texttt{photon\_eta} & \fdesc{Pseudorapidity $\eta$} \\
      \texttt{photon\_phi} & \fdesc{Azimuthal angle $\phi$} \\
      \texttt{photon\_e}   & \fdesc{Energy $E$} \\
      \cfield{photon\_charge} & \fdesc{Set to zero; aligns the slot with
                             the lepton block} \\
      \cfield{photon\_topoetcone40} & \fdesc{Topological $E_T$ isolation,
                             $\Delta R = 0.4$} \\
      \cfield{photon\_ptcone20} & \fdesc{Track isolation, $\Delta R = 0.2$} \\
      \fgroup{Hadronically decaying tau} \\
      \texttt{tau\_pt}     & \fdesc{Transverse momentum $p_T$} \\
      \texttt{tau\_eta}    & \fdesc{Pseudorapidity $\eta$} \\
      \texttt{tau\_phi}    & \fdesc{Azimuthal angle $\phi$} \\
      \texttt{tau\_e}      & \fdesc{Energy $E$} \\
      \cfield{tau\_charge} & \fdesc{Electric charge} \\
      \cfield{tau\_nTracks} & \fdesc{Charged tracks in the decay
                             ($1$- or $3$-prong)} \\
      \cfield{tau\_RNNJetScore} & \fdesc{RNN score, true tau against
                             fake-tau jets} \\
      \fgroup{Small-$R$ jet} \\
      \texttt{jet\_pt}     & \fdesc{Transverse momentum $p_T$} \\
      \texttt{jet\_eta}    & \fdesc{Pseudorapidity $\eta$} \\
      \texttt{jet\_phi}    & \fdesc{Azimuthal angle $\phi$} \\
      \texttt{jet\_e}      & \fdesc{Energy $E$} \\
      \cfield{jet\_btag\_quantile} & \fdesc{$b$-tagging score as a
                             working-point quantile} \\
      \cfield{jet\_jvt}    & \fdesc{Jet Vertex Tagger score (pile-up
                             suppression)} \\
      \fgroup{Large-$R$ jet} \\
      \texttt{largeRJet\_pt}  & \fdesc{Transverse momentum $p_T$} \\
      \texttt{largeRJet\_eta} & \fdesc{Pseudorapidity $\eta$} \\
      \texttt{largeRJet\_phi} & \fdesc{Azimuthal angle $\phi$} \\
      \texttt{largeRJet\_e}   & \fdesc{Energy $E$} \\
      \texttt{largeRJet\_m}   & \fdesc{Invariant mass} \\
      \cfield{largeRJet\_D2}  & \fdesc{$D_2$ substructure variable
                             ($W$/$Z$ tagger)} \\
    \end{tabular}
  \end{ruledtabular}
\end{table}
\endgroup

\FloatBarrier% flush all pending appendix floats before bibliography

\bibliography{bibliography}

\end{document}